\documentclass[a4paper,11pt]{article}
\usepackage{jcappub}

\pdfoutput=1

\usepackage{tabularx,booktabs}
\newcolumntype{Y}{>{\centering\arraybackslash}X}

\usepackage{fontawesome}
\usepackage[utf8]{inputenc}
\usepackage[font=small,labelfont=bf,tableposition=top]{caption}
\usepackage{float}
\usepackage{adjustbox}
\usepackage{arydshln}
\usepackage{empheq}

\usepackage{amsmath,amssymb,amsbsy,amstext, amsthm, simplewick, amsfonts, bm}
\usepackage{hyperref}
\usepackage[normalem]{ulem}
\usepackage{graphicx}
\usepackage{siunitx}
\usepackage{upgreek}
\usepackage{framed}
\usepackage{wrapfig}
\usepackage{multirow}
\usepackage{subfig}
\usepackage{bbm}
\usepackage[nameinlink, capitalise]{cleveref}
\usepackage[svgnames,dvipsnames,x11names]{xcolor}
\usepackage{wrapfig}
\usepackage{ulem}
\usepackage[htt]{hyphenat}
\usepackage[normalem]{ulem}
\allowdisplaybreaks
\usepackage{soul}
\usepackage[top=2.5cm,bottom=3.cm,right=2.5cm,left=2.5cm]{geometry}
\newcommand{\class}{{\sc class}}
\newcommand{\classoneloop}{{\sc class-OneLoop}}

\title{Effective Field Theory of Large Scale Structure and Newtonian Motion Gauges}

\author[a]{Christian Fidler}
\author[a]{, Julien Lesgourgues}
\author[a]{, Antonia Mattes}
\author[b]{, Azadeh Moradinezhad Dizgah}
\author[a]{, Simon Neuland}

\affiliation[a]{Institute for Theoretical Particle Physics and Cosmology (TTK), RWTH Aachen University, D-52056 Aachen, Germany}
\affiliation[b]{Laboratoire d’Annecy de Physique Theorique (LAPTh), CNRS/USMB, 99 Chemin de Bellevue BP110 - Annecy - F-74941 - France}

\emailAdd{fidler@physik.rwth-aachen.de}
\emailAdd{lesgourg@physik.rwth-aachen.de}
\emailAdd{antonia.mattes@rwth-aachen.de}
\emailAdd{azadeh.moradinezhad@lapth.cnrs.fr}

\abstract{The simplest flavor of the Effective Field Theory of Large Scale Structure (EFTofLSS) is based on Newtonian equations
and describes the nonlinear matter density and velocity using Einstein-de-Sitter (EdS) kernels. 
Even in the presence of massive neutrinos, for which these assumptions are clearly not valid, this simplest flavor has been argued to be sufficient for the analysis of data from Stage-III galaxy surveys like BOSS. In this paper, we show that there exists a simple way to extend the validity range of this framework to more complex problems with a scale-dependent growth factor, while incorporating linear general relativistic (GR) corrections as well. For a given cosmology, an Einstein-Boltzmann code can find the exact gauge transformation that brings the full linear equations of motion of the clustering matter components into a form where they are identical to Newtonian equations for a self-gravitating fluid with scale-independent growth. Non-linear clustering can be consistently computed in this gauge, and the results can be transformed back to the initial gauge in order to incorporate consistently GR and scale-dependent-growth effects. Redshift-space distortions can also be incorporated with a similar strategy. Our method does not incur any additional computational cost. As a showcase, we apply this method to cosmologies with massive neutrinos. 
For the real-space one-loop power spectrum, we find that the largest deviation between the accurate and standard methods remains below 0.7\% for $\sum m_\nu<0.30\,{\rm eV}$. However, in redshift space, it reaches $1.7\%$ for the one-loop quadrupole spectrum at $k=0.3\,h{\rm Mpc}^{-1}$ and $z=0$, with the largest contribution coming from the effect of the cosmological constant on the growth of the velocity field.
Our method could be applied to a much wider range of models with more significant scale-dependent growth, as long as a self-consistency condition evaluated by the Einstein-Boltzmann code (on the smallness of a gauge transformation field) is fulfilled.}

\usepackage{amsmath}
\usepackage{graphicx}

\begin{document}

\subheader{
\footnotesize
\vspace*{-3.7em}
\begin{flushright}
TTK-25-12\\
LAPTH-022/26
\end{flushright}
}

\date{\today}
\maketitle

\newpage

\section{Introduction}

Ongoing and forthcoming Stage‑IV galaxy surveys—Dark Energy Spectroscopic Instrument (DESI)\cite{DESI:2016fyo}, Euclid\cite{Euclid:2024yrr}, SPHEREx\cite{SPHEREx:2014bgr}, the Nancy Grace Roman Space Telescope \cite{Wang:2021oec}, the Prime‑Focus Spectrograph (PFS) \cite{PFSTeam:2012fqu}, and the Rubin Observatory LSST \cite{LSSTDarkEnergyScience:2012kar}—will deliver unprecedented volumes of high‑precision measurements of \emph{galaxy clustering} and \emph{weak lensing} statistics. These data will permit sensitive tests of physics beyond the minimal $\Lambda$ Cold Dark Matter ($\Lambda$CDM) paradigm, including the imprint of free‑streaming massive neutrinos, exotic dark‑sector interactions, and departures from General Relativity on cosmological scales.  Exploiting the full constraining power of these datasets demands theoretical predictions of the observables that remain accurate on mildly non‑linear scales yet fast enough for iterative Bayesian inference.  The challenge becomes acute when the linear growth factor is \emph{scale‑dependent}, invalidating the Einstein–de Sitter assumptions underlying the conventional formulation of the EFTofLSS \cite{Baumann:2010tm,Carrasco:2012cv,Perko:2016puo,Senatore:2017pbn,Cabass:2022avo}.  As we show below, the Newtonian Motion (NM) gauge framework \cite{Fidler:2015npa,Fidler:2016tir,Fidler:2017pnb} absorbs this scale dependence into a controlled gauge transformation, such that the dynamics can be evolved with standard Newtonian tools while retaining percent‑level accuracy in the observables of interest.

In recent years, the perturbative modeling of large‑scale structure within the EFTofLSS framework has reached a level of maturity that allows predictions for matter and biased‑tracer power spectra (and bispectra) to be computed with percent‑level accuracy on mildly nonlinear scales. By systematically including all symmetry‑allowed long‑wavelength operators, absorbing UV‐sensitive short‑scale contributions into a finite set of counterterms, and incorporating infrared resummation, nonlinear bias expansions, and redshift‑space distortions---together with efficient algorithms for the evaluation of loops \cite{McEwen:2016fjn,Schmittfull:2016jsw,Simonovic:2017mhp}---the EFTofLSS now underlies full‑shape analyses of galaxy survey data, yielding cosmological constraints from SDSS BOSS and eBOSS power spectrum \cite{Ivanov:2019pdj,Troster:2019ean,DAmico:2019fhj,Chen:2021wdi,Simon:2022csv} and bispectrum \cite{Philcox:2021kcw,DAmico:2022osl,Spaar:2023his} measurements and, most recently, from the DESI-DR1 galaxy power spectrum \cite{DESI:2024jis,Elbers:2025vlz} combined with the bispectrum \cite{Chudaykin:2025lww,Chudaykin:2025vdh,Chudaykin:2025aux,Ivanov:2026dvl,Chudaykin:2026nls} to constrain the $\Lambda$CDM model. 

Beyond the vanilla $\Lambda$CDM case, the same perturbative machinery extends naturally to several extended models. Modifications of only the background expansion or initial conditions---such as the $w$CDM model with constant equation of state of Dark Energy (DE) \cite{DAmico:2020kxu}, nonzero curvature \cite{Chudaykin:2020ghx}, warm or fuzzy Dark Matter (DM) \cite{Kousha:2023jzt}, and interacting DM-neutrinos scenario \cite{Mosbech:2024wxr}---require no new kernels, only appropriate linear inputs or specific counterterm values. Models with scale‑independent but non‑EdS growth \cite{Carrasco:2012cv,Donath:2020abv,Zhang:2021uyp}---most notably clustering quintessence \cite{Lewandowski:2016yce,DAmico:2020tty,Lu:2025gki} and quasi‑static Horndeski theories \cite{Cusin:2017wjg}---retain the standard EdS mode-coupling kernels but demand exact Green’s‑function time‑kernels and time‑dependent EFTofDE coefficients (with degenerate higher-order scalar-tensor theories (DHOST) \cite{Hirano:2022yss} even adding a few extra Poisson‑equation operators). Finally, some models break the EdS factorization and modify both the spatial mode‑coupling kernels and their time‑dependence. Two prime examples are: (i) screened modified‑gravity theories (e.g., $f(R)$, where a scale‑ and time‑dependent correction factor appears in the Poisson equation \cite{Rodriguez-Meza:2023rga,Aviles:2024zlw}), and (ii) massive‑neutrino cosmologies, where free‑streaming induces a scale‑dependent growth of structure and requires in principle a description in terms of a two-fluid system \cite{Senatore:2017hyk, Garny:2020ilv, Verdiani:2025jcf}. In both cases one must go beyond the building blocks of the EdS mode-coupling kernels with scale‑independent weights and instead incorporate fully temporal- and spatial-dependent kernels in the loop integrals. 

In principle, models featuring a scale-dependent growth at the linear level require a non-trivial treatment of the time and/or spatial dependence in loop integrals---yet the core EFTofLSS formalism remains the same. Our goal in this paper is to show that the NM‑gauge framework can be applied within the EFTofLSS in order to simplify the modeling of the one-loop matter power spectrum in such extensions of $\Lambda$CDM. We focus primarily on cosmologies with massive neutrinos as a showcase for our method.

Newtonian motion (NM) gauges provide a simplified framework for cosmological analysis by reducing the full relativistic equations to Newtonian dynamics. This allows, for example, to make a more consistent use of N-body simulations: one can define initial conditions, e.g., in the synchronous gauge, go to the NM gauge to run N-body simulations, and gauge-transform the results back to the synchronous gauge in order to capture general relativistic (GR) corrections\footnote{In these references and in the present work, when mentioning GR corrections, we always refer to the linear GR corrections coming from the presence of metric fluctuations in the Einstein and continuity equations, in addition to the Laplacian of the gravitational potential. These terms become important on scales comparable to (or bigger than) the Hubble radius, and are neglected in the Newtonian limit. We do not refer to the relativistic corrections to the harmonic or Fourier power spectrum of galaxy number count that are discussed in refs.~\cite{Yoo:2009au,Yoo:2010ni,Bonvin:2011bg,Challinor:2011bk,DiDio:2013bqa,Castorina:2021xzs,Foglieni:2023xca}. Our approach could in principle be extended to this observable. However, in this work, here we only target the calculation of the matter power spectrum in Fourier (and redshift) space.} with excellent accuracy \cite{Fidler:2015npa,Fidler:2016tir}. This approach requires that the coordinate displacement field accounting for the gauge transformation remains small, such that it can be captured by first-order perturbation theory. The method is then self-consistent and accurate, even if the matter perturbations go deep into the non-linear regime \cite{Fidler:2017pnb}.

The usefulness of the NM gauge can be pushed even further in the case of cosmologies with a scale-dependent linear growth factor (for instance, due to massive neutrinos, non-standard dark matter models, or modified gravity). Indeed, in many such cases, a small gauge transformation is also able to absorb the scale dependence. As a consequence, in the NM gauge, structure formation is described by equations which are not only Newtonian, but also scale-independent. Therefore, by going from the synchronous gauge to the NM gauge and back, one can infer accurate non-linear predictions for non-trivial cosmologies from simple N-body simulations featuring only standard cold dark matter (CDM) particles \cite{Fidler:2018bkg,Heuschling:2022rae}. As an example of application, several works have compared the non-linear power spectrum of cosmological models featuring massive neutrinos obtained either with this method or with more expensive simulations featuring both cold and hot dark matter particles. The results were found to be in excellent agreement \cite{Euclid:2022qde}. 

In this work, we apply NM gauges to the EFTofLSS rather than N-body simulations. The standard EFTofLSS framework assumes Newtonian gravity in an Einstein-de Sitter (EdS) universe, where time and scale dependencies are separable. NM gauges may overcome this limitation by absorbing the scale dependence induced by massive neutrinos, non-minimal dark matter, or modified gravity into the gauge definition, allowing the compute the EFTofLSS loops without added computational complexity. Instead of accounting for deviations from $\Lambda$CDM with a more complicated perturbative expansion and EFT formalism, we capture them with gauge transformations which are used, first, for defining appropriate initial conditions, and second, for post-processing the EFTofLSS results obtained in the NM gauge back to a more standard gauge suited to describe the results of observations.   

The remainder of the paper is structured as follows. Section \ref{sec:NMapp} fixes our notation, reviews the gauges most often used in cosmological perturbation theory, and shows how NM gauge perturbations can be constructed from a Newtonian description. Sections \ref{sec:apply_NM_massless} and \ref{sec:apply_NM_massive} embed NM gauges within the EFTofLSS for cosmologies with, respectively, massless and massive neutrinos. In section \ref{sec:rsd} we compare our NM gauge computation of the redshift‑space matter power spectrum with the standard EFTofLSS approach. Finally, section \ref{sec:conclusion} summarizes our results and outlines possible extensions of the method.

\section{Newtonian motion gauges and their applications}\label{sec:NMapp}

\subsection{Gauge definitions\label{sec:gauges}}

To follow this work, it is essential to understand the difference between a few gauges. Some  are widely known and used, like the Poisson gauge, the synchronous gauge and the comoving gauge. The Newtonian motion gauge, which plays a central role in this work, is less frequently used; The N-boisson gauge can be seen as intermediate step to understand the Newtonian motion gauge. In this subsection, we recall the definition of these gauges, specify the relations between them, and define our notations for metric fluctuations, gauges, and quantities in the non-relativistic limit.

In a general gauge, the perturbed FLRW metric, including only scalar perturbations and using conformal time as the temporal variable, is given by \cite{Bardeen:1980kt,Kodama:1984ziu,Hu:2003hjx,Malik:2008im,Durrer:2020fza}
\begin{align}
g_{00} &= -a^2 (1+2A)~,\\
g_{0i} &= -a^2 \hat{\nabla}_i B~,\\
g_{ij} &= a^2 \left[ \delta_{ij} (1+2H_{\rm L})
-2\left( \hat{\nabla}_i \hat{\nabla}_j + \frac{\delta_{ij}}{3} \right) H_{\rm T} \right]~,
\end{align}
where $\hat{\nabla}_i = - (-\nabla^2)^{-1/2} \nabla_i$ is the normalized gradient operator, given by $-ik_i/k$ in Fourier space. In this section we will often refer to two gauge-invariant combinations of these metric perturbations, the Bardeen potentials $\Psi$ and $\Phi$ \cite{Bardeen:1980kt}. We will also use the gauge-invariant curvature perturbation, $\zeta$, built from metric fluctuations and from the scalar degree of freedom contained in the total velocity field. In the notations of Ma \& Bertschinger (MB) \cite{Ma:1997cv} and of \class{} \cite{Blas:2011rf}, the latter is accounted for by the velocity divergence $\theta_{\rm tot}$. Then, the curvature perturbation reads in Fourier space
\begin{equation}
    \zeta 
    = H_{\rm L} + \frac{1}{3} H_{\rm T} - \frac{\cal H}{k^2} \theta_{\rm tot} + \frac{\cal H}{k} B~,
\end{equation}
with ${\cal H}=\dot{a}/a$, and a dot denotes a derivative w.r.t. conformal time, $\partial_\tau$.

In this work, we will refer to the following gauges:\vspace{-0.07in}
\begin{itemize}
\item {\bf Poisson gauge}: We use a superscript $^{\rm (P)}$ to denote quantities in this gauge, e.g., the density fluctuation of a species $x$ is referred to as $\delta^{\rm (P)}_x$. In this gauge, the metric is diagonal, $B^{\rm (P)}=H_{\rm T}^{\rm (P)}=0$, with perturbations $(\psi, \phi)=(A^{\rm (P)}, -H_{\rm L}^{\rm (P)})$ in the notations of MB and \class, and Bardeen potentials given by $(\Psi, \Phi) = (\psi, -\phi)=(A^{\rm (P)}, H_{\rm L}^{\rm (P)})$. In MB and  \class, this gauge is called the (conformal) Newtonian gauge. However, in this work, we refrain from using such a name, in order to avoid confusion with quantities defined in the full Newtonian limit of the theory.
\item {\bf Synchronous gauge} comoving with CDM: We use a superscript $^{\rm (S)}$ to denote quantities in this gauge, e.g., the density fluctuation of a species $x$ is referred to as $\delta^{\rm (S)}_x$. In that gauge, one imposes $A^{\rm (S)}=B^{\rm (S)}=0$ and $\theta^{\rm (S)}_{\rm c}=0$.  The remaining perturbations are $(H_{\rm L}^{\rm (S)}, H_{\rm T}^{\rm (S)})$, or in the notations of MB and \class: 
\begin{equation}
    (h,\eta)=\left(6 H_{\rm L}^{\rm (S)}\, , -H_{\rm L}^{\rm (S)} -\frac{1}{3} H_{\rm T}^{\rm (S)} \right)~.
\end{equation}
The Bardeen potentials are given in Fourier space by
\begin{equation}
    (\Psi, \Phi) = 
    \left( \frac{\dot{a}}{a} \alpha + \dot{\alpha}\,,
    \,\, - \eta + \frac{\dot{a}}{a} \alpha
    \right)
    =\left( \frac{\dot{a}}{a} \alpha + \dot{\alpha}\,,\,\, H_{\rm L}^{\rm (S)} + \frac{1}{3} H_{\rm T}^{\rm (S)} + \frac{\dot{a}}{a}  \alpha \right)~,
\end{equation}
with
\begin{equation}
\alpha = - \frac{\partial_\tau {H_T^{\rm (S)}}}{k^2} = \frac{\dot{h}+6\dot{\eta}}{2 k^2}~.
\end{equation}
In Fourier space, densities and velocities in the Poisson and synchronous gauge are related through
\begin{equation}
    \delta^{\rm (S)}_x = \delta^{\rm (P)}_x - \frac{{\dot{\bar{\rho}}_x}}{\bar{\rho}_x} \alpha ~, \qquad
    \theta^{\rm (S)}_x = \theta^{\rm (P)}_x - k^2  \alpha~.
\end{equation}
\item {\bf Comoving gauge}: We use a superscript $^{\rm (C)}$ to denote quantities in this gauge, e.g., the density fluctuation of a species $x$ is referred to as $\delta^{\rm (C)}_x$. In the comoving gauge, the total 3-velocity is orthogonal to equal-time hypersurfaces, $(\theta_{\rm tot}^{\rm (C)}-kB^{\rm (C)})=0$. To fix the gauge uniquely, one needs a second  condition. A common choice is to impose also $H_{\rm T}^{\rm (C)}=0$ \cite{Hu:2003hjx}.
An advantage of the comoving gauge is that it connects directly to observations. Observed quantities are by definition gauge-invariant. The directly observable density fluctuation of a field $x$ can be interpreted as a gauge-invariant quantity $D_x$ \cite{Yoo:2009au,Yoo:2010ni,Bonvin:2011bg,Challinor:2011bk,DiDio:2013bqa},
\begin{equation}
    D_x =
    \delta^{\rm (G)}_x 
    - \frac{\dot{\rho}_x}{\rho_x} k^{-2} \left(\theta_{\rm tot}^{\rm (G)}-k B^{\rm (G)}\right)~,
    \label{eq:def_D}
\end{equation}
where G denotes an arbitrary gauge. This $D_x$ coincides with the density fluctuation in the comoving gauge, $\delta^{\rm (C)}_x$. Similarly, it has been shown by the authors of refs.~\cite{Jeong:2011as,Bonvin:2011bg,Jeong:2014ufa} that the velocity field relevant for the calculation of redshift-space distortions is described by the gauge-independent velocity divergence
\begin{equation}
    V_x=\theta_x^{\rm (G)}-\dot{H}_{\rm T}^{\rm (G)}~, \label{eq:def_V}
\end{equation}
where G is an arbitrary gauge. It turns out that $V_x$ coincides with $\theta_x^{\rm (C)}$ (and also with $\theta_x^{\rm (P)}$, since $H_{\rm T}$ vanishes in both comoving and Poisson gauges). Thus, in the comoving gauge, both the density and velocity fields connect directly to observables. It is also worth stressing that in the special case where $x$ is non-relativistic, ${\dot{\rho}_x}/{\rho_x}=-3{\cal H}$, the relation between the comoving and synchronous gauge densities reads
\begin{equation}
    D_x = \delta^{\rm (C)}_x =
    \delta^{\rm (S)}_x 
    + 3 \frac{{\cal H}}{k^2} \theta_{\rm tot}^{\rm (S)}~.
    \label{eq:D_to_G_and_S}
\end{equation}
During matter domination, the second term on the right-hand side is extremely small compared to the fluctuation $\delta^{\rm (S)}_x$ of non-relativistic species, because $\theta_{\rm c}^{\rm (S)}=0$ and $\theta_{\rm tot}^{\rm (S)}$ only receives contribution from sub-dominant relativistic species and the relative baryon-to-CDM velocity. This relative velocity is only significant at very early times and/or very small scales, but not within the range of times and scales relevant for the calculation of EFTofLSS corrections. Thus we will often assume $\delta^{\rm (C)}_x\simeq\delta^{\rm (S)}_x$ for non-relativistic species.
\item {\bf N-boisson (Nb) gauge}: We use a superscript $^{\rm (Nb)}$ to denote quantities in this gauge, e.g., the density fluctuation of a species $x$ is referred to as $\delta^{\rm (Nb)}_x$ \cite{Fidler:2017pnb}. We impose in Fourier space
\begin{align}
    k B^{\rm (Nb)} &= \partial_\tau H_{\rm T}^{\rm (Nb)}~, \label{eq:Nboisson1}\\
    H_{\rm L}^{\rm (Nb)} &= {\cal H} \frac{\theta_{\rm tot}^{\rm (Nb)}}{k^2}~, \label{eq:Nboisson2}
\end{align}
where the first condition enforces the same temporal gauge fixing as in the Poisson gauge, while the latter implies a spatial gauge fixing condition such that $\zeta = \frac{1}{3} H_{\rm T}^{\rm (Nb)} + \frac{\cal H}{k^2} \partial_\tau H_{\rm T}^{\rm (Nb)}$. Additionally, in this gauge, there is an attractor solution on super-Hubble scales such that $H_{\rm T}^{\rm (Nb)}$ becomes constant and
\begin{equation} \label{eq:zetaNb}
\zeta = \frac{1}{3} H_{\rm T}^{\rm (Nb)}~.
\end{equation} 
In this gauge, one finds that the particle trajectories computed according to the full relativistic equations coincide with those derived from Newtonian equations whenever the density and pressure of radiation (photons and neutrinos) are neglected. In other words, in the no-radiation limit, velocities are the same in the Nb gauge and in the purely Newtonian framework. This gauge is thus implicitly assumed in N-body simulations.
\item {\bf Newtonian Motion (NM) gauges}: We use a superscript $^{\rm (NM)}$ to denote quantities in this gauge, e.g., the density fluctuation of a species $x$ is referred to as $\delta^{\rm (NM)}_x$ \cite{Fidler:2016tir,Fidler:2017pnb}. These gauges are constructed such that the trajectories of non-relativistic particles undergoing non-linear clustering (such as CDM and baryons), as inferred from either relativistic or Newtonian equations of motion, coincide with each other even when the density and pressure of radiation (and potentially of other smoothly-distributed relics such as massive neutrinos) are included. To achieve this, one needs to stick to the condition in eq. (\ref{eq:Nboisson1}), 
\begin{equation}
    k B^{\rm (NM)} = \partial_\tau H_{\rm T}^{\rm (NM)}~, \label{eq:NM1}\\
\end{equation}
while generalizing eq. (\ref{eq:Nboisson2}) to a differential equation that usually should be solved numerically \cite{Fidler:2016tir,Fidler:2017pnb}:
\begin{align}
    &(\partial_\tau^2 + {\cal H}\partial_\tau) H_{\rm T}^{\rm (NM)} +4 \pi G a^2 \bar{\rho}_{\rm cb} (3\zeta^{\rm (NM)} - H_{\rm T}^{\rm (NM)}) = S~,
    \label{eq:NM_condition} \\
    S &= 4 \pi G a^2 \delta \rho_{\rm other}^{\rm (NM)} + 3 \frac{{\cal H}}{k^2} (\bar{\rho}+\bar{p})_{\rm other} (\theta_{\rm tot}^{\rm (NM)} - \partial_\tau H_{\rm T}^{\rm (NM)}) + 8 \pi G a^2 \Sigma_{\rm other}^{\rm (NM)}, 
\end{align}
where `cb' refers to the CDM+baryon fluid, whose trajectories are meant to mimic Newtonian trajectories in this gauge, `other' are additional species (usually photons and massless or massive neutrinos) acting as a source in the gauge-fixing equation, `tot' stands for the total, and $ \Sigma= (\bar{\rho}+\bar{p})  \sigma$
is the trace-free scalar component of the spatial energy momentum tensor. In the next subsection, we will discuss concrete methods to find quantities in this gauge and relate them to other gauges. We note that eq. (\ref{eq:NM_condition}) admits different solutions that can be mapped to different choices of a time at which all quantities in the NM gauge coincide with those in the Nb gauge. One possible choice is to perform the matching at early times (in the context of N-body simulations, this means at the starting time of the simulation), which defines the forward NM gauge or \textit{forward method}. Alternatively, one can match the gauges today, resulting in the backward NM gauge or \textit{backward method}.
As outlined in \cite{Heuschling:2022rae}, NM gauges offer a streamlined approach for computing the non-linear evolution of the cold dark matter and baryon components consistently, even in cosmologies with massive neutrinos. If a prediction for the total matter spectrum is needed, the non-linear $\delta_{\rm cb}$ computed consistently with this method can be combined with the neutrino fluctuations $\delta_\nu$, which can be treated as linear for this purpose.\footnote{For neutrinos with a realistic mass, $\delta_\nu$ is indeed linear down to scales where the matter power spectrum is anyway strongly dominated by $\delta_{\rm cb}$.} 

A consistent solution to this approach exists as long as the solution of eq. (\ref{eq:NM_condition}) does not lead to large values of $|H_{\rm T}^{\rm (NM)}|$ that would break the assumption that linear perturbation theory holds for metric perturbations within the range of times and scales of interest. While this paper formally works to linear order in metric fluctuations, it has been demonstrated that NM gauges remain applicable  in a more general framework known as the \textit{weak-field limit} of general relativity~\cite{Fidler:2017pnb}. In this limit, a double expansion is assumed: metric perturbations are considered small, while matter perturbations—such as density and velocity—are only required to be perturbatively small on large scales. This approximation holds throughout most of the cosmological evolution of the Universe, except on the smallest scales where strong-field relativistic effects become significant.

The weak-field limit is well justified by the fact that, on large scales, even if local density contrasts become large, their total mass contribution remains conserved, leading to a negligible change of their gravitational influence on distant structures. This regime is fully consistent with the assumptions underlying cosmological $N$-body simulations, standard perturbation theory and the EFTofLSS.

Importantly, in this limit, the metric perturbation $H_{\rm T}^{\rm (NM)}$ remains consistent with linear theory, as it is effectively shielded from non-linear matter perturbations. In contrast, $H_{\rm L}^{\rm (NM)}$, while still perturbatively small, can deviate from its linear behavior because it is directly sourced by the non-linear matter density field. However, reference \cite{Fidler:2017pnb} checked that $|H_{\rm L}^{\rm (NM)}|$ is only slightly enhanced compared to $|H_{\rm T}^{\rm (NM)}|$, and that {\it all} metric perturbations remain small as long as $|H_{\rm T}^{\rm (NM)}|\ll 1$. We also checked this explicitly in a few cases in the present work. For simplicity, we use $|H_{\rm T}^{\rm (NM)}|\ll 1$ as the primary diagnostic for assessing the validity of the weak-field approximation.

Note that the temporal gauge-fixing condition is the same in the Poisson, Nb and NM gauges, since it is always given by eq. (\ref{eq:Nboisson1}). This implies that density perturbations are the same in all these gauges, $\delta^{\rm (P)}_x=\delta^{\rm (Nb)}_x=\delta^{\rm (NM)}_x$. However, velocity divergences are not:
\begin{equation}
\theta^{\rm (P)}_x = \theta^{\rm (Nb)}_x - 3 \dot{\zeta} = \theta^{\rm (NM)}_x - \dot{H}_{\rm T}~.
\end{equation}
\item {\bf Newtonian limit}: We use a superscript $^{\rm [Nl]}$ to denote quantities in the Newtonian limit, e.g., the Newtonian density fluctuation of a species $x$ is referred to as $\delta^{\rm [Nl]}_x$. We stress that these quantities are {\it not} the relativistic perturbations in the Newtonian gauge (actually called the Poisson gauge in this work), but the non-relativistic perturbations appearing in the approximate Newtonian equations used in standard perturbation theory and its EFT extension. The statement that particle trajectories in the NM gauge coincide with those in the Newtonian theory is equivalent to
\begin{equation}
    \theta^{\rm [Nl]}_{\rm cb} = \theta^{\rm (NM )}_{\rm cb}~.
\end{equation}
On the other hand, the respective densities are related by a geometric term \cite{Fidler:2016tir,Fidler:2017pnb}:
\begin{equation} \label{eq:counting}
    \delta^{\rm [Nl]}_{\rm cb} = \delta^{\rm (NM)}_{\rm cb}+3 H_{\rm L}^{\rm (NM)}~.
\end{equation}
One can understand this relation by noticing that local fluctuations of the inverse of the physical volume are given by $- 3 H_{\rm L}^{\rm (NM)}$. Adding $3 H_{\rm L}^{\rm (NM)}$ amounts to removing this relativistic correction from the density fluctuation $\delta^{\rm (NM)}_{\rm cb}$, such that $\delta^{\rm [Nl]}_{\rm cb}$ is a non-relativistic number density count.
\end{itemize}

\subsection{Inferring NM gauge perturbations from a Newtonian fluid}

There are two ways to compute the evolution of metric and matter perturbations in a NM gauge. First, one may solve jointly (within a Boltzmann code) the equations of conservation of energy and momentum for each species, two of the linearized Einstein equations, and finally the two constrain equations (\ref{eq:NM1}, \ref{eq:NM_condition}) that are necessary to close the differential system. Starting from an appropriate choice of initial conditions, this would give all the perturbations in one NM gauge, including $(\delta^{\rm (NM)}_{\rm cb}, \theta^{\rm (NM)}_{\rm cb})$. One could then check that 
\begin{equation}
(\delta^{\rm [Nl]}_{\rm cb}, \theta^{\rm [Nl]}_{\rm cb})
= (\delta^{\rm (NM)}_{\rm cb}+3H_{\rm L}^{\rm(NM)}, \theta^{\rm (NM)}_{\rm cb})\end{equation}
is a solution of the linearized Newtonian equations of motion for a self-gravitating `cb' fluid:
\begin{align}
    \dot{\delta}_{\rm cb}^{\rm[Nl]} &= - \theta_{\rm cb}^{\rm[Nl]}~, \label{eq:newton1}\\
    \dot{\theta}_{\rm cb}^{\rm[Nl]} &= -\frac{\dot{a}}{a} \theta_{\rm cb}^{\rm[Nl]} + k^2 \Psi^{\rm[Nl]}~,\label{eq:newton2}\\
    \Psi^{\rm[Nl]} &\equiv -\frac{3}{2} 
    \frac{{\cal{H}}^2}{k^2} (\Omega_{\rm c} + \Omega_{\rm b}) \, \delta_{\rm cb}^{\rm[Nl]}~.
    \label{eq:newton3}
\end{align}
This is automatically fulfilled, since eq.~(\ref{eq:NM_condition}) was derived in order to achieve such a behavior. Since these equations describe a self-gravitating fluid, the linear growth factor in such a gauge is scale-independent. Additionally, as long as $|H_{\rm T}^{\rm (NM)}|\ll 1$ and the weak-field approximation applies, reference \cite{Fidler:2017pnb} shows that the full non-linear equations of motion for $(\delta^{\rm (NM)}_{\rm cb}+3H_{\rm L}^{\rm(NM)}, \,\theta^{\rm (NM)}_{\rm cb})$, which include quadratic terms, match the non-linear Newtonian equations.

However, there is a second approach, which is simpler to implement in practice \cite{Heuschling:2022rae}. One may integrate all the equations (within a Boltzmann code) in a gauge $G$ of their choice, and simultaneously solve the Newtonian equations (\ref{eq:newton1}, \ref{eq:newton2}, \ref{eq:newton3}) to get $(\delta^{\rm [Nl]}_{\rm cb}, \theta^{\rm [Nl]}_{\rm cb})$ with appropriate initial conditions detailed in section \ref{sec:IC}. Then, for each time and each wavenumber, the comparison between $\delta^{\rm (G)}_{\rm cb}$ and $\delta^{\rm [Nl]}_{\rm cb}$ returns the fields providing the full (linear-order) gauge transformation from the $G$ gauge to one NM gauge, including large-scale relativistic corrections of any order in $(aH/k)$.

We implemented the second logic in a modified version of the \class{} code. To solve for $(\delta^{\rm [Nl]}_{\rm cb}, \theta^{\rm [Nl]}_{\rm cb})$, we define one extra fluid called the `Newtonian fluid', obeying the equations of motions in eqs. (\ref{eq:newton1}, \ref{eq:newton2}, \ref{eq:newton3}) and to initial conditions that we will specify in section \ref{sec:IC}. We then run the code either in the Poisson or synchronous gauge and infer the metric perturbation $H_{\rm T}^{\rm (NM)}$, which will be useful for the next steps, from a simple comparison between $\delta^{\rm [Nl]}_{\rm cb}$ and $\delta_{\rm cb}^{\rm (G)}$ in this gauge.

Let us start with the case of the Poisson gauge. By assumption, there is no temporal component in the gauge transformations from the Poisson gauge to NM gauges, which implies $\delta^{\rm (NM)}_{\rm cb}=\delta^{\rm (P)}_{\rm cb}$. Thus eq. (\ref{eq:counting}) immediately gives
\begin{equation} \label{eq:HLNM}
    3 H_{\rm L}^{\rm (NM)} = \delta^{\rm [Nl]}_{\rm cb} - \delta^{\rm (P)}_{\rm cb}~.
\end{equation}
To find $H_{\rm T}^{\rm (NM)}$, one can write the transformation of the components  $H_{\rm L}$ and $H_{\rm T}$ from the Poisson gauge to a NM gauge. Using the fact that the temporal component of the transformation vanishes and that $H_{\rm T}^{\rm (P)}=0$, one can eliminate the spatial part of the transformation and find
\begin{equation} \label{eq:HLP}
H_{\rm T}^{\rm(NM)} + 3 H_{\rm L}^{\rm(NM)}  = 3 H_{\rm L}^{\rm(P)} = -3\phi~.
\end{equation}
Combining the last two relations gives
\begin{equation} \label{eq:HTP}
    H_{\rm T}^{\rm (NM)} =  \delta_{\rm cb}^{\rm(P)} - \delta_{\rm cb}^{\rm[Nl]} - 3 \phi~.
    \end{equation}

A similar relation can be found in the case of the Synchronous gauge. We can use the well-known gauge-invariant quantity for non-relativistic species
\begin{equation}
    \delta_{\rm cb}^{\rm(S)} +3 H_{\rm L}^{\rm(S)} + H_{\rm T}^{\rm(S)} =
    \delta_{\rm cb}^{\rm(P)} +3 H_{\rm L}^{\rm(P)} + H_{\rm T}^{\rm(P)}  =
    \delta_{\rm cb}^{\rm(P)} - 3 \phi
    \end{equation}
and replace $\delta_{\rm cb}^{\rm(P)}$ using either eq.~(\ref{eq:HLP}) or eq.~(\ref{eq:HTP}) to get
\begin{align} 
    H_{\rm L}^{\rm (NM)} &= \frac{1}{3} \left( \delta_{\rm cb}^{\rm[Nl]} - \delta_{\rm cb}^{\rm(S)} \right) + \frac{\dot{a}}{a} \alpha~, \label{eq:HLsyn}\\
    H_{\rm T}^{\rm (NM)} &= \delta_{\rm cb}^{\rm(S)} - \delta_{\rm cb}^{\rm[Nl]} + 3 \left(H_{\rm T}^{\rm(S)}/3+H_{\rm L}^{\rm(S)}\right)
    = \delta_{\rm cb}^{\rm(S)} - \delta_{\rm cb}^{\rm[Nl]} - 3 \eta~. \label{eq:HTsyn}
\end{align}

\subsection{Forward and backward methods\label{sec:IC}}

A NM gauge retains a two-parameter freedom for specifying $H_{\rm T}$ and its derivative on one initial hypersurface. In our method this is equivalent to specifying the initial conditions for the density and velocity divergence of the Newtonian fluid \cite{Fidler:2017ebh}. Given that the N-boisson gauge represents a simple limit to the Newtonian motion gauges in the absence of massive neutrinos, we use the N-boisson gauge to specify these initial conditions, and match the Newtonian motion gauge to the N-boisson gauge. The only remaining freedom now is the time where this matching is performed. 
Since the N-boisson gauge is characterized by the relation in eq. (\ref{eq:zetaNb}) between $H_{\rm T}^{\rm (Nb)}$ and the gauge-invariant curvature perturbation $\zeta$, this freedom can be rephrased as a choice of time at which $H_{\rm T}^{\rm (NM)}$ is set to $3\zeta$.

In the {\it forward method}, we start evolving the Newtonian fluid at an initial time $\tau_{\rm ini}$ (or redshift $z_{\rm ini}$) of our choice, imposing that the NM gauge fluctuations are the same as in the N-boisson gauge at that time. Then, if the Boltzmann code is set to follow perturbations in the Poisson gauge, eq. (\ref{eq:HTP}) with  $H_{\rm T}^{\rm (NM)}=3\zeta$ gives the initial condition for the density of the Newtonian fluid,
\begin{equation} 
\delta_{\rm cb}^{\rm[Nl]} = \delta_{\rm cb}^{\rm(Nb)} + 3H_{\rm L}^{\rm(Nb)} =  \delta_{\rm cb}^{\rm(P)} - 3 \zeta - 3 \phi
\qquad {\rm at} \quad \tau=\tau_{\rm ini}~.
\label{eq:match_NM_Nb_P}
\end{equation}
If instead the Boltzmann code tracks the synchronous gauge variables, eq. (\ref{eq:HTsyn}) with  $H_{\rm T}^{\rm (NM)}=3\zeta$ gives
\begin{equation}
\delta_{\rm cb}^{\rm[Nl]} = \delta_{\rm cb}^{\rm(S)} - 3 \zeta - 3 \eta
\qquad {\rm at} \quad \tau=\tau_{\rm ini}~.
\label{eq:match_NM_Nb_S}
\end{equation}
Note that in the synchronous gauge $\zeta = -\eta - \frac{\cal H}{k^2} \theta_{\rm tot}^{\rm(S)}$ and the total velocity divergence is suppressed,\footnote{We recall that we are always referring to the synchronous gauge comoving with cold dark matter, $\theta_{\rm c}^{\rm (S)}=0$.} such that $\zeta \approx -\eta$. Then, the Newtonian fluid density is almost identical to the synchronous gauge density, at least at initial time. 

For the initial velocity, we want to pick up a pure growing mode, which is crucial as this is one of the main assumptions of perturbation theory and its EFT extension. The Newtonian solutions for $\delta_{\rm cb}^{\rm [Nl]}$ and $\theta_{\rm cb}^{\rm [Nl]}$ are separable and can be decomposed into growing and decaying solutions,
\begin{align}
    \delta_{\rm cb}^{\rm [Nl]}({\bf k},\tau) &= D_+(\tau) \, A({\bf k}) + D_-(\tau) \, B({\bf k})~,
    \nonumber \\
    \theta_{\rm cb}^{\rm [Nl]}({\bf k},\tau) & = f_+(\tau) \, A({\bf k}) + f_-(\tau) \, B({\bf k})~,
\end{align}
where $D_+$ and $f_+$ are the scale-independent growth factor and growth rate of the Newtonian growing mode. The functions $D_+(\tau)$ and $f_+(\tau)$ can be provided by a Bolztmann code. In \class{}, they are computed within the background module by solving the background equations
\begin{equation}
    \ddot{D}_+ +{\cal H}\,\dot{D}_+ +\frac{3}{2} {\cal H}^2 (\Omega_{\rm c}+\Omega_{\rm b}) \, D_+ =0~,
    \qquad f_+ = \frac{\dot{D}_+}{{\cal H} \, D_+}~.
  \label{eq:D}
\end{equation}
Note that the evolution equations for $D_+(\tau)$ and $f_+(\tau)$ are the same as those fulfilled by the Newtonian fluid variable $\delta_{\rm cb}^{\rm [Nl]}({\bf k},\tau)$, $\theta_{\rm cb}^{\rm[Nl]}({\bf k},\tau)$ at any ${\bf k}$. Since the former equations are solved by the Boltzmann code starting from an extremely early time (well before photon decoupling), the solutions ($D_+$, $f_+$) are guaranteed to account for the pure growing mode.
Thus, to select this mode, we define the initial velocity of the Newtonian fluid at $\tau=\tau_{\rm ini}$ using
\begin{equation}
\theta_{\rm cb}^{\rm[Nl]} = \frac{f_+}{D_+} \delta_{\rm cb}^{\rm [Nl]}~.
\label{eq:growing_mode}
\end{equation}
The solution to the linearized Newtonian equations (\ref{eq:newton1}, \ref{eq:newton2}, \ref{eq:newton3})
will then fulfill eq. (\ref{eq:growing_mode}) at all times.

In the {\it backward method}, we make a different choice: we impose that the NM and N-boisson gauge fluctuations coincide today, and that eqs. (\ref{eq:match_NM_Nb_P}, \ref{eq:match_NM_Nb_S}) hold at $z=0$. 
Again, we need to only pick a pure growing mode and ensure that eq. (\ref{eq:growing_mode}) holds. To achieve this, we first run the code like in the forward method. Then, we identify the factor by which $\delta_{\rm cb}^{\rm [Nl]}(k,\tau_0)$ should be multiplied in order to match our requirement at $z=0$. Finally, we post-process the results by multiplying $\delta_{\rm cb}^{\rm [Nl]}(k,\tau)$ and $\theta_{\rm cb}^{\rm [Nl]}(k,\tau)$ at all times with this factor. 

Since both the forward and backward method lead to possible NM gauges, they can be used indifferently in the calculations presented in the next sections. However, since they involve different gauge transformations, it is possible that the weak-field condition $|H_{\rm T}^{\rm (NM)}|\ll 1$ is fullfilled in one case and not the other, depending on the precise cosmological model. Having two prescriptions is useful for the purpose of checking the self-consistency of our approach and the accuracy of our results.

\section{Real-space power spectrum with massless neutrinos}\label{sec:apply_NM_massless}

\subsection{Standard approach}

The simplest flavor of the EFTofLSS relies on purely Newtonian (non-linear) equations for the perturbations of non-relativistic matter species: in this case, CDM and baryons. Still, it takes as an input the power spectrum of linear perturbations computed by a relativistic Boltzmann code in some gauge. 
The input is often chosen to be the matter power spectrum in the synchronous gauge, $P_{\rm cb,L}^{\rm (S)}(k,z) = \langle |\delta^{\rm (S)}_{\rm cb}(\mathbf{k},z)|^2\rangle$ (in a cosmology with massless neutrinos, we can use indifferently the subscript `m' or `cb' for total non-relativistic matter). The reason for sticking to the synchronous gauge in this context is that $\delta^{\rm (S)}_{\rm cb}$ is known to remain very close to the solution of purely Newtonian equations, $\delta^{\rm [Nl]}_{\rm cb}$, even in the limit $k\longrightarrow 0$. This would not be the case in the Poisson gauge: $\delta^{\rm (P)}_{\rm cb}$ deviates strongly from $\delta^{\rm [Nl]}_{\rm cb}$ on super-Hubble scales, by a factor of order $({\cal H}/k)^2$. Loop integrals based on $P_{\rm cb,L}^{\rm (P)}(k,z)$ could even diverge in the limit $k\longrightarrow 0$.

We already argued in section \ref{sec:gauges} that during matter domination and on scales of interest, the density fluctuation of CDM+baryons in the synchronous and comoving gauge coincide during matter domination, $\delta^{\rm (S)}_{\rm cb} \simeq \delta^{\rm (C)}_{\rm cb}$, up to negligible corrections in $\theta_{\rm tot}^{\rm (S)}$. To be precise, the quantity that \class{} uses as the density fluctuation in the final calculation of the linear matter power spectrum is the gauge-independent fluctuation $D_{\rm cb}$ defined in section \ref{eq:def_D}, or equivalently, $\delta^{\rm (C)}_{\rm cb}$.  
Thus, in \classoneloop, the input quantity for the calculation of EFTofLSS contributions is $P_{\rm cb,L}^{\rm (C)}(k,z)$. We make this distinction for the purpose of describing precisely our numerical implementation, but  $\delta^{\rm (S)}_{\rm cb}$ and $\delta^{\rm (C)}_{\rm cb}$ are so close to each other that the distinction between the synchronous and comoving gauge densities never plays any significant role in this work.

To get the non-linear power spectrum at a given redshift $z$, one can use an EFTofLSS algorithm in two ways: either computing the loops directly from 
$P_{\rm cb,L}^{\rm (C)}(k,z)$ at the same $z$, or computing them at a reference redshift (for instance, $z=0$) and rescaling each term with appropriate powers of the scale-independent growth factor $D_+(z)$ and growth rate $f_+(z)$ extracted from the Boltzmann code.\footnote{To be precise, the linear contribution to the power spectrum is always rescaled by the full growth factor $D_+(k,z)$, but the non-linear corrections are not.}

In theory, both schemes introduce a tiny approximation. In the first method, one still assumes EdS kernels in the loop integrals, which amounts in neglecting any `memory effect' related to the fact that at earlier times, the true perturbations are described by equations featuring relativistic corrections and gravitational couplings with relativistic species (photons, neutrinos), such that their precise expression is not exactly separable in time and wavenumber. In the second method, in addition to the EdS assumption, a second approximation is made: loop contributions are rescaled with a scale-independent growth factor, $D_+(z)$, and logarithmic growth rate, $f_+(z)$, while the actual growth factor and rate are slightly scale-dependent due to the impact of relativistic species (photons, massless neutrinos) and free-streaming species (like massive neutrinos, treated in the next section).

In practice, these approximations are expected to be safe for cosmologies which do not exhibit a considerable scale-dependence of the growth factor (as could be the case with modifications of gravity or exotic dark matter species), and in particular, for the $\Lambda$CDM model with only massless neutrinos.
As a matter of fact, the effect of relativistic corrections and gravitational couplings with relativistic particles is only significant on large scales and/or early time, that is, close to the Hubble radius. Instead, the non-linear growth of perturbations that the EFTofLSS tries to capture takes place deep in the sub-Hubble regime. In that regime, the Newtonian approximation is excellent, the solution of the full equation is almost separable, and the use of EdS kernels is justified. Thus, the non-linear power spectrum would get negligible corrections from these effects.

However, conceptually, it is very interesting that NM gauges offer an opportunity to check the accuracy of many of the previous assumptions. As long as we stick to a $\Lambda$CDM cosmology with massless neutrinos, this check is mainly of academic interest: for the reasons explained above, we know that the standard EFTofLSS implementation is sufficient for such models. For the sake of clarity, it is nevertheless useful to discuss this case first before moving to more interesting situations with massive neutrinos or more exotic ingredients.

\subsection{Newtonian Motion gauge approach without $\Lambda$ correction\label{sec:NM_massless}}

Using the NM gauge formalism, it is easy to confirm that neglecting GR and radiation effects in the EFTofLSS is a very good approximation at least in a $\Lambda$CDM model with massless neutrinos. Indeed, we know that the perturbations of relativistic species never grow and remain accurately described at the level of linear theory. Thus, we can use the following approach:

\begin{enumerate}
    \item Solve the full relativistic problem at linear order using a Boltzmann solver in an arbitrary gauge, and incorporate the method described in the last section (with either the forward or backward method) to define a NM gauge and infer the corresponding $H_{\rm T}^{\rm (NM)}(k,z)$. 
    \item Check that the condition $|H_{\rm T}^{\rm (NM )}|\ll 1$ is met at all relevant times and scales, in order to confirm that we are referring to a transformation to a NM gauge that remains consistent with the weak-field assumption. Previous works that used N-body simulations to validate this approach suggest that a sufficient consistency condition reads $|H_{\rm T}^{\rm (NM )}| \ll 10^{-3}$ \cite{Fidler:2017pnb}. Note that this condition holds at the level of the perturbation $H_{\rm T}^{\rm (NM )}(\mathbf{k},\tau)$, related to the transfer function $H_{\rm T}^{\rm (NM )}(k,\tau)$ through\footnote{We denote transfer functions with the same symbol as the field itself, but with a wavenumber instead of wavevector as argument.}
    \begin{equation}
        H_{\rm T}^{\rm (NM )}(\mathbf{k},\tau) = H_{\rm T}^{\rm (NM )}(k,\tau) \, {\cal R}(\mathbf{k})~.
    \end{equation}
    Since the primordial curvature fluctuations are of the order of $|{\cal R}(\mathbf{k})| \sim 10^{-5}$, the transfer functions $|H_{\rm T}^{\rm (NM )}(k,\tau)|$ could reach up to $\lessapprox 100$.
    \item Both the forward or backward methods provide solutions $\delta^{\rm [Nl]}_{\rm cb}(k,z)$ that obey exactly the Newtonian equations describing a self-gravitating `cb' fluid at the linear level. Thus, this $\delta^{\rm [Nl]}_{\rm cb}(k,z)$ coincides with the variable usually used in the EFTofLSS formalism. We can compute its non-linear evolution and its correlation functions (non-linear power spectrum $P$, bispectrum $B$, etc.) by applying the EFTofLSS formalism to the linear power spectrum $P_{\rm cb,L}^{\rm [Nl]}(k,z) = \langle |\delta^{\rm [Nl]}_{\rm cb}(\mathbf{k},z)|^2\rangle$ (and {\it not} to $P_{\rm cb,L}^{\rm (C)}(k,z)$ as in the standard approach). When using $P_{\rm cb,L}^{\rm [Nl]}(k,z)$, the use of scale-independent kernels is  justified. We denote the result as
    \begin{equation}
        P_{\rm cb,NL}^{\rm [Nl]}\left[P^{\rm [Nl]}_{\rm cb,L}(k,z)\right]~, \qquad
        B_{\rm cb,NL}^{\rm [Nl]}\left[P^{\rm [Nl]}_{\rm cb,L}(k,z)\right]~,
    \end{equation}
    where  $P_{\rm cb,NL}^{\rm [Nl]}[...]$, $B_{\rm cb,NL}^{\rm [Nl]}[...]$ are operators performing convolutions of the input linear spectrum with itself to provide a non-linear power spectrum or bispectrum.  
    \item We can then go back to an arbitrary gauge G  to estimate the non-linear correlation functions in that gauge. Let us assume that the density fluctuation in an arbitrary gauge G is related to its NM gauge counterpart through the time component of the coordinate displacement field $\xi^\mu(x^\alpha)$,
    \begin{equation}
        \delta^{\rm (G)}_{\rm cb} = \delta^{\rm (NM)}_{\rm cb} + {\cal H} \, \xi^0 =  \delta^{\rm [Nl]}_{\rm cb} + \Delta^{\rm (G)}~,
    \end{equation}
    where in the second equality we used eq. (\ref{eq:counting}) and defined  $\Delta^{\rm (G)} \equiv {\cal H} \, \xi^0 -3 H_L^{\rm (NM)}$. We know from eq. (\ref{eq:HTP}) that when G is the Poisson gauge,
    \begin{equation}
        \Delta^{\rm (P)} = H_{\rm T}^{\rm (NM)} + 3 \phi~,
    \end{equation}
    and from eq. (\ref{eq:HTsyn}) that for the synchronous gauge
    \begin{equation}
        \Delta^{\rm (S)} =  H_{\rm T}^{\rm (NM)} + 3 \eta~.
    \end{equation}
    Using the expression of the curvature perturbation in different gauges,
\begin{equation}
    \zeta= -\phi - \frac{{\cal H}}{k^2} \theta_{\rm tot}^{\rm (P)} = - \eta  - \frac{{\cal H}}{k^2} \theta_{\rm tot}^{\rm (S)}~,
\end{equation}
we see that $\Delta^{\rm (G)}$ with ${\rm G}= {\rm S}, {\rm P}$ is also given in both cases by
\begin{equation}
\Delta^{\rm (G)} = H_{\rm T}^{\rm (NM)} - 3 \zeta -  3 \frac{{\cal H}}{k^2} \theta_{\rm tot}^{\rm (G)}~.
\label{eq:delta_to_vel}
\end{equation}
Finally, using eq. (\ref{eq:D_to_G_and_S}), we find that the relation between $\delta^{\rm [Nl]}_{\rm cb}$ and the comoving gauge density $\delta^{\rm (C)}_{\rm cb}$ is simply given by $\delta^{\rm (C)}_{\rm cb}=\delta^{\rm [Nl]}_{\rm cb}+\Delta^{\rm (C)}$ with
\begin{equation}
\Delta^{\rm (C)} = H_{\rm T}^{\rm (NM)} - 3 \zeta~.
\label{eq:delta_to_c}
\end{equation}
We can now devise a general method to estimate correlation functions (such as the power spectrum $P_{\rm cb,NL}^{\rm (G)}$ and bispectrum $B_{\rm cb,NL}^{\rm (G)}$) in an arbitrary gauge G. In practice, we are mainly interested in performing such calculations in the comoving gauge, ${\rm G}={\rm C}$, but in the rest of this subsection we stick to an arbitrary gauge G to keep the discussion as general as possible.

For such a purpose, one should in principle evaluate the correlation functions with new loops involving convolutions of the linear field $\Delta^{\rm (G)}(\mathbf{k},z)$ with itself and with the linear density field $\delta^{\rm [Nl]}_{\rm cb,L}(\mathbf{k},z)$. For instance, for the power spectrum, one should start from the correlator
\begin{equation}
    P^{\rm (G)}_{\rm cb,NL} = \left\langle
    \left| \delta^{\rm [Nl]}_{\rm cb,NL} + \Delta^{\rm (G)} \right|^2 \right\rangle~,
\end{equation}
expand $\delta^{\rm [Nl]}_{\rm cb,NL}$ at a given order in $\delta^{\rm [Nl]}_{\rm cb,L}$ (while sticking to linear order in $\Delta^{\rm (G)}$), and express the result in terms of loop integrals involving only the Gaussian fields $\Delta^{\rm (G)}$ and $\delta^{\rm [Nl]}_{\rm cb,L}$. We performed this expansion consistently, keeping all the terms that contribute to the one-loop power spectrum. The result is presented in eq.~(\ref{eq:Pcb_full}) of Appendix~\ref{app:full}. We implemented it in \classoneloop{} as one possible way to compute $P^{\rm (G)}_{\rm cb,NL}$ in the NM gauge approach. However, the calculation of $P^{\rm (G)}_{\rm cb,NL}$ can also be addressed in a simpler manner, because it features a separation of scales:
\begin{itemize}
    \item On small scales, $\delta^{\rm [Nl]}_{\rm cb}$ grows non-linear while $\Delta^{\rm (G)}$ remains linear and becomes totally negligible compared to $\delta^{\rm [Nl]}_{\rm cb,NL}$, such that
\begin{equation}
    P^{\rm (G)}_{\rm cb,NL}(k,z) \simeq \left\langle \left| \delta^{\rm [Nl]}_{\rm cb,NL}(\mathbf{k},z) \right|^2 \right\rangle 
    \equiv P_{\rm cb,NL}^{\rm [Nl]}(k,z)~.
\end{equation}
\item On large scales, $\Delta^{\rm (G)}$ and $\delta^{\rm [Nl]}_{\rm cb}$ both remain linear and can be expressed in terms of transfer functions,
\begin{align}
    \delta^{\rm [Nl]}_{\rm cb}(\mathbf{k},z) &= \delta^{\rm [Nl]}_{\rm cb}(k,z) \,\, {\cal R}(\mathbf{k})~, 
    \nonumber \\
    \Delta^{\rm (G)}(\mathbf{k},z) &= \Delta^{\rm (G)}(k,z) \,\, {\cal R}(\mathbf{k})~,
    \label{eq:ad1}
\end{align}
where ${\cal R}$ stands for primordial curvature perturbations. Then, the power spectrum reads
\begin{align}
    P^{\rm (G)}_{\rm cb,L}(k,z) &= 
    \left(
    \delta^{\rm [Nl]}_{\rm cb}(k,z)
    + \Delta^{\rm (G)}(k,z) \right)^2
    P_{\cal R}(k)
    \nonumber \\
    &
= P_{\rm cb,L}^{\rm [Nl]}(k,z) + 2 \sqrt{P_{\Delta^{\rm (G)}}(k,z)} \sqrt{P_{\rm cb,L}^{\rm [Nl]}(k,z)} + P_{\Delta^{\rm (G)}}(k,z)~,
\end{align}
where $P_{{\Delta}^{\rm (G)}}$ is the linear power spectrum of $\Delta^{\rm (G)}$. 
\item On intermediate scales, the effect of  ${\Delta}^{\rm (G)}$ and of non-linear corrections are both negligible. This statement is not obvious {\it a priori} and is potentially model-dependent. However, for the range of neutrino masses that we are exploring, this conclusion can be drawn very clearly from our results. In all the residual plots shown in the next sections, corrections due to ${\Delta}^{\rm (G)}$ are visible on large scales, non-linear corrections on small scales, and in between, there is a range of scales affected by neither of them.
\end{itemize}
\end{enumerate}

\noindent Thus, the result for the non-linear power spectrum in a gauge G is captured on all scales by the expression
    \begin{equation}
    P^{\rm (G)}_{\rm cb,NL}(k,z) 
\simeq P_{\rm cb,NL}^{\rm [Nl]}(k,z) + 2 \sqrt{P_{\Delta^{\rm (G)}}(k,z)} \sqrt{P_{\rm cb,L}^{\rm [Nl]}(k,z)} + P_{\Delta^{\rm (G)}}(k,z)~.
\label{eq:Pcb}
\end{equation}
We checked numerically that the simple expression in eq.~(\ref{eq:Pcb}) and the full one-loop expression in eq.~(\ref{eq:Pcb_full}) yield results that are indistinguishable (up to 0.001\% level). We implemented eq.~(\ref{eq:Pcb}) as our default method to get $P_{\rm cb,NL}^{\rm (C)}$ in the NM motion approach. This method to estimate the relativistic one-loop power spectrum including linear GR and radiation effects is easy and straightforward to implement, since the linear power spectra $P_{\Delta^{\rm (G)}}$ and $P_{\rm cb,L}$ involved in eq.~(\ref{eq:Pcb}) can be computed by a Boltzmann code very efficiently. 

\subsection{Newtonian Motion gauge approach with $\Lambda$ correction\label{sec:NM_lambda}}

The growth of perturbations in the Newtonian limit (or equivalently in the NM gauge) can be described with very compact equations after defining a doublet with components $\psi_1=\delta_{\rm cb}^{\rm [Nl]}$, $\psi_2=-\theta_{\rm cb}^{\rm [Nl]}/({\cal H} f)$ and using the logarithm of the linear growth factor $D$ as a time variable, see, e.g., \cite{Garny:2022fsh}. (From now on, for simplicity, we drop the subscript $+$ to denote
that the growth rate/factor are defined from the pure growing mode.) Then, the non-linear equations of motion for the Newtonian (or NM gauge) variables read
\begin{equation}
\partial_{\ln D} \, \psi_a(\mathbf{k},\tau) + \Omega_{ab}(\tau) \, \psi_b(\mathbf{k},\tau)=
\int_{\mathbf{k}_1,\mathbf{k}_2}\!\!\!\!\!\delta_D(\mathbf{k}-\mathbf{k}_1-\mathbf{k}_2)
\,
\gamma_{abc}(\mathbf{k},\mathbf{k}_1,\mathbf{k}_2) 
\, \psi_b(\mathbf{k}_1,\tau) 
\, \psi_c(\mathbf{k}_2,\tau)~,
\label{eq:eqmopsi}
\end{equation}
where the kernels $\gamma_{abc}$ are given, e.g., in \cite{Garny:2022fsh}, while the matrix $\Omega_{ab}(\tau)$ reads
\begin{equation}
\Omega_{ab} =
\left(
    \begin{tabular}{cc}
    $0$ & $-1$ \\
    $-\frac{3}{2} \frac{\Omega_{\rm m}}{f^2}$ & $\frac{3}{2} \frac{\Omega_{\rm m}}{f^2} - 1$\\
    \end{tabular}
\right)~.    
\end{equation}
The condition for having a separable solution of the form $\psi_a(\mathbf{k},\tau)=\sum_{n=1}^\infty \, D^n(\tau) \, \psi_a^{(n)}(\mathbf{k})$ is that the coefficients $\Omega_{ab}$ are independent of time.

In the absence of massive neutrinos, both $f^2$ and $\Omega_{\rm m}$ are equal to one during matter domination and decrease at nearly the same rate during $\Lambda$ domination. Thus, $\Omega_{\rm m}/f^2$ can be approximated as one. This so-called EdS approximation is used in the standard approach, and also in our `NM gauge approach without $\Lambda$ correction'. In the latter case, using $\Omega_{\rm m}/f^2=1$ even during $\Lambda$ domination is actually the only approximation performed in the calculation of the one-loop matter power spectrum (as long as the weak-field limit holds).

In reality, $\Omega_{\rm m}/f^2$ grows during $\Lambda$ domination and reaches $\sim 1.14$ at $z=0$ in a $\Lambda$CDM cosmology with $\Omega_{\rm m} \simeq 0.31$.
Bernardeau \cite{Bernardeau:1993qu} derived an exact (but lenghty) expression for the kernels that takes this effect into account and allows to calculate the spectrum up to the one-loop order. Garny \& Taule \cite{Garny:2020ilv} developped a (numerically expensive) code that integrates the full kernels over time. Their method agrees with the analytical approach of Bernardeau (see their Appendix A) and shows
that the EdS approximation leads to an underestimation of one-loop corrections to the real-space matter power spectrum of the order of $\sim 0.7\%$ at $z=0$ (see the blue curves in their figure 2). Another systematic way to incorporate the exact time-dependence of scale-independent kernels, based on Green's functions, is presented e.g. in \cite{Carrasco:2012cv,Lewandowski:2017kes,Donath:2020abv}.\footnote{This method is implemented in the code PyBird \cite{DAmico:2020kxu} as the option {\tt exact\_time}. When this option is de-activated, PyBird defaults to EdS kernels but still computes the loops out of the linear power spectra computed at the requested redshift $z$ rather than rescaling them from $z=0$. We thank Guido D'Amico and Pierre Zhang for this precision.}

It is not possible to absorb the time-dependence of $\Omega_{\rm m}/f^2$ through a gauge transformation since this dependence is present on all scales, while the impact of a linear gauge transformation is always negligible in the sub-Hubble limit. But at least, the NM gauge approach guarantees that the coefficient $\Omega_{ab}$ are strictly functions of time and not of $k$, even when GR corrections and gravitational couplings with radiation are taken into account. Then, the time dependence of the function
\begin{equation}
\nu(t)\equiv \Omega_{\rm m}(\tau)/f^2(\tau)
\end{equation}
can be treated consistently with time-dependent kernels following, e.g., the exact results of Bernardeau \cite{Bernardeau:1993qu}. Given the very complicated form of these kernels at order three, $F_3$ and $G_3$, we derived a new approximation scheme for the time-dependent kernels, explained in Appendix~\ref{app:kernels}. In a nutshell, the idea is to derive first the exact shape of the kernels when $\nu$ is different from one but constant in time, see Appendix~\ref{app:constant_nu}. We then stick to the same ansatz for the kernels, but promote each coefficient $\nu$ to be a function of time. We finally derive and solve the differential equations that these functions should obey in order to fulfill the angle-averaged recurrence relation between the kernels. Our method leads to tractable kernel expressions up to order three, see Appendix~\ref{app:time_nu}, and would be straightforward to generalize to higher order. It provides results essentially indistinguishable from those obtained with the exact kernels of Bernardeau \cite{Bernardeau:1993qu} and with the numerical method of Garny \& Taule~\cite{Garny:2020ilv}, see Appendix~\ref{app:Bernardeau}.

Once the time-dependent kernels are incorporated in our NM gauge approach, all effects in the calculation of the $\Lambda$CDM one-loop matter power spectrum are consistently taken into account. Later, we refer to this method as the `NM gauge approach with $\Lambda$ correction'.

\subsection{Results and comparison \label{sec:MasslessComp}}

 \begin{figure}
    \centering
    \begin{minipage}{\textwidth}
        \includegraphics[width=\linewidth]{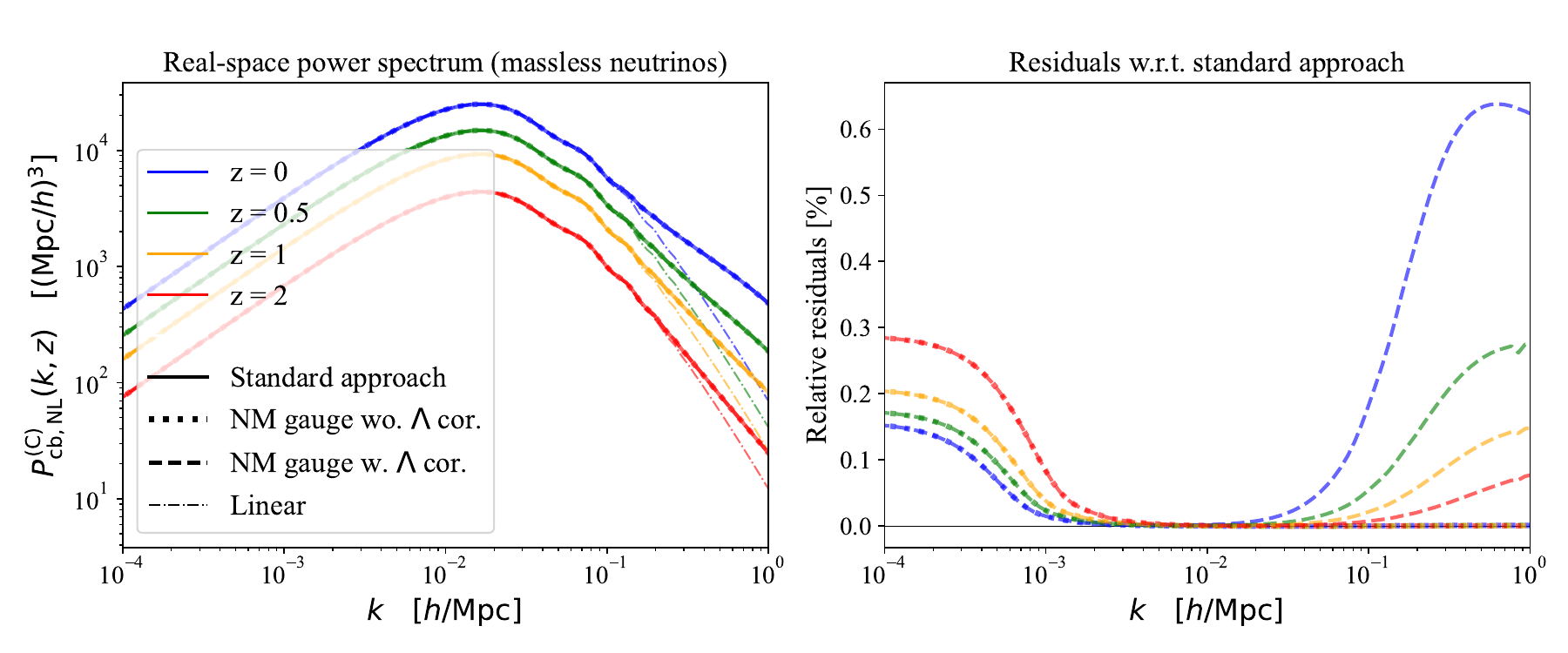}
    \end{minipage}
    \caption{(Left) Non-linear CDM+baryon power spectrum of 
  comoving gauge density, $P^{\rm (C)}_{\rm cb,NL}(k,z)$, in a $\Lambda$CDM cosmology with massless neutrinos. The solid lines show the prediction from the EFTofLSS applied directly to the linear comoving gauge
  power spectrum $P^{\rm (C)}_{\rm cb,L}$ (standard approach), while the dotted and dashed lines show the results from our approach (EFTofLSS applied the NM gauge power spectrum $P^{\rm [Nl]}_{\rm cb,L}$ inferred from the backward method and transformed back to the comoving gauge), either with EdS kernels (NM gauge wo. $\Lambda$ cor.) or kernels accounting for $\Lambda$ corrections (NM gauge w. $\Lambda$ cor.). The dotted-dashed line shows the linear prediction $P^{\rm (C)}_{\rm cb,L}$ for comparison. (Right) Percentage difference between the non-linear power spectra obtained from our NM approach and the standard approach. The dashed lines correspond to the NM gauge approach with $\Lambda$ corrections. The case without $\Lambda$ corrections (dotted lines) coincides with the dashed line on large scales and with 1 on small scales.
    \label{fig:massless}}
\end{figure}

To test the standard approach and its EdS approximation, we compare the non-linear power spectra $P_{\rm cb,NL}^{\rm (C)}(k,z)$ obtained directly using the standard EFTofLSS method with the one inferred from loop calculations in a NM gauge. For this comparison, we customized a version of the \class{} code and \classoneloop{} module to implement the relations specific to the NM gauge that we described in previous sections, 
with or without the $\Lambda$ correction. The difference between the three approaches can be summarized as follows:\\
\mbox{}\\
\noindent{\bf Standard approach:}
\begin{enumerate}
\item Run the \class{} background and linear perturbation modules in the P or S gauge, and infer $P_{\rm cb,L}^{\rm (C)}(k,z)$;
\item Run the \class{} one-loop module to get $P_{\rm cb,NL}^{\rm (C)}\left[P_{\rm cb,L}^{\rm (C)}(k,z) \right]$, using EdS kernels.
\end{enumerate}
\noindent{\bf NM gauge approach without $\Lambda$ correction}:
\begin{enumerate}
\item Run the \class{} background and linear perturbation modules in the P or S gauge including also the Newtonian fluid (normalized with the backward or forward method), and infer $P_{\rm cb,L}^{\rm [Nl]}(k,z)$ and $P_{\Delta^{\rm (G)}}(k,z)$;
\item Run the \class{} one-loop module to get $P_{\rm cb,NL}^{\rm [Nl]}\left[P_{\rm cb,L}^{\rm [Nl]}(k,z) \right]$ (that is, in the NM gauge), using EdS kernels; 
\item Infer $P_{\rm cb,NL}^{\rm (C)}$ through a gauge transformation, using eq.~(\ref{eq:Pcb}) with ${\rm G}={\rm C}$.
\end{enumerate}
\noindent{\bf NM gauge approach with $\Lambda$ correction}:
\begin{enumerate}
\item[\textcolor{white}{\textbullet}] Same as the previous approach, using however in step 2 the time-dependent kernels described in Appendix~\ref{app:time_nu} with $\nu(\tau)=\Omega_{\rm m}(\tau)/f^2(\tau)$.
\end{enumerate}
Note that in the first two approaches, computing the loop integrals either directly from the linear power spectrum evaluated at the requested redshift or with a rescaling of the loop integrals evaluated at $z=0$ does not lead to any significant difference. The error introduced by the rescaling in the standard approach is orders of magnitude smaller than the other effects discussed in this work, such that any of the two strategies can be used indifferently. In the third approach, by construction, the loops must be computed at each $z$.

We perform the comparison between these approaches assuming Planck 2018 best-fit values for the cosmological parameters of a $\Lambda$CDM cosmology with massless neutrinos, and an EFT counter-term $c_{\rm s}^2=0$. The various spectra and their relative differences are shown in figure~\ref{fig:massless} at redshifts $z=0, 0.5, 1,$ and $2$.\\ 

\noindent {\bf Results without $\Lambda$ correction.} The percentage difference between the one-loop matter power spectra derived from the `NM gauge without $\Lambda$ correction' and `standard' approaches is plotted in the right panel of figure~\ref{fig:massless} with dotted lines. These lines are masked by the dashed ones on large scales ($k \leq 4 \times 10^{-3}$Mpc$^{-1}$) and indistinguishable from zero on smaller scales. As expected, for a $\Lambda$CDM cosmology with massless neutrinos, the difference between the two approaches is negligible, ranging from 0.3\% for $k \sim 10^{-4} h {\rm Mpc}^{-1}$ at $z=2$ to less than 0.001\% for $k > 10^{-2}$Mpc$^{-1}$. 

The small difference at large scales ($k \leq 4 \times 10^{-3}$Mpc$^{-1}$) is dominated by the fact that GR corrections and gravitational couplings with relativistic species are neglected in the standard approach. As a matter of fact, a similar difference was found by comparing NM gauge predictions with Newtonian N-body simulations \cite{Adamek:2017grt}. The relative difference grows with redshift simply because the matter power spectrum is smaller at higher redshift: we checked that the absolute difference is almost redshift-independent. 

As argued in previous sections, the main reason for which the standard approach is so accurate in the massless neutrino case is that the quantity $\Delta^{\rm (C)}=H_{\rm T}^{\rm (NM)}-3\zeta$ that gives the relation between the Newtonian density, $\delta^{\rm [Nl]}_{\rm cb}$, and the comoving gauge density, $\delta^{\rm (C)}_{\rm cb}$, remains negligible at all  times and scales of interest, even in the limit $k\rightarrow 0$. Note that the difference $H_{\rm T}^{\rm (NM)}-3\zeta$ quantifies the difference between the N-boisson and NM gauges. In a $\Lambda$CDM cosmology with massless neutrinos, NM gauges have no significant scale-dependent growth to absorb, and remain extremely close to the N-boisson one. Thus, at the linear level, one gets $P^{\rm (C)}_{\rm cb,L}(k,z)\simeq P^{\rm [Nl]}_{\rm cb,L}(k,z)$ up to negligible corrections. At the non-linear level, since the solution for both $\delta^{\rm (C)}_{\rm cb}$ and $\delta^{\rm [Nl]}_{\rm cb}$ is nearly a separable function of time and wavenumber in the sub-Hubble limit, the use of scale-independent kernels, which is exact in the NM gauge approach, is still very accurate in the case of the standard method, such that $P^{\rm (C)}_{\rm cb,NL}(k,z)
\simeq P^{\rm [Nl]}_{\rm cb,NL}(k,z)$. For this reason, in a $\Lambda$CDM model with massless neutrinos, one can safely apply the EFTofLSS expansion directly to the field $\delta^{\rm (C)}_{\rm cb}$ (or $\delta^{\rm (S)}_{\rm cb}$), and use the output Newtonian spectrum as an excellent approximation to the relativistic spectrum $P^{\rm (C)}_{\rm cb,NL}$ (or $P^{\rm (S)}_{\rm cb,NL}$). This is what codes like CLASS-PT, \classoneloop, PyBird, Velocileptor, and others do in practice.

Note that there would be no reason to expect that the spectra derived from the standard and NM gauge methods remain close to each other in other gauges, such as the Poisson gauge. In that case, the comparison would even be difficult to perform. As a matter of fact, the non-linear spectrum $P^{\rm (P)}_{\rm cb,NL}(k,z)$ could be evaluated using the NM gauge approach, but not using the standard approach, since on very large scales $\delta^{\rm (P)}_{\rm cb}$ is enhanced with respect to $\delta^{\rm (C)}_{\rm cb} \simeq \delta^{\rm (S)}_{\rm cb}$ by a factor $({\cal H}/k)^2$, which leads to divergent integrals.

We tested two possible implementations of the NM gauge method, based either on the forward or backward method from section \ref{sec:IC}. These methods lead to different NM gauges. We found that the spectrum derived from the two methods agree to better than 0.1\% in the massless neutrino case. We checked that the quantity $H_{\rm T}^{\rm (NM)}-3\zeta$ is much smaller when using the backward method, which is therefore by far the most accurate. The results shown in figure~\ref{fig:massless} are all based on this method.\\

\noindent {\bf Results with $\Lambda$ correction.} In the right panel of figure~\ref{fig:massless}, the dashed lines show the percentage difference between the one-loop matter power spectra derived from the `NM gauge with $\Lambda$ correction' and `standard' approaches. In this case, the standard approach still assumes EdS kernels, while the NM gauge approach features time-dependent kernels accounting for the evolution of $\nu(\tau)=\Omega_{\rm m}(\tau)/f^2(\tau)$. At $z=2$, this makes little difference, since the universe is still strongly matter-dominated. The amplitude of the loop corrections is enhanced during $\Lambda$ domination. At $z=0$ and around $k\sim 0.3 \, h {\rm Mpc}^{-1}$, the $\Lambda$ correction increases the non-linear power spectrum by $\sim 0.6\%$. The shape and the amplitude of the enhancement is highly consistent with the results of \cite{Garny:2020ilv}, based on a full integration of the kernels over time.
The very good level of agreement is best seen by comparing the right panel of our figure~\ref{fig:massless} with the blue lines in the two panels of figure~2 of \cite{Garny:2020ilv}, obtained at $z=0$ and $z=0.5$ while assuming $\Lambda$CDM parameter values close to ours.

\section{Real-space power spectrum with massive neutrinos\label{sec:apply_NM_massive}}

The situation is more interesting in the context of a $\Lambda$CDM model with massive neutrinos, which become non-relativistic at late time and start to experience gravitational collapse after that time. In this case, it is far from obvious that the difference between the density in the comoving gauge $\delta^{\rm (C)}_{\rm cb}$ and the Newtonian limit density $\delta^{\rm [Nl]}_{\rm cb}$ remains completely negligible. Moreover, the presence of massive neutrinos influences the gravitational potential and induces a scale-dependent growth rate at late time, such that the solution of the linearized Newtonian equations is no longer separable in time and scale. Various authors have argued that this effect is small and that separable solutions can still be used in very good approximation. The NM formalism actually offers a new opportunity to accurately check this statement. 

\subsection{Scale-dependent growth of linear perturbations\label{sec:scale_dep}}

In cosmologies with massive neutrinos, there are three non-relativistic matter components in the late universe: CDM, baryons, and neutrinos. Already at the linear level, the perturbations of each species grow at different rates. However, on midly non-linear scales and at times relevant for structure formation, CDM and baryons share the same density fluctuations $\delta_{\rm cb}$, while neutrinos have distinct fluctuations $\delta_\nu$. In the Poisson, synchronous or comoving gauges and on sub-Hubble scales, the linear growth of $\delta_{\rm cb}$ is governed by the equation
\begin{equation}
    \ddot{\delta}_{\rm cb} + {\cal H} \, \dot{\delta}_{\rm cb}
    + \frac{3}{2} {\cal H}^2 (\Omega_{\rm cb}\, \delta_{\rm cb} + \Omega_\nu \, \delta_\nu )=0~,
    \label{eq:growth_cbn}
\end{equation}
where we neglected additional contributions from photon perturbations and sub-dominant corrections from metric fluctuations. After the neutrino non-relativistic transition, the fact that $\delta_\nu=\delta_{\rm cb}$ on large scales (compared to the neutrino free-streaming scales) while $\delta_\nu/\delta_{\rm cb}$ gradually tends towards zero in the small-scale limit is responsible for the scale-dependent growth of `cb' fluctuations. The purely growing solution of eq.~(\ref{eq:growth_cbn}) normalized to one at $z=0$ defines the scale-dependent linear growth factor $D(k,z)$, while the scale-dependent growth rate is given by $f(k,z)=\dot{D}(k,z)/[{\cal H} \, D(k,z)]$ (from now on, for simplicity, we drop the subscript $+$ to denote that the growth rate/factor are defined from the pure growing mode).
\begin{figure}
    \centering
    \includegraphics[width=0.99\linewidth]{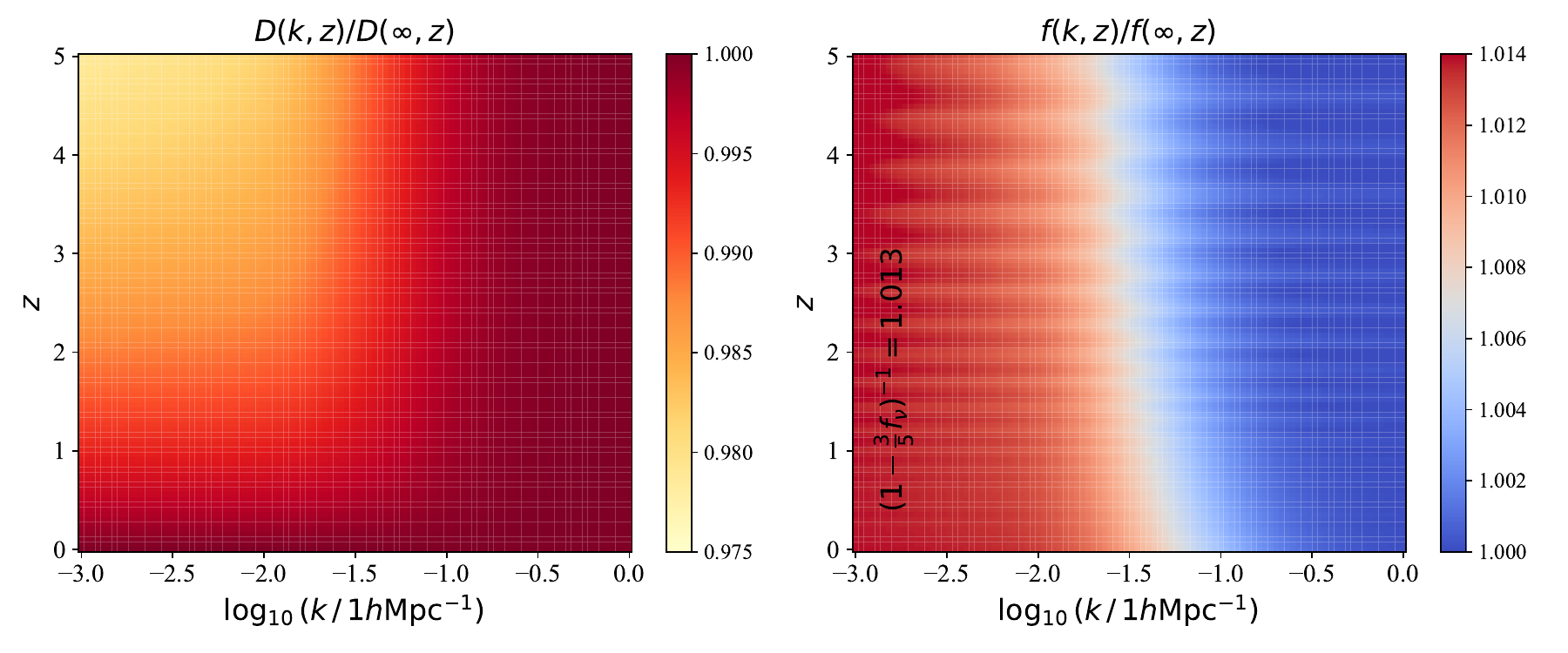}
    \caption{Growth rate $D(k,z)$ and growth factor $f(k,z)$ compared to their small-scale limits $D(\infty,z)$, $f(\infty,z)$, in a $\Lambda$CDM cosmology with three degenerate massive massive neutrinos of individual mass $m_\nu=0.12$~eV. 
    \label{fig:growth-rate-ratio}}
\end{figure}

Figure \ref{fig:growth-rate-ratio} shows that these quantities have distinct behaviors on large and small scales. Since $D(k,z)$ grows faster on large scales, and since the left plot shows the ratio 
\begin{equation}
    \frac{D(k,z)}{D(\infty,z)} = \frac{\delta_{\rm cb}^{\rm (C)}(k,z)}{\delta_{\rm cb}^{\rm (C)}(k,0)}
    \,\, \frac{\delta_{\rm cb}^{\rm (C)}(\infty,0)}{\delta_{\rm cb}^{\rm (C)}(\infty,z)}~,
\end{equation}
which is normalized to one by construction at both $z=0$ and $k\rightarrow \infty$, we observed a depletion of this ratio at large scales and high redshift. The value of $f(k,z)$ is  different on large and small scales. During matter domination, the growth rate $f(k,z)$ can be approximated as 1 on large scales and $(1-\frac{3}{5} f_\nu)$ on small scales \cite{Bond:1980ha,Lesgourgues:2006nd,Lesgourgues:2013sjj}; during $\Lambda$ domination, the overall rate is reduced, but the ratio between the large- and small-scale limits remains close to $(1-\frac{3}{5} f_\nu)$. For both $D$ and $f$, the transition between the small- and large-scale behaviors takes place at the neutrino free-streaming scale, which decreases by a very small amount (such that the associated wavenumber increases) when time flows from $z=5$ to $z=0$. For neutrinos with an individual mass of $m_\nu=0.12$~eV, the figure shows that $D$ and $f$ deviate from their small-scale behavior for $k < 0.08\,h$Mpc$^{-1}$, which is consistent with the expression of the free-streaming scale in eq. (5.93) of reference \cite{Lesgourgues:2013sjj},
\begin{equation}
k_\mathrm{fs}=0.776 \, (1+z)^{-2} \frac{H(z)}{H_0} \left( \frac{m_\nu}{1\,\mathrm{eV}}\right) h \mathrm{Mpc}^{-1}~.
\end{equation}

The scale-independent growth factors/rates can be inferred from the purely growing solution of the background equations
\begin{equation}
  \ddot{D}+{\cal H}\,\dot{D}+\frac{3}{2} {\cal H}^2 \Omega_x \, D =0~,
  \qquad
  f = \dot{D}/({\cal H}D)~.
  \label{eq:D2}
\end{equation}
The Hubble rate is always inferred from the total energy density. Instead, the fractional density $\Omega_x(z)$ in the third term could be defined in two ways:
\begin{itemize}
    \item if we use $\Omega_x=\Omega_{\rm c}+\Omega_{\rm b} + \Omega_\nu$, the growth rate $D$ and the associated growth factor $f$ will reflect the evolution of $\delta_{\rm cb}^{\rm (C)}(k,z) \simeq \delta_{\rm cb}^{\rm (S)}(k,z)$ on large scales, where neutrinos do not free stream and cluster like baryons and CDM.
    \item if we use $\Omega_x=\Omega_{\rm c}+\Omega_{\rm b}$, $D$ and $f$ will reflect the evolution of $\delta_{\rm cb}^{\rm (C)}(k,z)\simeq\delta_{\rm cb}^{\rm (S)}(k,z)$ on small scales, where neutrinos free stream and slow down the clustering of baryons and CDM. This second choice reduces $f$ by a factor $(1-\frac{3}{5} f_\nu)$.
\end{itemize}
The definition of the scale-independent growth rate/factor in \class{} and the non-linear calculations performed in \classoneloop{} are based on the second choice. By comparing the standard \classoneloop{} implementation with a NM gauge approach, we will cross-check that this is a well-motivated choice.

\subsection{Standard approach\label{sec:SA}}

In cosmologies with massive neutrinos, a consistent model of structure formation requires a multi-fluid formalism to account for the CDM, baryon, and neutrino components. A two‐fluid EFTofLSS treatment of CDM and baryons was developed in \cite{Lewandowski:2014rca,Braganca:2020nhv}, and the a two‐fluid extension including dark matter and massive neutrinos was first formulated in \cite{Senatore:2017hyk}. An alternative “hybrid’’ scheme based on a full Boltzmann hierarchy at high redshift and an effective two-fluid description (consisting of CDM+baryon and neutrino fluids coupled gravitationally) at low redshift was developed in \cite{Blas:2014hya,Garny:2020ilv,Garny:2022fsh}. 

On times and scales relevant for the analysis of galaxy clustering, CDM and baryons share identical fluctuations while the neutrino fluctuations $\delta_\nu$ remain very small. Therefore, a one‑fluid EFTofLSS for the CDM+baryon component--using the linear power spectrum $P_{\rm cb}$ as input for the loop calculation--captures the relevant non‑linear dynamics, while neutrino perturbations can be treated as linear. Weak‑lensing measurements, which probe smaller scales where baryonic effects become important, benefit from a two‑fluid treatment of baryons and CDM \cite{Braganca:2020nhv}, but even there neutrino clustering is well described by linear theory.  Consequently, the standard one‑loop matter power spectrum in massive‑neutrino cosmologies can be obtained by applying the EFTofLSS to the CDM+baryon fluid alone and adding the neutrino contribution linearly. For this, one can express the total-matter power spectrum in the comoving gauge as
\begin{equation}
P^{\rm (C)}_{\rm m,NL} = \left\langle \left| (1-f_\nu) \, \delta^{\rm (C)}_{\rm cb,NL} + f_\nu \,\delta^{\rm (C)}_\nu \right|^2 \right\rangle = (1-f_\nu)^2 \left\langle \left| \delta^{\rm (C)}_{\rm cb,NL} + \frac{f_\nu}{1-f_\nu}  \delta^{\rm (C)}_\nu \right|^2 \right\rangle~.
\end{equation}
After expanding $\delta^{\rm (C)}_{\rm cb,NL}$ up to third order in perturbations (while sticking to linear order for $\delta^{\rm (C)}_\nu$), we get for the one-loop total-matter power spectrum
\begin{equation}
P^{\rm (C)}_{\rm m,NL} = (1-f_\nu)^2 
\left(
    P^{\rm (C)}_{\rm cb,NL}
    + \left[2 R^{\rm (C)}_\nu + \left(R^{\rm (C)}_\nu \right)^2 \right] P^{\rm (C)}_{\rm cb,L}
    + 2 R^{\rm (C)}_\nu
    P^{\rm (C)}_{\rm cb,13}
    \right)~,
    \label{eq:P_tot}
\end{equation} 
where $R_\nu^{\rm (C)}$ is a ratio of transfer functions,
\begin{equation}
    R^{\rm (C)}_\nu(k,z) \equiv  
    \frac{f_\nu}{(1-f_\nu)} \,
    \frac{\delta^{\rm (C)}_\nu (k,z)}{\delta^{\rm (C)}_{\rm cb,L}(k,z)}~,
\end{equation}
while $f_\nu(z) = \rho_\nu(z)/\rho_{\rm m}(z)$ is the neutrino fraction, and $P^{\rm [Nl]}_{\rm cb,13}$ is the loop contribution to the $P^{\rm [Nl]}_{\rm cb,NL}$ spectrum coming from the correlator $\left\langle \left| \delta_{\rm cb}^{(1)} \delta_{\rm cb}^{(3)} \right| \right\rangle$. Once multiplied by $R_\nu^{\rm (C)}$, this term actually accounts for the correlator $\left\langle \left| \delta_\nu^{(1)} \delta_{\rm cb}^{(3)} \right| \right\rangle$, which contributes to the same order in perturbations. Note that, instead of eq.~(\ref{eq:P_tot}), many previous works use the approximation
\begin{equation}
P^{\rm (C)}_{\rm m,NL}(k,z) = (1-f_\nu)^2 P^{\rm (C)}_{\rm cb,NL}(k,z) + 2 f_\nu (1-f_\nu) \sqrt{P^{\rm (C)}_{\rm cb,L}(k,z)}
\sqrt{P^{\rm (C)}_{\nu,{\rm L}}(k,z)}
+ f_\nu^2 
P^{\rm (C)}_{\nu,{\rm L}}(k,z)~,
\label{eq:P_tot_approx}
\end{equation}
which misses the term proportional to $P^{\rm (C)}_{\rm cb,13}$ in (\ref{eq:P_tot}) and is otherwise equivalent.\footnote{Some works use a version of eq.~(\ref{eq:P_tot_approx}) where $P^{\rm (C)}_{\rm cb,L}$ is replaced by $P^{\rm (C)}_{\rm cb,NL}$ in the cross-term. This is not justified in the context of a one-loop expansion.} One issue is that the term $2 R_\nu^{\rm (C)} P^{\rm (C)}_{\rm cb,13}$, that is, the correlator $2\left\langle \left| \delta_{\nu}^{(1)} \delta_{\rm cb}^{(3)} \right| \right\rangle$, contains an infrared divergence which would normally cancel out when combining this term with $\left\langle \left| \delta_{\nu}^{(2)} \delta_{\rm cb}^{(2)} \right| \right\rangle$. We neglect the latter term because it contains second-order neutrino density fluctuations, but to be consistent we should subtract from the former an infrared divergence term proportional to $k^2 R_\nu(k) \, P_\mathrm{cb,L}(k)$. In practice, this does not matter, since this contribution is suppressed on large scales by the $k^2$ factor and on small scales by the $R_\nu$ factor. With our choice of infrared cut-off scale, the last term in eq.~(\ref{eq:P_tot}) changes the final result at most by a negligible 0.02\%.

We have already stressed that, in the presence of massive neutrinos, the density fluctuation $\delta_{\rm cb}^{\rm (C)}(k,z)\simeq \delta_{\rm cb}^{\rm (S)}(k,z)$ is no longer a separable function of wavenumber and redshift. Then, one must track the full time- and scale-dependence of the fluctuations. The hybrid approach of \cite{Garny:2020ilv,Garny:2022fsh} addresses this by matching the exact linear Boltzmann solution to a non-linear two-fluid EFT at a chosen redshift. In this method, one solves the coupled continuity/Euler equations for both the CDM+baryon fluid (with EFT counterterms) and the neutrino fluid (truncated to its lowest moments with an effective sound speed), and then integrates the exact Eulerian Green’s functions so that each nonlinear kernel carries the full scale- and time-dependence. The resulting precision, however, comes at substantial computational cost. As a faster alternative, a one-fluid EFTofLSS method developed in \cite{Aviles:2020wme,Aviles:2021que,Noriega:2022nhf} works with a single continuity/Euler system for the CDM+baryon component only, taking into account neutrino fluctuations in the gravitational source term of the Poisson equation (and approximating $\delta_\nu$ as $\delta_{\rm cb}$ rescaled by a ratio of linear transfer functions, as suggested in \cite{Lesgourgues:2009am,Blas:2014hya}). The exact time- and scale-dependent Eulerian kernels are obtained by mapping the Lagrangian-PT displacement solution into Eulerian space. It was further shown in \cite{Noriega:2022nhf} that, to percent-level accuracy, one may approximate those kernels by the standard EdS building blocks weighted by the exact, scale-dependent logarithmic growth rate. These ``fk-kernels'' reduce all loop integrals to FFTLog-friendly sums of power laws, yielding a fast yet accurate one-loop prediction in neutrino cosmologies. This picture is also consistent with the broader mixed-dark-matter EFTofLSS analysis of \cite{Verdiani:2025jcf}, which finds that the ordinary massive-neutrino case lies in the regime where an effective single-fluid description of the cold sector is appropriate. The method of \cite{Noriega:2022nhf} is implemented in the code FOLPS\footnote{\url{https://github.com/cosmodesi/FolpsD}} and has been used in the analysis of DESI full-shape data. Compared to a naive approach based on pure EdS kernels, this more accurate method improves the sensitivity to the neutrino mass by approximately 15\% \cite{DESI:2026haa}.

The results from these improved attempts to accurately model matter and galaxy clustering in presence of multiple components and scale-dependent growth rates suggest that in a $\Lambda$CDM cosmology with ordinary massive neutrinos, the prediction of a simple one-fluid EFTofLSS with EdS kernels is sufficiently accurate to fit current data. The authors of~\cite{Garny:2020ilv,Garny:2022fsh,Aviles:2020wme,Noriega:2022nhf} only find sub-percent differences on midly non-linear scales between the standard treatment with EdS kernels and their more refined approaches. Thus, for the sake of computational efficiency, codes such as CLASS-PT \cite{Chudaykin:2020aoj}, PyBird \cite{DAmico:2020kxu}, Velocileptors \cite{Chen:2020fxs}, or \classoneloop \cite{Linde:2024uzr} stick to this simple method, which we refer to as the ``standard approach".

In this standard approach, to predict the non-linear power spectrum at a given redshift, $P_{\rm cb,NL}^{\rm (C)}(k,z)$, one can choose either to perform a dedicated loop calculation based on $P_{\rm cb,L}^{\rm (C)}(k,z)$ (but with EdS kernels), or to compute the loops at a fixed redshift (still with EdS kernels), for instance $z=0$, and rescale the various contributions with appropriate powers of the scale-independent linear growth rate and growth factor. 
As already mentioned in section \ref{sec:scale_dep}, \classoneloop{} infers these factors in such way that they match the small-scale limit of the scale-dependent factors. With both massless or massive neutrinos, we find that these two strategies can be used indifferently. The difference between them is negligible compared to the effect of other corrections discussed in the next sections.

\subsection{Newtonian Motion gauge approach with/without $\Lambda$ correction \label{sec:NM_massive}}

We can safely apply the approach described in the section \ref{sec:NM_massless} to a $\Lambda$CDM model with massive neutrinos. The transformation from the synchronous or comoving gauge to a NM gauge is not the same as in the massless neutrino limit, with a new solution for $H_{\rm T}^{\rm (NL)}(k,z)$ that absorbs the effect of massive neutrinos. As long as the condition 
$|H_{\rm T}^{\rm (NL)}(k,z)|\ll 1$ is met at all wavenumbers and redshifts of interest, this transformation brings us to a NM gauge  in which the weak-field approximation holds while the trajectory of baryons and CDM is the same as in the Newtonian problem described at the linear level by eqs.~(\ref{eq:newton1}) - (\ref{eq:newton3}), which can also be written as:
\begin{align}
&\dot{\delta}_{\rm cb}^{\rm [Nl]} + \theta_{\rm cb}^{\rm [Nl]}=0~,
\label{eq:ContNl_massive}\\
&\dot{\theta}_{\rm cb}^{\rm [Nl]} + {\cal H} \, \theta_{\rm cb}^{\rm [Nl]} + \frac{3}{2} {\cal H}^2 \Omega_{\rm m} (1-f_\nu) \, \delta_{\rm cb}^{\rm [Nl]}=0~,
\label{eq:EulerNl_massive}
\end{align}
with time-dependent factors ${\cal H}(z)$ and $\Omega_{\rm m}(z) (1-f_\nu(z))=\Omega_{\rm c}(z)+\Omega_{\rm b}(z)$.
We thus obtain a closed differential system with coefficients that depend on time but not Fourier modes $k$.
These equations apply indifferently to ($\delta_{\rm cb}^{\rm [Nl]}$, $\theta_{\rm cb}^{\rm [Nl]}$) or ($\delta_{\rm cb}^{\rm (NM)}+3H_{\rm L}^{\rm (NM)}$, $\theta_{\rm cb}^{\rm (NM)}$). We conclude that the NM gauge allows us to deal with scale-independent linearized equations, like in the absence of massive neutrinos. The gauge transformation from an arbitrary gauge to the NM gauge has absorbed the $k$-dependence of the gravitational source term, which was visible for instance in eq.~(\ref{eq:growth_cbn}). As a consequence, these equations define a scale-independent linear growth factor $D$ and growth rate $f$ which exactly coincide with the small-scale limit of the scale-dependent growth rate/factor found in the synchronous or comoving gauge, that is, to the $D(z)$ and $f(z)$ used in \class{}, which follow from eq.~(\ref{eq:D2}) with $\Omega_x=\Omega_{\rm c}+\Omega_{\rm b}$.

Using the doublet notation $\psi_a$, the non-linear equation of motion for the Newtonian (or NM gauge) variables is given by eq.~(\ref{eq:eqmopsi}) with the same kernels $\gamma_{abc}$ as in the massless neutrino case, but a slightly different matrix\footnote{When comparing our matrix $\Omega_{ab}$ with that of eq.~(2.13) in \cite{Garny:2022fsh}, one should keep in mind that our $(D, f)$ are those of the NM gauge, coinciding with the small-scale $(D, f)$ of the synchronous or comoving gauge, while \cite{Garny:2022fsh} define their $(D,f)$ as those in a massless neutrino universe, coinciding with the large-scale $(D, f)$ in the synchronous or comoving gauge.} 
\begin{equation}
\Omega_{ab} =
\left(
    \begin{tabular}{cc}
    $0$ & $-1$ \\
    $-\frac{3}{2} \frac{\Omega_{\rm m}}{f^2}(1-f_\nu)$ & $\frac{3}{2} \frac{\Omega_{\rm m}}{f^2}(1-f_\nu) - 1$\\
    \end{tabular}
\right)~.  
\end{equation}
The NM gauge approach allowed us to absorb the scale-dependence of the growth rate and to formulate the problem with an exactly scale-independent matrix $\Omega_{ab}$, even in the presence of massive neutrinos. However, the time-dependence of $\Omega_{ab}$ (induced by the variation of $\Omega_{\rm m}/f^2$ during $\Lambda$ domination) cannot be absorbed by the gauge since it is present on all scales, while linear gauge transformations only have an impact on large scales. Like in the massless neutrino case, we need to take it into account differently.

\begin{figure}
    \centering
    \begin{minipage}{\textwidth}
        \includegraphics[width=0.95\linewidth]{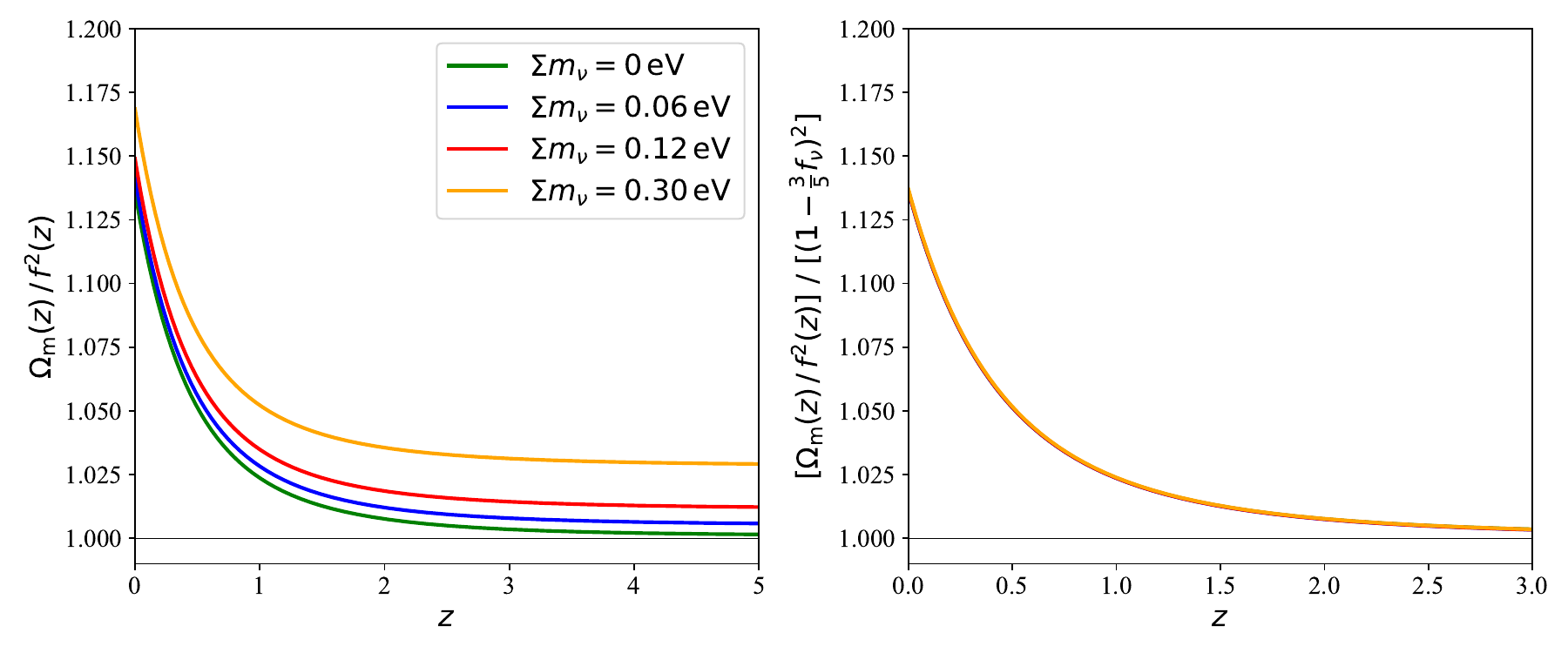}
    \end{minipage}
    \caption{(Left) Evolution of the ratio $\Omega_{\rm m}/f^2$ as a function of redshift the end of matter domination and during $\Lambda$ domination, for different values of the summed neutrino mass $\Sigma m_\nu$. (Right) After rescaling by a factor $(1-\frac{3}{5} f_\nu)^{-2}$, this ratio exhibits the same evolution as in a massless neutrino universe (and the four curves become indistinguishable). 
    \label{fig:Of}}
\end{figure}

In a cosmology with massive neutrinos, $\Omega_{\rm m}/f^2$ is rescaled at all times by a factor $(1-\frac{3}{5} f_\nu)^{-2}$. This is obvious during matter domination, since $\Omega_{\rm m}=1$ and  $\delta_{\rm cb} \propto a^{1-\frac{3}{5} f_\nu}$ (which would hold only in the small-scale limit for $\delta_{\rm cb}^{\rm (S)}$ or $\delta_{\rm cb}^{\rm (C)}$, while it holds at any scale for $\delta_{\rm cb}^{\rm [NL]}$). With a Boltzmann code, it is trivial to check that this remains true during $\Lambda$ domination. Figure~\ref{fig:Of} shows that, at any $z$, the approximation $\Omega_{\rm m}(z)/f^2(z)=(1-\frac{3}{5} f_\nu)^{-2}$ is just as good for $f_\nu\neq0$ as in a massless universe. Thus, we can define two schemes:
\begin{itemize}
\item If we want to neglect the effect of $\Lambda$ domination, we can use the approximation $\Omega_{\rm m}/f^2 \simeq (1-\frac{3}{5} f_\nu)^{-2}$. This means that the loop corrections can be computed with kernels $F_2$, $G_2$, $F_3$, $G_3$ involving a constant factor 
\begin{equation}
   \nu 
   \equiv 
   \frac{\Omega_{\rm m}}{f^2}(1-f_\nu)
   \simeq   
   \frac{1-f_\nu}{(1-\frac{3}{5} f_\nu)^2}~, 
   \label{eq:nuwithout}
\end{equation}
given in Appendix~\ref{app:constant_nu}.
We call this method the `NM gauge approach without $\Lambda$ corrections'. We expect this approach to improve over the standard one through its fully self-consistent treatment of massive neutrinos and of their impact on the growth of structure. On the other hand, it approximates $\Lambda$-domination effects exactly in the same way as the standard approach.
\item To incorporate $\Lambda$-domination effects, we compute the loop corrections with the time-dependent kernels of Appendix~\ref{app:time_nu} with 
\begin{equation}
   \nu(\tau) \equiv \frac{\Omega_{\rm m}(\tau)}{f^2(\tau)} (1-f_\nu)~.
   \label{eq:nuwith}
\end{equation}
This method accounts for all the (relativistic and non-relativistic) physical effects that affect the one-loop matter power spectrum in a $\Lambda$CDM cosmology with massive neutrinos.
\end{itemize}
We implemented these two schemes in \classoneloop.
Then, applying the standard EFTofLSS calculations to $P_{\rm cb,L}^{\rm [Nl]}$ (like we did before for massless neutrinos, but now with the $\nu$-dependent kernels) is justified and accurate. The effect of massive neutrinos are encoded, firstly, in the initial conditions for $\delta^{\rm [Nl]}_{\rm cb}(k,\tau_{\rm ini})$, which are implemented as described in section \ref{sec:IC};
secondly, in the evolution of linear perturbations, which results in a slower growth (e.g., $\delta^{\rm [Nl]}_{\rm cb}(\mathbf{k},\tau) \propto a^{1-\frac{3}{5} f_\nu}$ during matter domination); thirdly, in the $\nu$-dependence of the kernels in loop integrals;
and finally, in the final gauge transformation from the NM gauge to a more common gauge G (Poisson, synchronous or comoving), performed using using equation (\ref{eq:Pcb}). These four steps all `know' about the presence of massive neutrinos.

Finally, if we are interested in the total-matter one-loop power spectrum in an arbitrary gauge G, we need to expand 
\begin{equation}
    P^{\rm (G)}_{\rm m,NL} = \left\langle \left| (1-f_\nu) \, \delta^{\rm (G)}_{\rm cb,NL} + f_\nu \,\delta^{\rm (G)}_\nu \right|^2 \right\rangle = (1-f_\nu)^2 \left\langle \left| \delta^{\rm [Nl]}_{\rm cb,NL} + \Delta^{\rm (G)} + \frac{f_\nu}{1-f_\nu}  \delta^{\rm (G)}_\nu \right|^2 \right\rangle~.
\end{equation}
We can follow similar steps as in the calculation of Appendix~\ref{app:full}, treating both the gauge transformation term $\Delta^{\rm (G)}$ from eq.~(\ref{eq:delta_to_vel}) and the neutrino overdensity $\delta^{\rm (G)}_\nu$ as linear fields. We define the ratio of transfer functions 
\begin{equation}
    R^{\rm (G)}_{\rm m}(k,z) \equiv  
    \frac{\Delta^{\rm (G)}(k,z)+\frac{f_\nu}{(1-f_\nu)} \,\delta^{\rm (G)}_\nu (k,z)}{\delta^{\rm [Nl]}_{\rm cb,L}(k,z)}~,
\end{equation}
which is computed by the Bolztmann code. Finally, we can express the one-loop total-matter power spectrum as\footnote{As argued in section \ref{sec:SA}, one should in principle subtract an infrared divergence from the term $P^{\rm [Nl]}_{\rm cb,13}$, but this is not relevant in practice since this divergence is suppressed by some $k^2$ and $R_\mathrm{m}(k)$ factors. With our choice of infrared cut-off scale, the last term in eq.~(\ref{eq:Pm_full}) brings a negligible correction to total matter power spectrum.}
\begin{equation}
    P^{\rm (G)}_{\rm m,NL} = (1-f_\nu)^2 \left( 
    P^{\rm [Nl]}_{\rm cb,NL}
    + \left[2 R^{\rm (G)}_{\rm m} + \left(R^{\rm (G)}_{\rm m} \right)^2 \right] P^{\rm [Nl]}_{\rm cb,L}
    + 2 R^{\rm (G)}_{\rm m}
    P^{\rm [Nl]}_{\rm cb,13} \right)~.
    \label{eq:Pm_full}
\end{equation} 

\begin{figure}
    \centering
    \begin{minipage}{\textwidth}
        \includegraphics[width=0.95\linewidth]{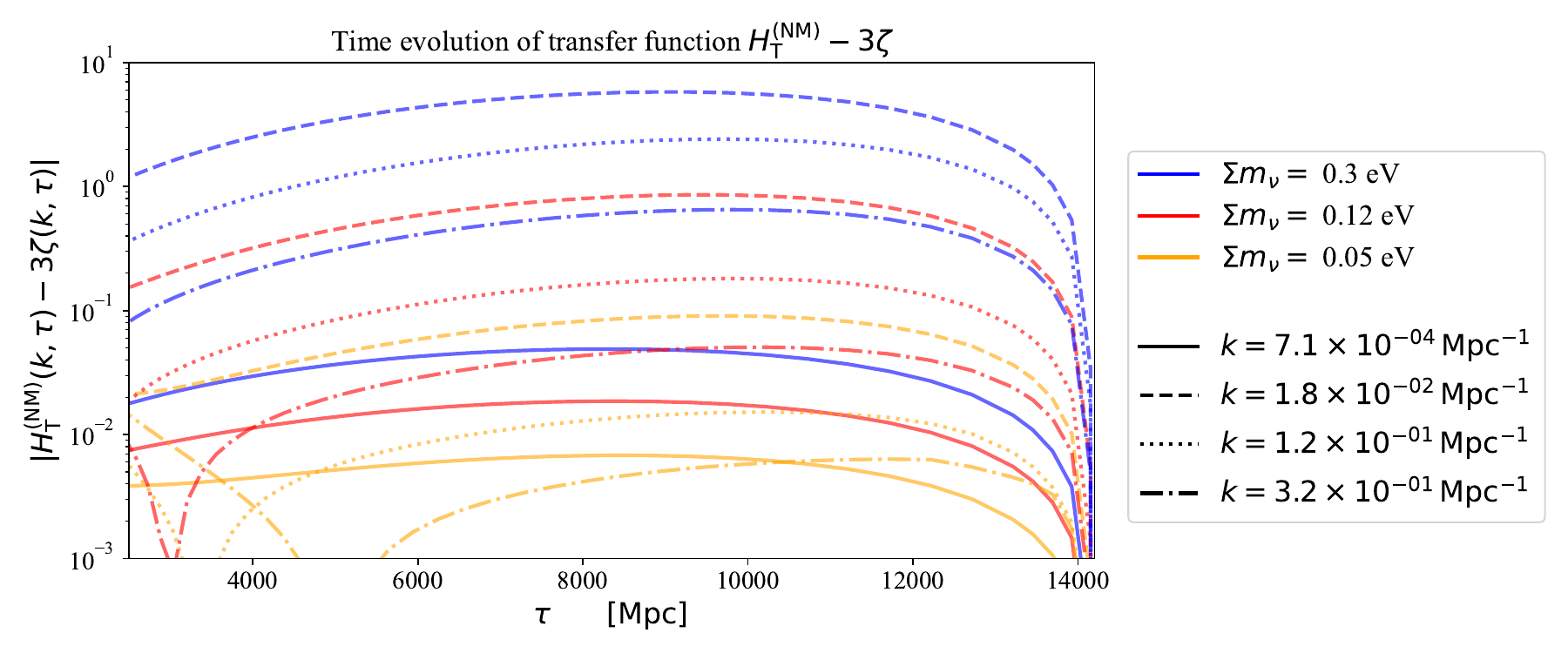}
    \end{minipage}
    \caption{Transfer function of the difference $H_{\rm T}-3\zeta$ as a function of conformal time $\tau$ (expressed in natural units of Mpc) for four wavenumbers $k$ and three neutrino masses $\Sigma m_\nu$. This difference quantifies the amplitude of the displacement field between the coordinates of the N-boisson and NM gauge, where the latter is defined using the backward method. Note that the actual field $H_{\rm T}-3\zeta$ is five orders of magnitude smaller than the displayed transfer function.
    \label{fig:HT}}
\end{figure}

\subsection{Results and comparison}

We run our modified version of \classoneloop{}  for a few $\Lambda$CDM models with three degenerate massive neutrinos, using either the standard approach or the NM gauge method to compute the non-linear power spectrum in the comoving gauge, $P^{\rm (C)}_{\rm cb,NL}(k,z)$. The summed neutrino mass is set to either $\Sigma m_\nu=0.05$, 0.12, or 0.3~eV.\\

\noindent {\bf Self-consistency check.} For each neutrino mass and using the backward method, Figure \ref{fig:HT} shows the transfer function of the difference $H_{\rm T}-3\zeta$ as a function of time at various wavenumbers. As discussed in previous sections, this difference quantifies the amplitude of the displacement field between the coordinates of the N-boisson and NM gauge, which is designed to absorb the effect of the scale-dependent growth factor. In this figure, we refer to the NM gauge defined using the backward method, such that all cases converge by construction towards $H_{\rm T}-3\zeta=0$ at the present time. We recall that the actual fields are five orders of magnitude smaller than the displayed transfer functions.

We find that the field $H_{\rm T}-3\zeta$ remains smaller than $10^{-4}$ on both small and large scales. As a matter of fact, on large scales, neutrinos are indistinguishable from CDM, while on small scales they free-steam and their perturbations get erased. Thus, in these two limits, there is no need to absorb any massive neutrino effect, and the perturbations in the NM gauge remain very close to their counterpart in the N-boisson gauge. On intermediate scales, neutrinos play a role and lead to an enhancement of $H_{\rm T}-3\zeta$. For $\Sigma m_\nu=0.3$~eV, the maximum value of the transfer function is reached for $k\sim 0.05$~Mpc$^{-1}$ and $\tau \sim 10^4$~Mpc. We also checked that $H_{\rm T}$ alone also remains perturbatively small, such that in each case the NM gauge is fully compatible with the weak-field approximation.

We checked that $H_{\rm T}-3\zeta$ becomes larger (by about one order of magnitude) when using the forward method from section~\ref{sec:IC}. We thus expect this method to be less accurate; As a matter of fact, when computing the one-loop matter power spectrum using the two methods, we find differences of the order of 0.05\% on mildly non-linear scales. In appendix~\ref{app:nu0}, we present a test that proves that the backward method does not introduce any error (up to the $\sim 0.001\%$ level) in the calculation of the one-loop power spectrum, while the forward method does. In what follows, our results are always derived from the backward method.\\

\begin{figure}
    \centering
    \begin{minipage}{\textwidth}
        \includegraphics[width=\linewidth]{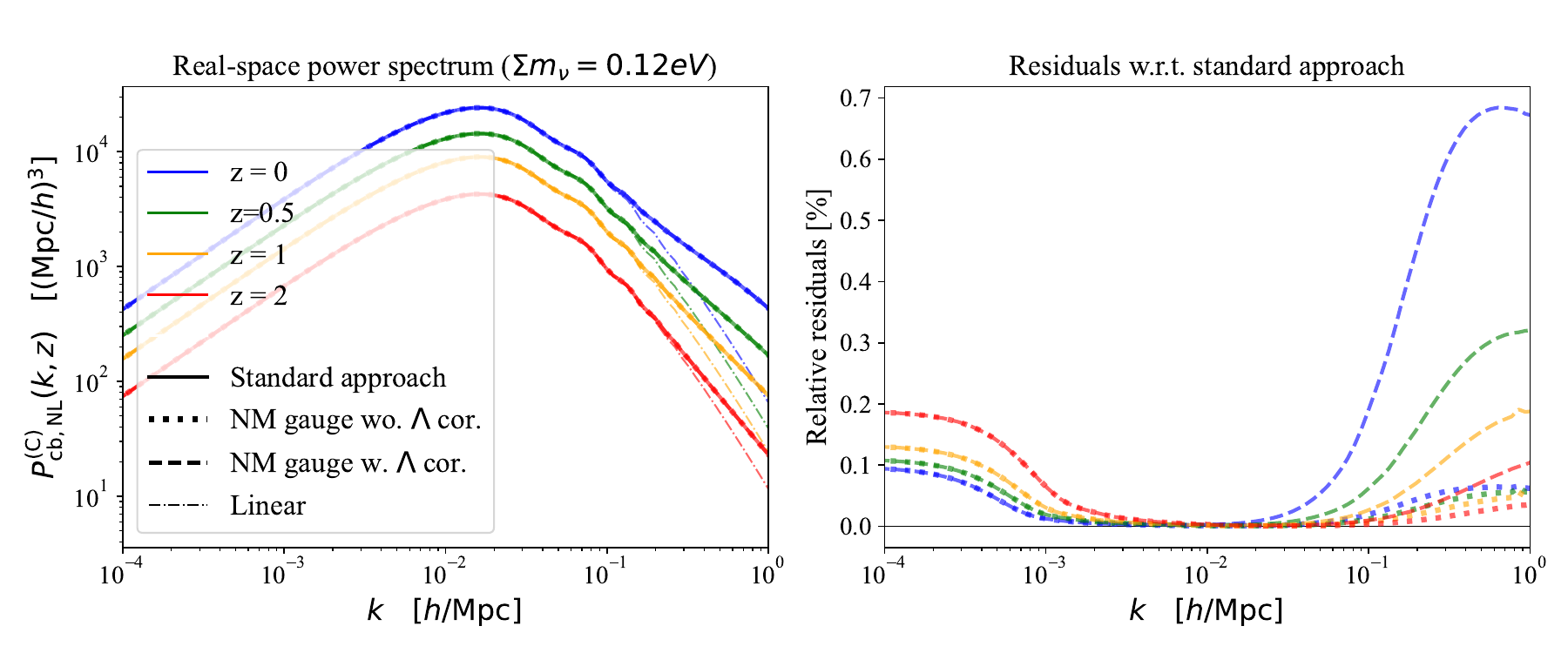}
    \end{minipage}
    \caption{(Left) Non-linear CDM+baryon power spectrum of comoving gauge density, $P^{\rm (C)}_{\rm cb,NL}(k,z)$, at $z=0, 0.5, 1, 2$, in a $\Lambda$CDM cosmology with three degenerate massive neutrinos and $\Sigma m_\nu=0.12$~eV, computed with different approaches: EFTofLSS applied to the linear comoving gauge power spectrum $P^{\rm (C)}_{\rm cb,L}$ (standard approach, solid lines); and EFTofLSS applied to the NM gauge power spectrum $P^{\rm [Nl]}_{\rm cb,L}$ inferred from the backward method and transformed back to the comoving gauge, either without (dotted lines) or with (dashed lines) the $\Lambda$ correction. We also show for reference the linear spectrum (dashed-dotted lines). (Right) Percentage difference in the non-linear spectrum from each method relative to that from the standard approach. On large scales, the dotted and dashed lines coincide.
    \label{fig:massive}}
\end{figure}

\noindent {\bf Results for fixed neutrino mass and different redshifts.} In figure \ref{fig:massive}, we compare the non-linear power spectrum $P^{\rm (C)}_{\rm cb,NL}(k,z)$ obtained from the standard or NM gauge approach  for a $\Lambda$CDM model with $\Sigma m_\nu=0.12$~eV, at redshifts $z=0, 0.5, 1$ or $2$. 

We first focus on the dotted lines in the right panel, which shows the percent difference between the `NM gauge without $\Lambda$ correction' and `standard' approaches:
\begin{itemize}
\item On large scales $k<10^{-2} h {\rm Mpc}^{-1}$, where the dotted lines exactly coincide with the dashed ones, the large-scale deviations are found to be exactly the same as for massless neutrinos. These deviations still comes from GR corrections and gravitational couplings with relativistic species. 
\item Instead, on small scales, we find some deviations of the order of $\sim 0.1\%$ caused specifically by massive neutrinos. The sign of the effect is consistent with expectations. The standard approach assumes the same matrix $\Omega_{ab}$ as if there would be massless neutrinos, and thus, quickly-growing perturbations $\delta_{\rm cb}$. The NM gauge approach features a $\nu$-dependent matrix $\Omega_{ab}$ with $\nu$ given by eq.~(\ref{eq:nuwithout}), which describes a reduced growth rate of perturbations. However, in both method, the non-linear power spectrum is inferred from the linear one normalized {\it today}. Thus, the standard approach implicitly assumes that perturbations were smaller in the past than they should, and predicts smaller non-linear corrections. The amplitude of this small-scale effect is however extremely small compared, e.g., to the sensitivity of Stage-IV surveys: only $\sim 0.06\%$ at $z=0$, and even smaller at higher $z$ (mainly because non-linear
corrections are overall smaller at high redshift compared to the linear power spectrum). The smallness of these corrections is consistent with previous claims in the literature, advocating that the standard method is sufficient to account for massive neutrino effects.\footnote{The standard method takes into account the reduced growth rate induced by massive neutrinos simply by using the correct linear power spectrum at $z=0$ as an input for loop calculations. It does not take this effect into account at the level of the kernels, which means that it has a ``wrong memory of the past''. However, our results show that this is irrelevant up to 0.06\% corrections for a summed neutrino mass of 0.12~eV.} 
\end{itemize}

We then consider the dashed lines in the right panel, which shows the percent difference between the `NM gauge with $\Lambda$ correction' and `standard' approaches. The $\Lambda$ correction enhances loop contributions by an additional
0.54\% at $z=0$. Thus, in total, we find a $\sim 0.60\%$
enhancement at $k=0.3\,h\,{\rm Mpc}^{-1}$ and $z=0$ in the true one-loop power spectrum compared to the standard EdS one. This is highly consistent with the results of \cite{Garny:2020ilv} (figure 7, comparison between the dashed green and dashed red lines) obtained with full time and scale-dependent kernels. The $\Lambda$ correction decreases at high $z$: at $z>2$, the total correction is dominated by neutrinos.\\

\begin{figure}
    \centering
    \begin{minipage}{\textwidth}
        \includegraphics[width=\linewidth]{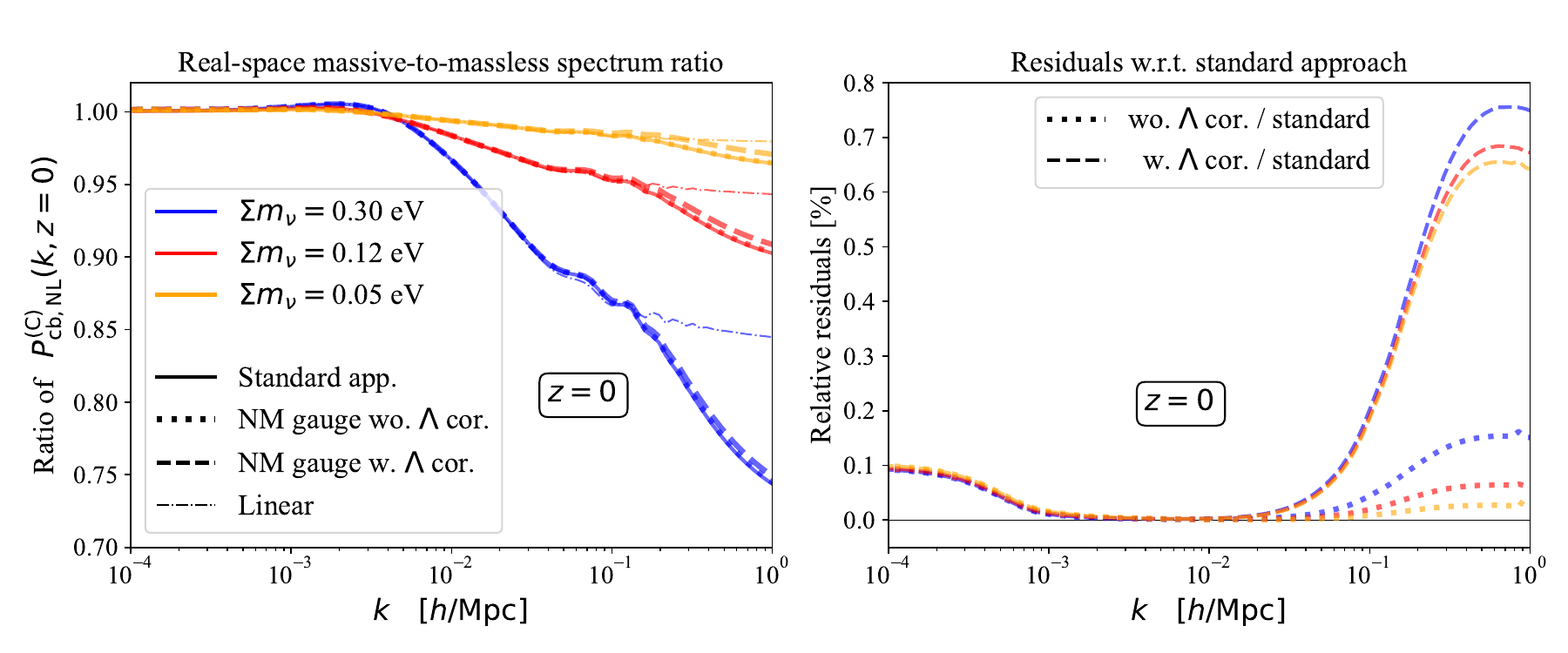}
        \includegraphics[width=\linewidth]{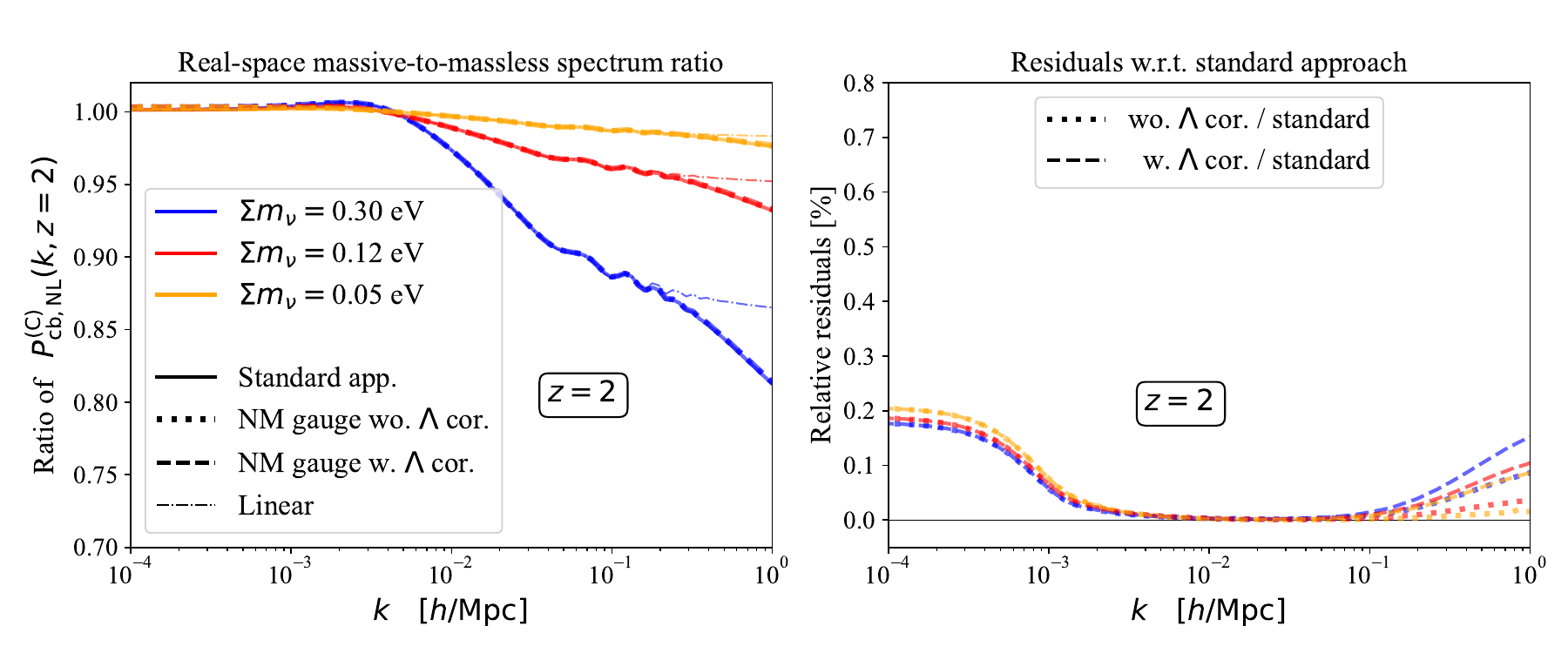}
    \end{minipage}
    \caption{(Left panels) Ratio of massive-to-massless CDM+baryon power spectra of comoving gauge density, $P^{\rm (C)}_{\rm cb,NL}(k,z)$, in a $\Lambda$CDM cosmology with three degenerate massive neutrinos and $\Sigma m_\nu=0.05, \, 0.12, \, 0.3\,$eV, computed at $z=0$ (top) or $z=2$ (bottom) with different approaches: EFTofLSS applied to the comoving gauge power spectrum $P^{\rm (C)}_{\rm cb,L}$ (standard approach, solid); EFTofLSS applied to the NM gauge power spectrum $P^{\rm [Nl]}_{\rm cb,L}$ and transformed back to the comoving gauge (NM gauge), either without (dotted lines) or with (dashed lines) the $\Lambda$ correction; and linear theory (dot-dashed). (Right) For each summed neutrino mass, percentage difference between the power spectrum of the NM gauge and standard approach, at $z=0$ (top) or $z=2$ (bottom).
    \label{fig:different-masses}}
\end{figure}

\noindent {\bf Results for different neutrino masses.} In figure \ref{fig:different-masses}, the left panels show the ratio of massive-to-massless neutrino power spectra computed using the various approaches for three different values of the total neutrino mass $\Sigma m_\nu$, while the right panels show the percentage difference between NM gauge and standard predictions in each case. The top panels correspond to $z=0$ and the bottom ones to $z=2$. All massive and massless models share the same value of $h$, $\omega_{\rm b}$ and $\omega_{\rm m}=\omega_{\rm b}+\omega_{\rm c}+\omega_\nu$. 

On large scales, the difference between the standard and NM gauge depends very weakly on neutrino masses, but on small scales it grows with $\Sigma m_\nu$.
At $z=2$, the difference is only of $\sim 0.06\%$ at $k = 0.3\,h{\rm Mpc}^{-1}$ and for $\Sigma m_\nu=0.30 \, {\rm eV}$, with equal contribution from neutrino and $\Lambda$ effects
(as can be checked by comparing the dotted line, without $\Lambda$ corrections, with the dashed lines, with $\Lambda$ corrections). At $z=0$, the difference is significantly bigger, reaching $\sim 0.66\%$ at $k = 0.3\,h{\rm Mpc}^{-1}$ and for the same mass. In that case, it is dominated by $\Lambda$ corrections, since neutrino-only effects do not exceed 0.14\%.\\

\noindent {\bf Role of the growth factor.}
In figures \ref{fig:massive}, \ref{fig:different-masses}, the loop corrections of the `standard' and `NM gauge without $\Lambda$ correction' approaches  are computed at $z=0$ and rescaled to $z=2$ using the scale-independent growth factor $D(z)$.\footnote{Instead, the linear power spectrum is always rescaled consistently with $D(k,z)$.} We already stressed that in the NM gauge, there is no ambiguity on the fact that this growth factor should be the small-scale one, coming from eq. (\ref{eq:D2}) with $\Omega_x=\Omega_\mathrm{c}+\Omega_\mathrm{b}$, and scaling like $a^{1-\frac{3}{5} f_\nu}$ during matter domination. As a matter of fact, in this particular gauge, it describes the growth of $\delta_{\rm cb}$ on {\it all} scales. We also argued that the same choice should be performed when using the standard approach, since this growth factor is the one that applies approximately to the scales over which non-linear corrections are computed. In the standard approach, rescaling the loops with the large-scale $D(z)$ would introduce a very large error. This is illustrated by the right panel of figure~\ref{fig:rescaling} in appendix~\ref{app:rescaling}. With $\Sigma m_\nu=0.12\,{\rm eV}$, rescaling the loops from $z=2$ to $z=0$ using the large-scale $D(z)$ leads to a 1.5\% error at $k= 0.3 \, h\,{\rm Mpc}^{-1}$. 

\subsection{Total matter spectrum}

\begin{figure}
    \centering
    \begin{minipage}{\textwidth}
        \includegraphics[width=\linewidth]{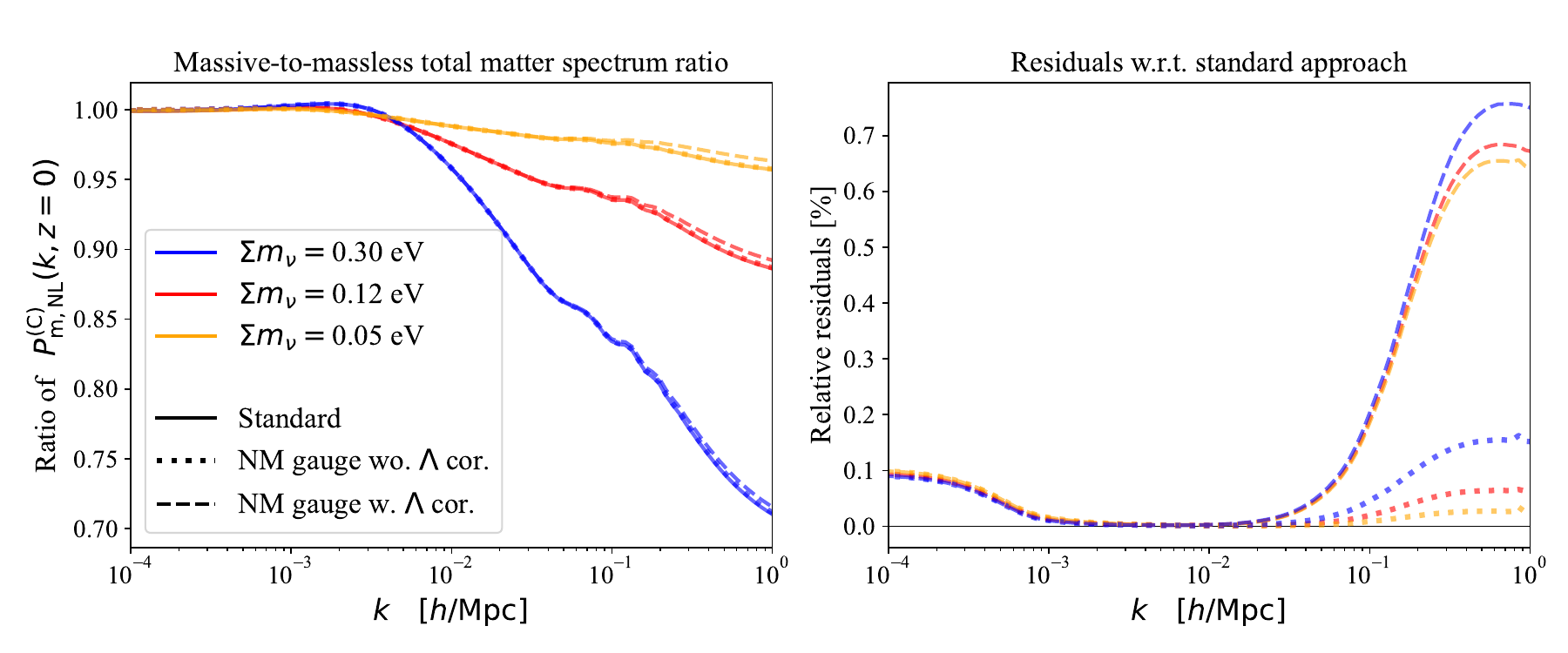}
    \end{minipage}
    \caption{(Left) Ratio of massive-to-massless matter power spectra of the comoving gauge density at $z=0$, $P^{\rm (C)}_{\rm m,NL}(k,z=0)$, in a $\Lambda$CDM cosmology with three degenerate massive neutrinos and $\Sigma m_\nu=0.05, \,0.12, \,0.3\,$eV, computed with different approaches: EFTofLSS applied to the linear comoving gauge power spectrum $P^{\rm (C),L}_{\rm cb}$, with neutrino perturbations added linearly at the end (standard approach, solid); and EFTofLSS applied to the NM gauge power spectrum $P^{\rm [Nl]}_{\rm cb,L}$ and transformed back to the comoving gauge, with neutrino perturbations added linearly at the end, using kernels without (dotted) or with (dashed) the $\Lambda$ correction. (Right) Percentage difference between the total matter power spectrum of the NM gauge and standard approaches at $z=0$.
    \label{fig:total-matter}}
\end{figure}

Since the NM gauge is constructed in such way that $\delta_{\rm cb}^{\rm (NM)}+3H_{\rm L}^{\rm (NM)}=\delta_{\rm cb}^{\rm [Nl]}$ obeys to Newtonian equations of evolution, there is no ambiguity on the fact that, in this method, the EFTofLSS contributions must be computed from $P_{\rm cb,L}^{\rm [Nl]}$ rather than $P_{\rm m,L}^{\rm [Nl]}$. If needed, the total-matter power spectrum in the comoving gauge, $P_{\rm m,NL}^{\rm (C)}$, can then be inferred from eq.~(\ref{eq:P_tot}). Similarly, in the standard approach, one should compute the loops corrections (and other EFTofLSS contributions) starting from $P_{\rm cb,L}^{\rm (C)}$ rather than $P_{\rm m,L}^{\rm (C)}$, since the results of the previous sections demonstrate that $\delta_{\rm cb}^{\rm (C)}$ remains very close to $\delta_{\rm cb}^{\rm [Nl]}$, and thus, close to a self-gravitating fluid with ~Newtonian evolution. The total matter power spectrum can then be inferred from eq. (\ref{eq:P_tot}) -- but with $P_{\rm cb,NL}^{\rm (C)}$ evaluated directly, instead of going through eq.~(\ref{eq:Pcb}).

In figure~\ref{fig:total-matter}, we compare the total matter power spectrum obtained using the different methods at $z=0$. The level of agreement between the `standard' and `NM gauge' approaches is very similar for $P_{\rm m,NL}^{\rm (C)}$ and $P_{\rm cb,NL}^{\rm (C)}$, as can be seen by comparing figure~\ref{fig:total-matter} with the top panels of figure~\ref{fig:different-masses}. To further illustrate that computing the loop contributions from $P_{\rm cb,L}^{\rm (C)}$ rather than $P_{\rm m,L}^{\rm (C)}$ in the standard approach is the right choice, we show in figure \ref{fig:total-matter-wrong} of appendix~\ref{app:total_matter} the result obtained when using the latter. The wrong prescription introduces a 1.2\% error at  $k=0.3\,h {\rm Mpc}^{-1}$ for $\Sigma m_\nu=0.30\,{\rm eV}$.

\subsection{Summary of real-space results}

In conclusion of section \ref{sec:apply_NM_massive}, we find that at the level of either the CDM+baryon or total matter power spectrum in real space, the conventional EFT analysis remains accurate for cosmologies with massive neutrinos, provided that it is applied correctly (starting from the linear `cb' spectrum in the conformal or synchronous gauge, using the small-scale growth factor, and adding neutrinos linearly). A comparison with the more consistent NM gauge approach confirms that there is no need to include a more complex treatment of massive neutrinos, given the current observational bounds on the total mass of neutrinos. As a matter of fact, if we restrict ourselves to non-linear scales, neutrinos introduce an approximately scale-independent linear growth rate (e.g., $D\propto a^{1-\frac{3}{5}f_\nu}$ during matter domination), which is consistent with the assumption of scale-independent kernels; and the deviation of these kernels from the EdS ones has a sub-percent impact on the final result. The accurate one-loop power spectrum predicted by the NM gauge approach deviates from the standard approach by at most $\sim 0.66\%$ for $z=0$, $k=0.3\,h{\rm Mpc}^{-1}$ and $\sum m_\nu = 0.30\,{\rm eV}$, with about $\sim 0.14\%$ coming from neutrino effects and $\sim 0.51\%$ from $\Lambda$ effects. On the other hand, on large scales, linear theory remains accurate and GR corrections are typically of order $\sim 0.1\%$. 

\section{Redshift-space power spectrum with massive neutrinos\label{sec:rsd}}

\subsection{Standard approach}

In the standard EFTofLSS approach, redshift-space distortions are built from a purely Newtonian point of view. The starting point is the transformation from the densities in ordinary Fourier space, $\delta(\mathbf{k})$, to those in the configuration space of the past light-cone, 
$\delta_r(\mathbf{k})$, given by \cite{Kaiser:1987qv}
\begin{equation}
    \delta_r(\mathbf{k})
    =
    \delta(\mathbf{k})
    + \int d^3\mathbf{x}\,\,
    e^{-i \mathbf{k} \cdot \mathbf{x}}
    \left(
    e^{-i \frac{k_z}{\cal H} u_z(\mathbf{x})}
    -1\right)
    (1+\delta(\mathbf{x}))~,
    \label{eq:rsd_transf}
\end{equation}
where ($k_z$, $u_z$) are respectively the line-of-sight components of the wavevector $\mathbf{k}$ and peculiar velocity field $\mathbf{u}$. This transformation can be plugged into the expansion of correlation functions (like the power spectrum) at a given order in perturbation theory. The result can be expressed as a series depending on the Newtonian density and velocity divergence fields ($\delta_{\rm cb}^{\rm [Nl]}$, $\theta_{\rm cb}^{\rm [Nl]}$). Finally, the redshift-space power spectrum can be expressed schematically (up to counter-terms) as
\begin{equation}
    P_{\rm cb,NL}^{\rm [Nl]}(k,\mu,z)=
    \left\langle \left| \delta_{\rm cb}^{\rm [Nl]} - \frac{\mu^2}{\cal H} \theta_{\rm cb}^{\rm [Nl]} \right|^2 \right\rangle
    + \Delta P_{\rm loops}
    \left[
    \delta_{\rm cb}^{\rm [Nl]},
    \theta_{\rm cb}^{\rm [Nl]}
    \right]~,
    \label{eq:newt_rsd}
\end{equation}
where $\mu$ is the cosine of the line-of-sight angle, while the loop corrections $\Delta P_{\rm loops}$ include higher-order moments of the fields ($\delta_{\rm cb}^{\rm [Nl]}$, $\theta_{\rm cb}^{\rm [Nl]}$). 
While the aim of the standard EFTofLSS approach is to work with purely Newtonian variables, we have seen that in practice, standard implementations rely on the statistics of the linear density contrast in the comoving or synchronous gauge, $\delta_{\rm cb}^{\rm (C)}\simeq\delta_{\rm cb}^{\rm (S)}$, as computed by Boltzmann codes. However, the divergence of the Newtonian velocity, $\theta_{\rm cb}^{\rm [Nl]}$, is approximated with the growing mode (GM) velocity $-\dot{\delta}_{\rm cb, GM}^{\rm (C)} \simeq -\dot{\delta}_{\rm cb, GM}^{\rm (S)}$, rather than the synchronous gauge velocity, $\theta_{\rm cb}^{\rm (S)}$, which is vanishingly small by definition.\footnote{We recall that the synchronous gauge used in Boltzmann codes is assumed to be comoving with CDM, $\theta_{\rm c}^{\rm (S)}=0$. Thus, $\theta_{\rm cb}^{\rm (S)}$ is only seeded by corrections from the relative velocity between baryons and CDM, which are only present at very early times and small scales, and play a negligible role in this problem.} In the EdS approximation, which assumes separable solutions and a scale-independent growth factor, the relation  $\dot{\delta}_{\rm cb, GM}^{\rm (C)}(k,z) = f(z)\, {\cal H}(z) \, \delta_{\rm cb}^{\rm (C)}(k,z)$ holds. Under all these approximations, the leading-order term in eq. (\ref{eq:newt_rsd}) simplifies,
\begin{align}
    P_{\rm cb,NL}^{\rm [Nl]}(k,\mu,z) &\simeq \left(1+f \mu^2 \right)^2
    \left\langle \left| \delta_{\rm cb}^{\rm (C)} \right|^2 \right\rangle
    + \Delta P_{\rm loops}
    \left[
    \delta_{\rm cb}^{\rm (C)}
    \right] \nonumber\\
    &= \left(1+f \mu^2 \right)^2
    P_{\rm cb,L}^{\rm (C)}(k,z)
    + \Delta P_{\rm loops}
        \left[
    \delta_{\rm cb}^{\rm (C)}
    \right]~,
    \label{eq:Pcb_rsd_standard}
\end{align}
and one recovers the famous Kaiser formula at tree level \cite{Kaiser:1987qv}.

The overall goal of this work is to put the standard Newtonian calculation back in a relativistic framework and to compute the loops in a gauge where the the solution for linear perturbations are separable functions of $k$ and $\tau$. It has been shown by the authors of refs.~\cite{Jeong:2011as,Bonvin:2011bg,Jeong:2014ufa} that in a general relativistic context, 
the observable power spectrum in redshift space takes the form of eq.~(\ref{eq:newt_rsd})
provided that densities and velocities are expressed in a gauge-independent way, using the gauge-independent density $D_x$ defined in eq.~(\ref{eq:def_D}) and velocity divergence defined in eq.~(\ref{eq:def_V}). It turns out that $D_x$ coincides with $\delta_x^{\rm (C)}$, while $V_x$ coincides with $\theta_x^{\rm (C)}=\theta_x^{\rm (P)}$. Thus, in a relativistic context, we may express the observable (gauge-independent) redshift-space power spectrum as
\begin{align}
    P_{\rm cb,NL}^{\rm (C)}(k,\mu,z)
        &=
    \left\langle \left| D_{\rm cb} - \frac{\mu^2}{\cal H} V_{\rm cb} \right|^2 \right\rangle
    + \Delta P_{\rm loops}
    \left[
    D_{\rm cb},
    V_{\rm cb}
    \right] \label{eq:P_gi_rsd0}
    \\
    &=
    \left\langle \left| \delta_{\rm cb}^{\rm (C)} - \frac{\mu^2}{\cal H} \theta_{\rm cb}^{\rm (C)} \right|^2 \right\rangle
    + \Delta P_{\rm loops}
    \left[
    \delta_{\rm cb}^{\rm (C)},
    \theta_{\rm cb}^{\rm (C)}
    \right]~.
    \label{eq:P_gi_rsd}
\end{align}
Using the continuity equation for non-relativistic species in the synchronous gauge, 
\begin{equation}
\dot{\delta}_{\rm cb}^{\rm (S)}+\theta_{\rm cb}^{\rm (S)}=-3 \dot{H}_{\rm L}^{\rm (S)}~,
\end{equation}
in combination with $\theta_{\rm cb}^{\rm (P)}=\theta_{\rm cb}^{\rm (S)}-\dot{H}_{\rm T}^{\rm (S)}$ and with eq. (\ref{eq:D_to_G_and_S}), one gets
\begin{equation}
\theta_{\rm cb}^{\rm (C)} =
  \theta_{\rm cb}^{\rm (P)} = -  \dot{\delta}_{\rm cb}^{\rm (C)} 
  + \left[  
  - \dot{H}_{\rm T}^{\rm (S)}
  - 3 \dot{H}_{\rm L}^{\rm (S)}
  + 3 \frac{\cal H}{k^2} \theta_{\rm tot}^{\rm (S)}
  \right]
  = - \dot{\delta}_{\rm cb}^{\rm (C)} 
  + 3 \left[  
  \dot{\eta}
  + \frac{\cal H}{k^2} \theta_{\rm tot}^{\rm (S)}
  \right] 
  = - \dot{\delta}_{\rm cb}^{\rm (C)}  - 3 \zeta~.
\end{equation}
In the standard approach for the computation of the redshift-space power spectrum, there are three approximations: the contribution of the term $-3 \zeta$ to $\theta_{\rm cb}^{\rm (C)}$ is neglected, $\dot{\delta}_{\rm cb}^{\rm (C)}$ is replaced by $f \, {\cal H} \, \delta_{\rm cb}^{\rm (C)}$ with a scale-independent $f$, and $\delta_{\rm cb}^{\rm (C)}$ is treated as separable in $k$ and $\tau$, with a growth described by EdS kernels.

The comparison with results from the NM gauge approach in real space already showed that the last of these three approximations was accurate enough for $\Lambda$CDM cosmologies with massless or massive neutrinos. We will now generalize this approach to redshift space and check whether the first assumptions are equally accurate.

\subsection{Newtonian Motion gauge approach with/without $\Lambda$ correction\label{sec:NM_method_massless}}

We want to evaluate the gauge-invariant power spectrum of eq. (\ref{eq:P_gi_rsd0}) without making any approximation on the growth rate of perturbations. According to the discussion in section~\ref{sec:NM_massive}, EFTofLSS calculations  with separable functions of $k$ and $\tau$ and scale-independent kernels are fully justified only in NM gauges, that is, when we compute the moments of the fields 
\begin{equation}
(\delta_{\rm cb}^{\rm [Nl]}, \theta_{\rm cb}^{\rm[Nl]})=(\delta_{\rm cb}^{\rm (NM)}+3 H_{\rm L}^{\rm (NM)}, \theta_{\rm cb}^{\rm(NM)})~,
\end{equation}
related to the fields appearing in eq.~(\ref{eq:P_gi_rsd0}) or (\ref{eq:P_gi_rsd}) through
\begin{align}
    \delta_{\rm cb}^{\rm(C)} &= 
    \delta_{\rm cb}^{\rm[Nl]} + \Delta^{\rm (C)} = \delta_{\rm cb}^{\rm[Nl]}
    +H_{\rm T}^{\rm (NM)}-3\zeta~, \\
    \theta_{\rm cb}^{\rm (C)} &=
    \theta_{\rm cb}^{\rm[Nl]}
    - \dot{H}_{\rm T}^{\rm(NM)}~,
    \label{eq:thetacbnl}
\end{align}
where in the last equality we used the expression of $V_{\rm cb}$ in the comoving and NM gauges. Thus, we can express the observable power spectrum of eq. (\ref{eq:P_gi_rsd0}) as
\begin{equation}
    P_{\rm cb,NL}^{\rm (C)}(k,\mu,z)
    =
    \left\langle \left| \delta_{\rm cb}^{\rm [Nl]} - \frac{\mu^2}{\cal H} \theta_{\rm cb}^{\rm [Nl]} 
    +\Delta^{\rm (RSD)}
    \right|^2 \right\rangle
    + \Delta P_{\rm loops}
    \left[
    \delta_{\rm cb}^{\rm [Nl]}+\Delta^{\rm (C)},
    \theta_{\rm cb}^{\rm [Nl]}-\dot{H}_{\rm T}^{\rm (NM)}
    \right]~,
    \label{eq:P_expand_rsd}
\end{equation}
where we defined
\begin{equation}
   \Delta^{\rm (RSD)}(k,\mu,z) \equiv \Delta^{\rm (C)} (k,z) 
   + \frac{\mu^2}{\cal H} \dot{H}_{\rm T}^{\rm (NM)}(k,z) = H_{\rm T}^{\rm (NM)}
   -3\zeta + \frac{\mu^2}{\cal H} \dot{H}_{\rm T}^{\rm (NM)}~.
   \label{eq:rsd}
\end{equation}
When the linear power spectrum $P_{\rm cb,L}^{\rm [Nl]}$ is passed to \classoneloop{}, the module computes the one-loop redshift-space power spectrum  $P_{\rm cb, NL}^{\rm [Nl]}$ assuming consistently $\theta_{\rm cb}^{\rm [Nl]} = - f \, {\cal H} \, \dot{\delta}_{\rm cb}^{\rm [Nl]}$ with a scale-independent $f$. This relation is exact, unlike in the standard approach based on comoving-gauge variables. Like for the real-space power spectrum, the loop calculation is performed with the $\nu$-dependent kernels that account for the modified growth of density and velocity perturbations caused by massive neutrinos, and possibly also by $\Lambda$ domination: we can use eq.~(\ref{eq:nuwithout}) to define the kernels $F_2$, $G_2$, $F_3$, $G_3$ in the NM gauge approach without $\Lambda$ corrections, or eq.~(\ref{eq:nuwith}) in the approach with $\Lambda$ corrections.

To compute the spectrum $P_{\rm cb,NL}^{\rm (C)}(k,\mu,z)$ at a given order using eq.~(\ref{eq:P_expand_rsd}), we would need to compute the moments of the linear and gaussian fields ($\delta_{\rm cb}^{(1)}$,
$\theta_{\rm cb}^{(1)}$,
$\Delta^{\rm (C)}$,
$\dot{H}_{\rm T}^{\rm (NM)}$) at this order, where the non-linear fields ($\delta_{\rm cb}^{\rm [Nl]}$,
$\theta_{\rm cb}^{\rm [Nl]}$) are expanded in powers of $\delta_{\rm cb}^{(1)}$,
$\theta_{\rm cb}^{(1)}$.
The moments that do not contain  ($\Delta^{\rm (C)}$,
$\dot{H}_{\rm T}^{\rm (NM)}$) are the ones computed by standard EFTofLSS algorithms when the linear power spectrum of the Newtonian fluid $P_{\rm cb,L}^{\rm [Nl]}$ is passed in input. The sum of these moments provides the non-linear power spectrum in redshift space, $P_{\rm cb,NL}^{\rm [Nl]}$, not accounting for relativistic and scale-dependent growth effects. In principle, 
the additional moments containing the fields ($\Delta^{\rm (C)}$, $\dot{H}_{\rm T}^{\rm (NM)}$)
could be computed explicitly, noting however that $\Delta^{\rm (C)}$ and $\dot{H}_{\rm T}^{\rm (NM)}$ are not exactly separable functions of time and wavenumber. However, like in section~\ref{sec:NM_massless}, we find that it is not necessary to perform the full explicit calculation in order to capture leading-order effects, because the problem features a separation of scales: 
\begin{itemize}
    \item On small scales, $\delta^{\rm [Nl]}_{\rm cb}$ and $\theta^{\rm [Nl]}_{\rm cb}$ grow non-linear, while $\Delta^{\rm (G)}$ and $\dot{H}_{\rm T}^{\rm (NM)}$ remain linear and totally negligible compared to these fields, such that
\begin{equation}
    P_{\rm cb,NL}^{\rm (C)}(k,\mu,z)
    \simeq 
    \left\langle \left| \delta_{\rm cb}^{\rm [Nl]} - \frac{\mu^2}{\cal H} \theta_{\rm cb}^{\rm [Nl]} 
    \right|^2 \right\rangle
    + \Delta P_{\rm loops}
    \left[
    \delta_{\rm cb}^{\rm [Nl]},
    \theta_{\rm cb}^{\rm [Nl]}
    \right]=P_{\rm cb,NL}^{\rm [Nl]}(k,\mu,z)~.
\end{equation}
\item On large scales, all fields remain linear and $\Delta P_{\rm loops}$ is negligible. Then, the power spectrum reads
\begin{equation}
    P_{\rm cb,L}^{\rm (C)}(k,\mu,z)
\simeq P_{\rm cb,L}^{\rm [Nl]}(k,\mu,z) + 2 \sqrt{P_{\Delta^{\rm (RSD)}}(k,\mu,z)} \sqrt{P_{\rm cb,L}^{\rm [Nl]}(k,\mu,z)} + P_{\Delta^{\rm (RSD)}}(k,\mu,z)~,
\end{equation}
with
\begin{equation}
    P_{\rm cb,L}^{\rm [Nl]}(k,\mu,z)
    =
    \left\langle \left| \delta_{\rm cb}^{\rm [Nl]} - \frac{\mu^2}{\cal H} \theta_{\rm cb}^{\rm [Nl]} 
    \right|^2 \right\rangle~,
\qquad
    P_{{\Delta}^{\rm (RSD)}}(k,\mu,z)
    =
    \left\langle \left| 
    \Delta^{\rm (RSD)}
    \right|^2 \right\rangle~.
\end{equation}
\item On intermediate scales, the effect of the gauge transformation and of non-linear corrections are both negligible.
\end{itemize}
Thus, to a very good approximation, the result for the observable power spectrum in redshift space is captured on all scales by the expression
    \begin{equation}
    P_{\rm cb,NL}^{\rm (C)}(k,\mu,z) 
\simeq P_{\rm cb,NL}^{\rm [Nl]}(k,\mu,z) + 2 \sqrt{P_{\Delta^{\rm (RSD)}}(k,\mu,z)} \sqrt{P_{\rm cb,NL}^{\rm [Nl]}(k,\mu,z)} + P_{\Delta^{\rm (RSD)}}(k,\mu,z)~.
\label{eq:Pcb_RSD}
\end{equation}
We stress that, like in real space, this expression applies even in presence of massive neutrinos. Thanks to the use of the NM gauge for the calculation of EFTofLSS contribution, it captures general-relativistic effects, gravitational couplings with relativistic species, and scale-dependent growth due to free-streaming species. These effects are implemented through the peculiar initial conditions and background equations of the Newtonian fluid used for the calculation of $P_{\rm cb,L}^{\rm [Nl]}$, and by corrections in $P_{\Delta^{\rm (RSD)}}$ that account for the transformation from the NM gauge to gauge-independent variables accounting for observable quantities. When the kernels are defined using eq.~(\ref{eq:nuwith}), the effect of $\Lambda$ domination is also incorporated in the calculation of the one-loop power spectrum.

\begin{figure}
    \centering
    \begin{minipage}{\textwidth}
        \includegraphics[width=\linewidth]{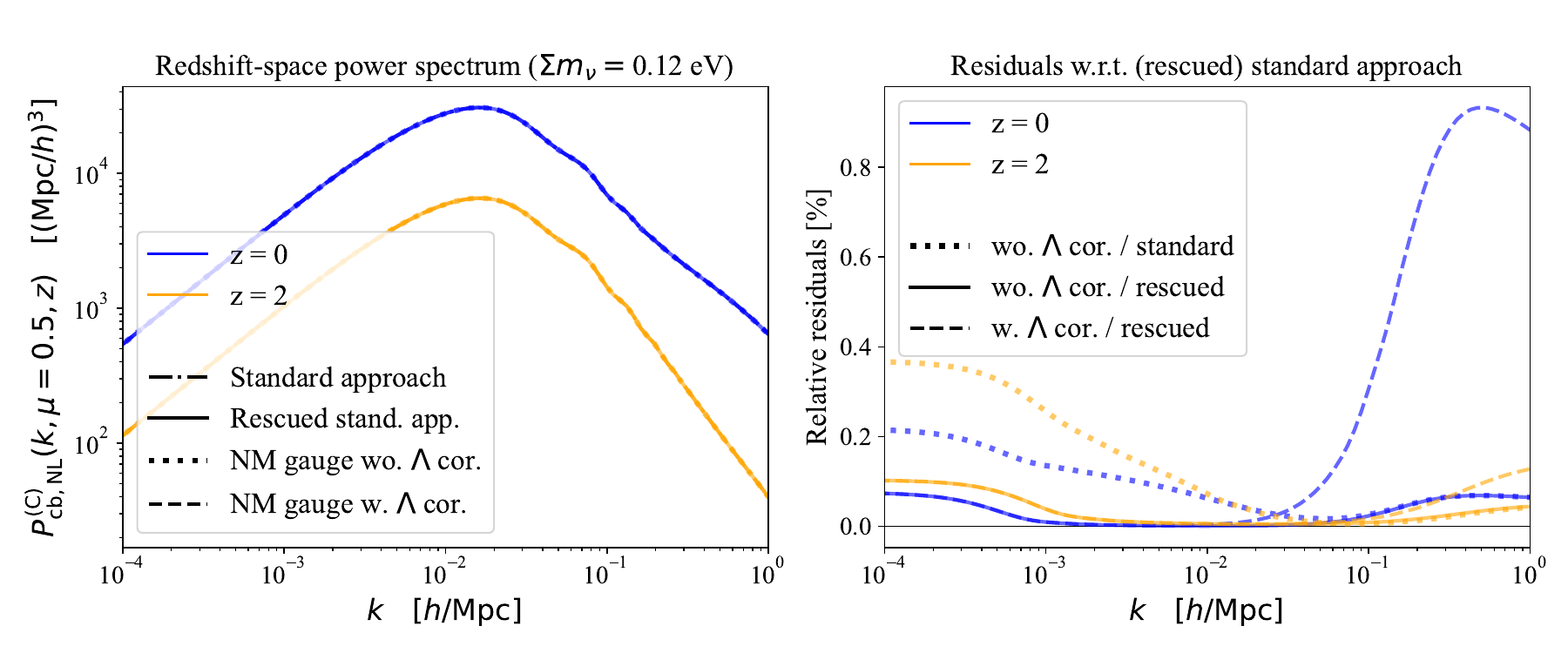}
    \end{minipage}
    \caption{(Left) One-loop redshift-space CDM+baryon power spectrum $P_{\rm cb,NL}^{\rm (C)}(k,\mu,z)$ at $\mu=0.5$ and $z=0,2$, in a $\Lambda$CDM cosmology with three degenerate massive neutrinos and $\Sigma m_\nu=0.12$~eV, computed with three different approaches:  EFTofLSS applied to the linear comoving gauge power spectrum $P^{\rm (C)}_{\rm cb,L}$ (standard approach, dot-dashed); standard method `rescued' by using the scale-dependent growth rate $f(k,z)$ in the Kaiser term (solid); and EFTofLSS applied to the NM gauge power spectrum $P^{\rm [Nl]}_{\rm cb,L}$ and transformed back to the comoving gauge, either without (dotted) or with (dashed) $\Lambda$ correction. (Right) Percentage difference between the non-linear power spectra obtained from the NM gauge approach (with/without $\Lambda$ corrections) and the standard approach (with/without a scale-dependent growth rate in the Kaiser term).
    \label{fig:massiveRSD}}
\end{figure}

\begin{figure}
    \centering
    \begin{minipage}{\textwidth}
        \includegraphics[width=\linewidth]{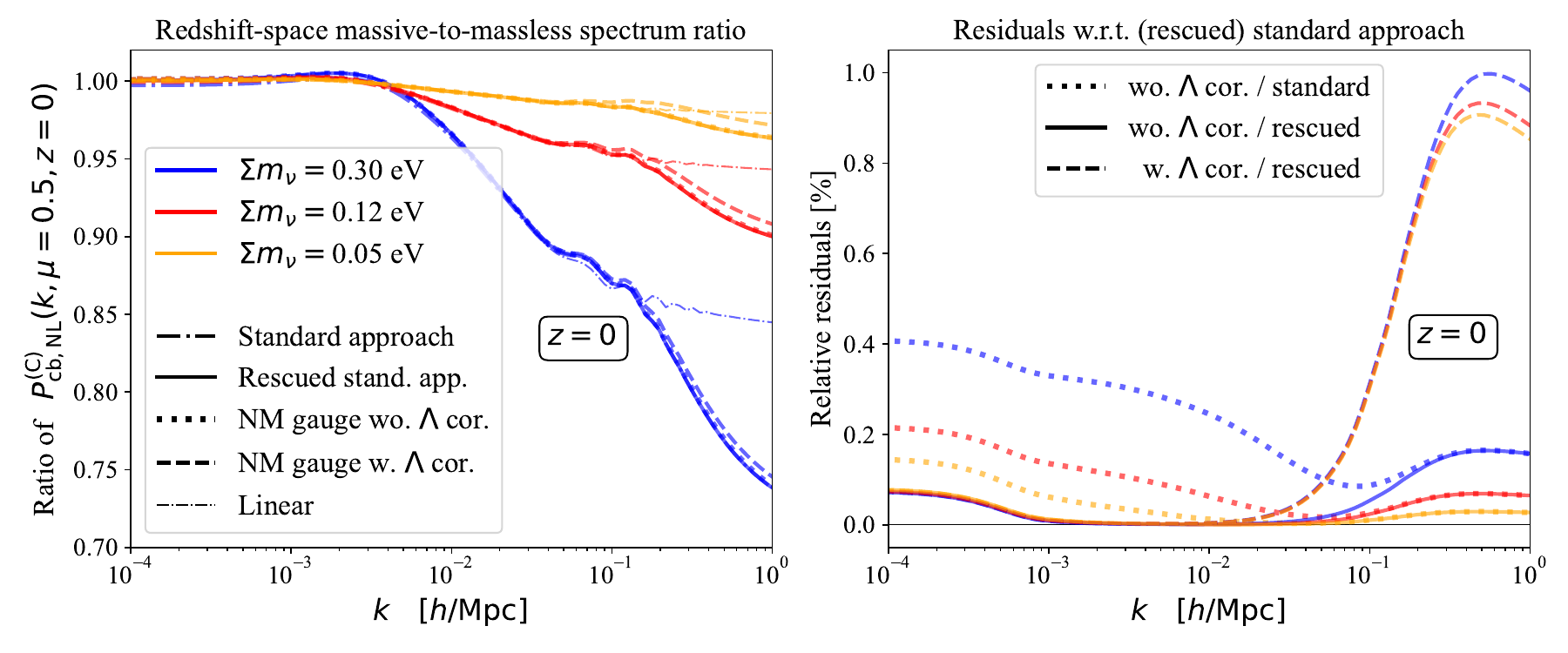}
        \includegraphics[width=\linewidth]{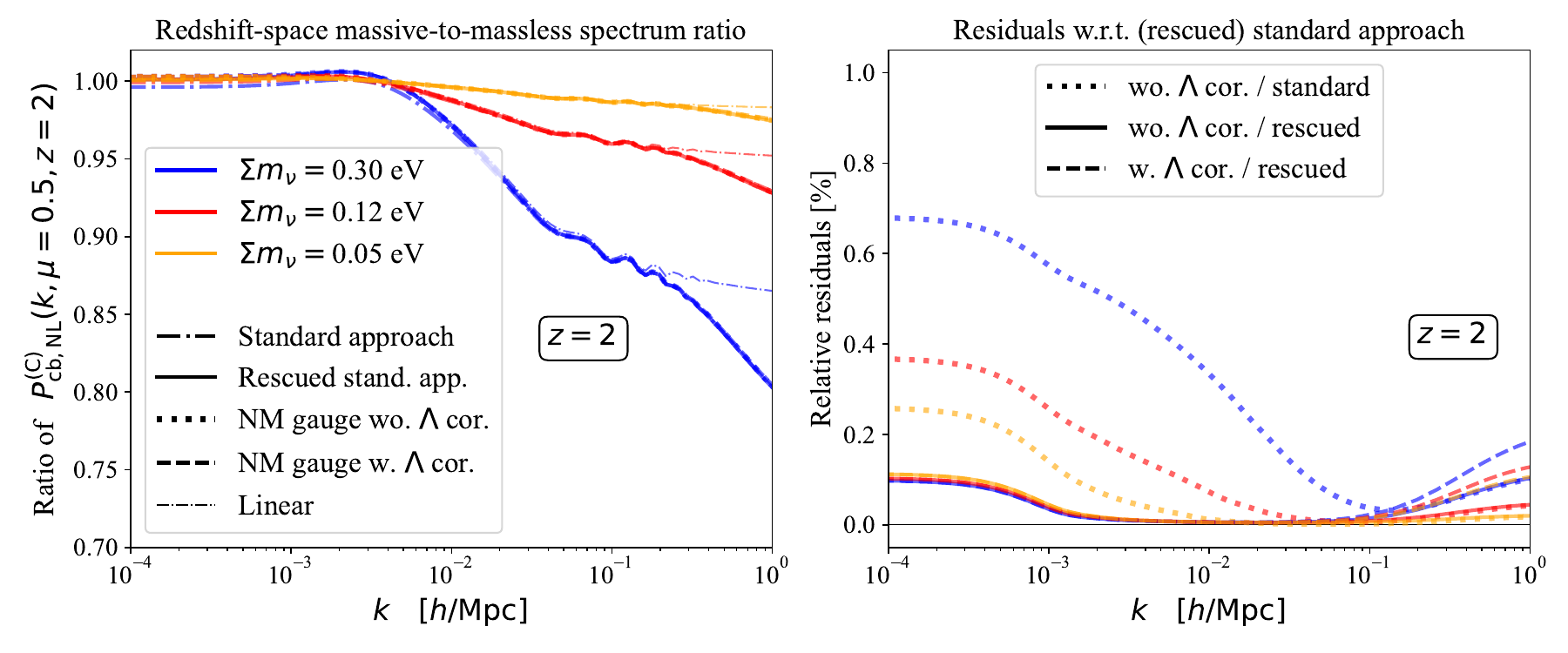}
    \end{minipage}
    \caption{(Left) Ratio of massive-to-massless power spectra in redshift space, $P_{\rm cb,NL}^{\rm (C)}(k,\mu,z)$, for $\mu=0.5$ and either $z=0$ (top) or $z=2$ (bottom), in a $\Lambda$CDM cosmology with three degenerate massive neutrinos and $\Sigma m_\nu=0.05, \,0.12, \,0.3\,$eV, computed with different approaches: EFTofLSS applied to the linear comoving gauge power spectrum $P^{\rm (C)}_{\rm cb,L}$ (standard approach, dot-dashed);
    standard approach `rescued' by using the scale-dependent growth rate $f(k,z)$ in the Kaiser term (solid); and EFTofLSS applied to the NM gauge power spectrum $P^{\rm [Nl]}_{\rm cb,L}$ and transformed back to the comoving gauge, either without (dotted) or with (dashed) $\Lambda$ corrections. (Right) Percentage difference between the NM gauge approach (with/without $\Lambda$ corrections) and the standard approaches (with/without a scale-dependent growth rate in the Kaiser term).
    \label{fig:nwtfl-rsd}}
\end{figure}

\subsection{Results and comparison}

\noindent {\bf Results for fixed neutrino mass and different redshifts.} In figure~\ref{fig:massiveRSD}, we compare the redshift-space power spectrum computed using the standard and NM gauge approaches at $z=0$ and 2, for $\mu=0.5$, with degenerate neutrinos of mass $\Sigma m_\nu=0.12\,$eV, and assuming a counter-term $c_s^2=0$. 

We first focus on the dotted lines in the right panel, accounting for the difference between the `NM gauge without $\Lambda$ correction approach' and the `standard approach' in its simplest flavor. In this case, we do find a substantial difference, reaching 0.3\% on large scales for $\Sigma m_\nu=0.3\,$eV and $z=2$. However, the dominant contribution to this error is obvious. In the simplest version of the standard method, the growth factor $f$ is assumed to be scale-independent, because this assumption is required for the self-consistency of loop calculations. However, the growth factor does not appear only in the normalization of the loops. It is also involved in the normalization of the tree-level power spectrum through the Kaiser term, that is, the multiplicative factor in eq.~(\ref{eq:Pcb_rsd_standard}). This term is important on all scales. The differences observed in the dotted line of the right panel of  figure~\ref{fig:massiveRSD} is dominated by the fact that the standard approach uses the small-scale $f$ also in the Kaiser term on large scales.

To make the standard approach more accurate, we can in principle use the full $f(k,z)$ in the tree-level Kaiser term, while sticking to the small-scale factor $f(z)$ in the calculation of loop corrections. We call this the `rescued standard approach'. This correction is also implemented in other codes such as CLASS-PT. In the right panel of figure \ref{fig:massiveRSD}, the solid lines compare the `NM gauge without $\Lambda$ correction approach' to the `rescued standard approach'. The redshift-space spectra are now in excellent agreement up to the $\sim 0.1\%$ level, like for the real-space spectra. On intermediate scales, $10^{-3}h\,{\rm Mpc}^{-1}<k<3\times 10^{-2}h\,{\rm Mpc}^{-1}$, the agreement is even better, which is non-trivial and remarkable since the two spectra are computed in very different ways: in the rescued standard approach, with a scale-dependent $f(k,z)$ in the tree-level contribution; and in the NM gauge approach, with a scale-independent $f(z)$ in all contributions, but appropriate corrections induced by the gauge transformation.

Finally, the dashed lines in the right panel of figure \ref{fig:massiveRSD} compare the `NM gauge with $\Lambda$ correction approach' to the `rescued standard approach'. For $\mu=0.5$ and $z=0$, the $\Lambda$ correction enhances loops corrections by a larger amount than in real space, about  $+0.8\%$ at $k=0.3\,h{\rm Mpc}^{-1}$. The reason is that $\Lambda$ effects have a larger impact on the velocity field than on the density field, as already established by the results of \cite{Garny:2020ilv}.\\

\noindent {\bf Results for different neutrino masses.} In figure \ref{fig:nwtfl-rsd}, the left panels show the massive-to-massless power spectrum ratio predicted by each approach for $\mu=0.5$ and at $z=0$ (top) or $z=2$ (bottom), while the right panel directly compares the power spectrum of the NM gauge approach (with/without $\Lambda$ corrections) to the standard or rescued standard approach. The dashed lines in the right panels compare the accurate `NM gauge with $\Lambda$ correction approach' to the `rescued standard approach'. In this case, the results are qualitatively similar in redshift and real space, with small GR corrections on large scales (at the 0.1\% level) and larger corrections on small scales. The small-scale corrections grow with $\sum m_\nu$. They are dominated by neutrino effects at $z\geq2$ and the $\Lambda$ correction at $z<2$. For $\sum m_\nu=0.30 \, {\rm eV}$ and $\mu=0.5$, the correction at $k=0.3\,h {\rm Mpc}^{-1}$ grows from $\sim 0.10\%$ at $z=2$ to $\sim 0.93\%$ at $z=0$. \\

\begin{figure}
    \centering
    \begin{minipage}{\textwidth}
        \includegraphics[width=\linewidth]{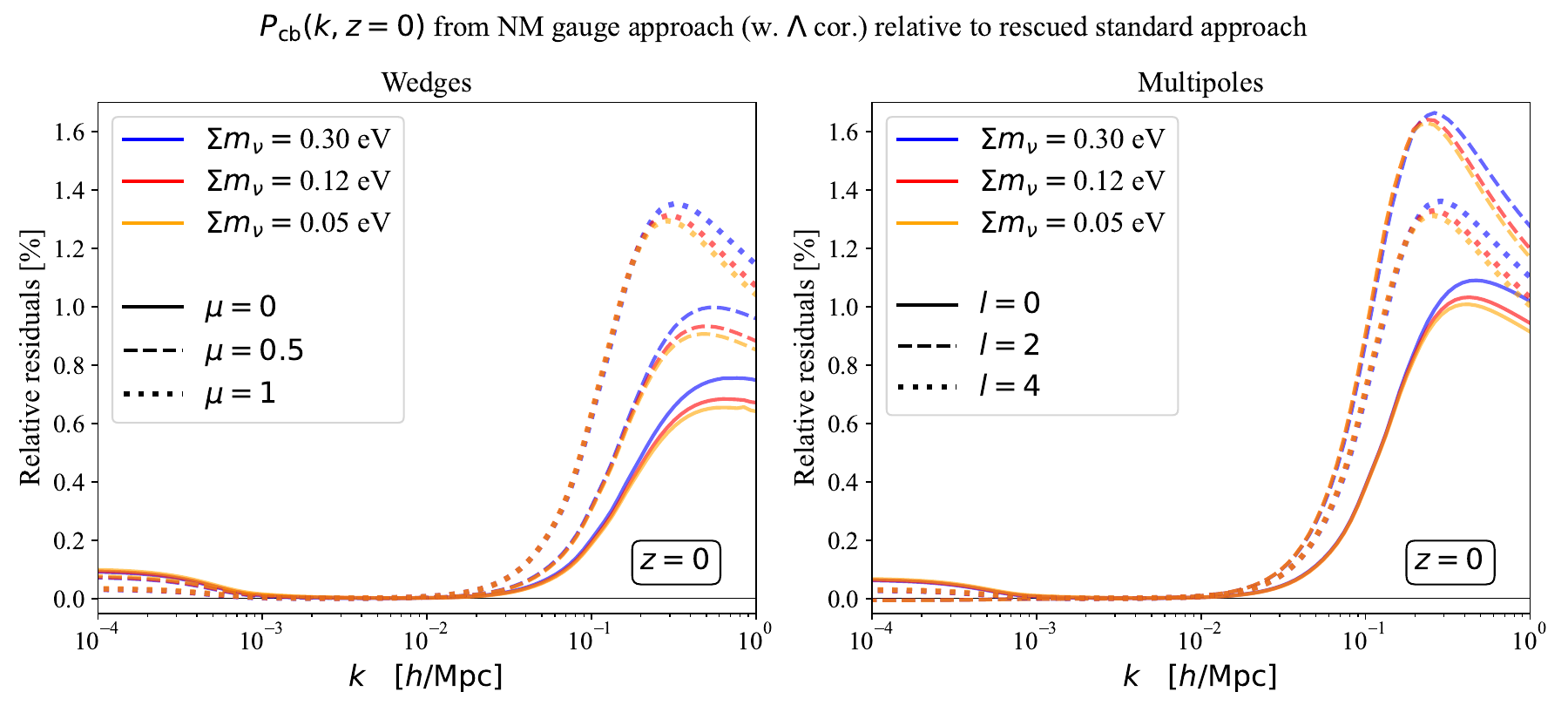}
        \includegraphics[width=\linewidth]{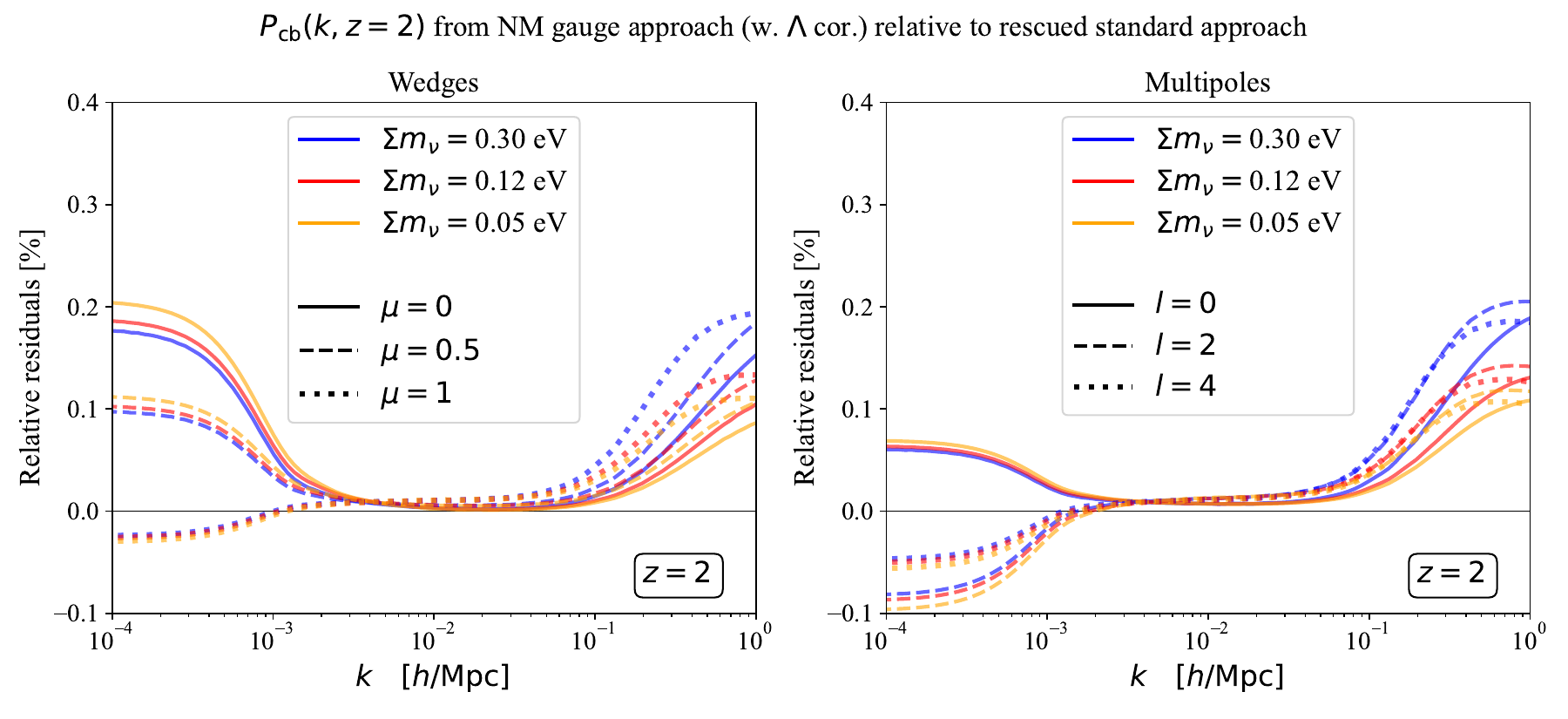}
    \end{minipage}
    \caption{(Left) Percentage difference between the wedge power spectra in redshift space, $P_{\rm cb,NL}^{\rm (C)}(k,\mu,z)$, at $z=0$ (top) or $z=2$ (bottom), inferred from the `NM gauge with $\Lambda$ correction approach' or from the `rescued standard approach'. We assume a $\Lambda$CDM cosmology with three degenerate massive neutrinos and $\Sigma m_\nu=0.05, \,0.12, \,0.3\,$eV, and show the results for three different values of $\mu=0, 0.5, 1$ ($\mu=0$ corresponds to power spectra in real space). (Right) Same for the power spectrum multipoles, $P_{\rm cb,NL}^{\rm (C)}(k,\ell)$, with $\ell=0,2,4$.
    \label{fig:real-rsd}}
\end{figure}

\noindent {\bf Comparison between results in real and redhsift space.} The residual differences have roughly the same shape and amplitude in real and redshift space, but are not exactly identical since we are now applying the gauge transformation and the time-dependent kernels both to the density and velocity fields. We can compare the real-space and redshift-space residuals in the left panel of figure~\ref{fig:real-rsd}, which displays the percentage difference between the `NM gauge with $\Lambda$ correction' and `rescued standard' approaches for three different cosines: $\mu=0, 0.5, 1$. The case $\mu=0$ corresponds to the real space power spectra, while increasing values of $\mu$ give more and more weight to the corrections induced by the velocity field.

On large scales, $k<10^{-3}h\,{\rm Mpc}^{-1}$, the correction induced by the gauge transformation is smaller for larger $\mu$. The reason is that in the expression of $\Delta^{\rm (RSD)}$, see eq.~(\ref{eq:rsd}), the terms $\dot{H}_{\rm T}^{\rm (NM)}$ and $\Delta^{\rm (C)}$ have opposite sign. On small scales, $k>10^{-2}h\,{\rm Mpc}^{-1}$, the effect is reversed: when $\mu$ increases, the correction induced by massive neutrinos and $\Lambda$ corrections in the NM gauge approach starts earlier in $k$-space and reaches higher values. The explanation is that on non-linear scales, the velocity field is significantly more affected by these corrections than the density field.

This observation is fully consistent with the full time and scale-dependent treatment of Garny \& Taule \cite{Garny:2022fsh}, who show in their figure 3 that the one-loop velocity power spectrum is more enhanced by massive neutrino and $\Lambda$ effects than the density power spectrum.\footnote{The dashed lines in figure~3 of \cite{Garny:2022fsh} show a suppression rather than an enhancement of the density and velocity power spectra, because they correspond to the ratio of the inaccurate over accurate method, while we usually display the opposite ratio.} In the wedge power spectrum, the velocity field contributes the most for $\mu=1$. In this case, we find that for $\sum m_\nu=0.30\,{\rm eV}$, $z=0$ and $k=0.3\,h{\rm Mpc}^{-1}$, the total enhancement reaches $\sim 1.35\%$, compared to just $\sim 0.66\%$ for $\mu=0$ (that is, compared to real space).

In the right panel of figure~\ref{fig:real-rsd}, we compare the multipoles $\ell=0,2,4$ of the CDM+baryon power spectrum, $P_{\rm cb,NL}^{\rm (C)}(k,\ell,z)$, derived from the NM gauge and rescued standard approaches. On large scales, the corrections induced by the gauge transformation and $\Lambda$ domination are the largest for $\ell=0$, and get smoothed by the integral over $\mu$ for $\ell=4$ and even more for $\ell=2$. On small scales, these effects are the largest for the quadrupole $\ell=2$, which gives a lot of weight to the velocity field. For $\sum m_\nu=0.30\,{\rm eV}$ and $z=0$, the enhancement of the quadrupole peaks at $k=0.3\,h{\rm Mpc}^{-1}$, where it reaches $\sim 1.67\%$. In comparison, the octopole is enhanced by $\sim 1.36\%$ and the monopole by $\sim 1.03\%$.

\subsection{Summary of redshift-space results}
In this section, we first found that the most naive version of the standard approach, in which the scale-independent growth factor $f(z)$ defined at small scales is used both in the (tree-level) Kaiser term and for rescaling loop corrections, is inaccurate -- by, e.g., $\sim 0.3\%$ for $\sum m_\nu = 0.30\,{\rm eV}$. The situation improves a lot when considering a `rescued standard approach' in which this rate $f(z)$ is used for loop corrections, while the Kaiser term is obtained from the full $f(k,z)$. The agreement between this method and our more accurate NM gauge approach is very good, with differences which are roughly of the same order as for the real-space power spectrum, but depend on the angle $\mu$ (in the wedge power spectrum) or $\ell$ (in the multipole power spectrum).

In the wedge power spectrum, the NM gauge leads to corrections which decrease with $\mu$ on large scales but increase with $\mu$ on small scales, see the panel of figure~\ref{fig:real-rsd}, essentially because the exact treatment of massive neutrino effects on the growth of perturbations has more impact on the velocity field than on the density field. For $\sum m_\nu = 0.30\,{\rm eV}$, the correction reaches $\sim 1.35\%$ at $k=0.3\,h{\rm Mpc}^{-1}$, $\mu=1$, $z=0$. In the multipole power spectrum, the corrections of the NM gauge approach are slightly reduced on large scales compared to the ones in real space, but significantly larger on small scales in the case of the quadrupole, as shown by the right panel of figure~\ref{fig:real-rsd}. For the same neutrino mass, the correction reaches $\sim 1.67\%$ at $k=0.2\,h{\rm Mpc}^{-1}$, $\ell=2$, $z=0$.

\section{Summary and outlook}\label{sec:conclusion}

The NM gauge approach allows to perform loop calculations in standard perturbation theory or its extensions (such as the EFTofLSS) in a fully consistent framework, incorporating GR corrections on large scales, and allowing to stick to scale-independent kernels even when the assumed cosmological model does feature a scale-dependent growth. The approach is easy to implement by modifying a Boltzmann code, mainly at the linear level. Once this is implemented, the calculation of EFTofLSS contributions do not take more time in the NM approach than in the standard approach. 

For the real-space and redshift-space power spectra of CDM+baryons or total matter, we find that the corrections coming from the more accurate NM gauge approach are at most of the order of 0.2\% on large scales $k<10^{-2}h{\rm Mpc}^{-1}$, even in presence of massive neutrinos. 
On mildly non-linear scales, they can be substantially larger. At a scale of $k=0.3\,h{\rm Mpc}^{-1}$, redshift $z=0$, and for a summed mass of $\sum m_\nu=0.30\,{\rm eV}$, the correction to the real-space power spectrum caused by massive neutrinos remains as small as $\sim 0.15\%$. However, in combination with $\Lambda$-domination effects, it reaches $\sim 0.7\%$. Additionally, the correction gets significantly amplified in redshift space, due to the behavior of the velocity field. It reaches $\sim 1.0\%$ for the
monopole power spectrum, and $\sim 1.7\%$ for quadrupole power spectrum at $k=0.2\,h{\rm Mpc}^{-1}$. 

In the analysis of future redshift surveys, the impact of these corrections might be mitigated by the marginalization over EFT parameters. However, given the negligible computational cost of our method, we believe that it is worth including it in data analysis pipelines.

We have computed these corrections with respect to an optimized version of the standard approach, in which we stick to EdS kernels but perform a few choices that are important in presence of massive neutrinos: first, the non-linear corrections are always computed for the cold dark matter + baryon power spectrum, starting from the linear spectrum $P_{\rm cb,L}(k,z)$ computed in the synchronous or comoving gauge; second, one should use the scale-independent growth factor $D(z)$ and rate $f(z)$ computed in the small-scale limit; third, when computing the redshift-space power spectrum, the scale-dependent growth rate $f(k,z)$ should be used in the Kaiser term; and finally, when the total matter power spectrum is requested, the non-linear $P_{\rm cb,NL}(k,z)$ spectrum can be combined with the linear neutrino spectrum in the fashion of eq.~(\ref{eq:P_tot}) or eq.~(\ref{eq:P_tot_approx}).

Our work can be expanded in several interesting directions. 
First, we could generalize the NM gauge approach to the level of biased tracers. Second, this method opens the possibility to study models with a stronger scale-dependent growth than $\Lambda$CDM cosmologies with massive neutrinos. This includes extended gravity models, as well as models featuring hot relics heavier than neutrinos (such as thermal axions), decaying dark matter or interacting dark matter. In some of these cases, the differences with respect to the standard approach might be larger than found in this work, even at the level of the real-space power spectrum.

\acknowledgments
We thank Alejandro Aviles, Emanuele Castorina, Guido D'Amico, Mathias Garny, Oliver Philcox and Pierre Zhang for very useful feedback on this work. The work of AMD is supported by the Agence Nationale de la Recherche (ANR) under grant No. ANR-23-CPJ1-0160-01.

\bibliographystyle{utphys}
\bibliography{refs}

\newpage
\appendix

\def\q{{\bm q}}
\def\k{{\bm k}}

\section{Full one-loop CDM+baryon power spectrum in real space\label{app:full}}

The power spectrum in a gauge G reads
\begin{equation}
    P^{\rm (G)}_{\rm cb,NL} = \left\langle \left| \delta^{\rm [Nl]}_{\rm cb,NL} + \Delta^{\rm (G)} \right|^2 \right\rangle~.
\end{equation}
To compute it at the one-loop order, we need to expand the nonlinear density field  up to order three, $\delta^{\rm [Nl]}_{\rm cb,NL}=\delta^{(1)}_{\rm cb}+\delta^{(2)}_{\rm cb}+\delta^{(3)}_{\rm cb}$, where $\delta^{(1)}_{\rm cb}\equiv\delta^{\rm [Nl]}_{\rm cb,L}$ and $\delta^{(n)}_{\rm cb}$ is of order $\left(\delta^{(1)}_{\rm cb}\right)^n$. Instead, we should assume that $\Delta^{\rm (G)}$ remains linear since we work in the weak-field limit. Then, we get
\begin{equation}
    P^{\rm (G)}_{\rm cb,NL} = \left\langle \left| \delta^{\rm [Nl]}_{\rm cb,L} \right|^2 \right\rangle
    + 2 \left\langle \left| \delta^{(1)}_{\rm cb} \Delta^{\rm (G)} \right| \right\rangle
    + \left\langle \left| \Delta^{\rm (G)} \right|^2 \right\rangle
    + \left\langle \left| \delta^{(2)}_{\rm cb} \right|^2 \right\rangle
    + 2 \left\langle \left| \delta^{(1)}_{\rm cb} \delta^{(3)}_{\rm cb} \right| \right\rangle
    + 2 \left\langle \left| \delta^{(3)}_{\rm cb} \Delta^{\rm (G)} \right| \right\rangle~,
\end{equation}
where the first three terms account for the tree-level power spectrum while the last three terms contain all non-vanishing one-loop contributions. The first term is the linear power spectrum of the Newtonian fluid,  $P^{\rm [Nl]}_{\rm cb,L}$. The fourth and fifth term represent its usual non-linear corrections and can be denoted as $P^{\rm [Nl]}_{\rm cb,22} + 2\,P^{\rm [Nl]}_{\rm cb,13}$. Altogether, these three terms form the one-loop power spectrum of the Newtonian fluid, $P^{\rm [Nl]}_{\rm cb,NL}$, which is returned by \classoneloop. The power spectrum in a gauge G thus reads
\begin{equation}
    P^{\rm (G)}_{\rm cb,NL} = 
    P^{\rm [Nl]}_{\rm cb,NL}
    + 2 \left\langle \left| \delta^{(1)}_{\rm cb} \Delta^{\rm (G)} \right| \right\rangle
    + \left\langle \left| \Delta^{\rm (G)} \right|^2 \right\rangle
    + 2 \left\langle \left| \delta^{(3)}_{\rm cb} \Delta^{\rm (G)} \right| \right\rangle~.
\end{equation} 
In presence of adiabatic initial conditions, all linear perturbations can be expressed in terms of transfer functions, see eqs.~(\ref{eq:ad1}). We can define the ratio of transfer functions
\begin{equation}
    R^{\rm (G)}_{\rm cb}(k,z) \equiv  
    \frac{\Delta^{\rm (G)}(k,z)}{\delta^{\rm [Nl]}_{\rm cb,L}(k,z)}~.
\end{equation}
We finally express the one-loop power spectrum in a gauge G in terms of the linear and non-linear power spectrum of the Newtonian fluid,
\begin{equation}
    P^{\rm (G)}_{\rm cb,NL} = 
    P^{\rm [Nl]}_{\rm cb,NL}
    + \left[2 R^{\rm (G)}_{\rm cb} + \left(R^{\rm (G)}_{\rm cb}\right)^2 \right] P^{\rm [Nl]}_{\rm cb,L}
    + 2 R^{\rm (G)}_{\rm cb}
    P^{\rm [Nl]}_{\rm cb,13}~.
    \label{eq:Pcb_full}
\end{equation} 
This expression has two obvious limits.

On small scales, $\delta^{\rm [Nl]}_{\rm cb,L}$ is enhanced by structure formation while $\Delta^{\rm (G)}$ remains very small (since the overall consistency of our approach requires that we stay in the weak-field regime). Then, $|R^{\rm (G)}_{\rm cb}|\ll 1$ and \begin{equation}
P^{\rm (G)}_{\rm cb,NL} 
\xrightarrow[k \to \infty]{}
P^{\rm [Nl]}_{\rm cb,NL}+ 2 R^{\rm (G)}_{\rm cb}
    P^{\rm [Nl]}_{\rm cb,13}~.
\end{equation}
A priori, it is not obvious that the last term can be neglected. Indeed, the loop correction $P^{\rm [Nl]}_{\rm cb,22} + 2\,P^{\rm [Nl]}_{\rm cb,13}$ remains of the same order of magnitude as the linear power spectrum on mildly non-linear scales, but the individual terms $P^{\rm [Nl]}_{\rm cb,22} \simeq - 2\,P^{\rm [Nl]}_{\rm cb,13}$ are larger by about one order of magnitude. Thus, the last term contains the product of a small factor  $R^{\rm (G)}_{\rm cb}$ and a potentially large factor $P^{\rm [Nl]}_{\rm cb,13}$. However, we find numerically that this term is always subdominant, especially in the backward method which leads to extremely suppressed values of $\Delta^{\rm (G)}$ on sub-Hubble scales.

In the large-scale limit, the non-linear correction to $\delta_{\rm cb}^{\rm [Nl]}$ remains small while $R^{\rm (G)}_{\rm cb}$ may become significant, such that 
\begin{equation}
    P^{\rm (G)}_{\rm cb,NL} 
    \xrightarrow[k \to 0]{}
    \left[1 + 2 R^{\rm (G)}_{\rm cb} + \left(R^{\rm (G)}_{\rm cb}\right)^2 \right] P^{\rm [Nl]}_{\rm cb,L}~.
\end{equation} 

We can compare the prediction of eq.~(\ref{eq:Pcb_full}) with that of eq.~(\ref{eq:Pcb}), which was obtained under a separation-of-scale ansatz instead of a full one-loop expansion. The two expressions are equivalent up to the fact that eq.~(\ref{eq:Pcb}) does not include the last term of (\ref{eq:Pcb_full}), that is, the contribution of the correlator $\left\langle \left| \delta^{(3)}_{\rm cb} \Delta^{\rm (G)} \right| \right\rangle$ to one-loop corrections. For all neutrino masses considered in this work, we checked numerically that this difference shifts the power spectrum by less than 0.001\% (excepted in the less accurate forward method, for which there is a $\sim$0.01\% shift on small scales). We conclude that the separation-of-scale ansatz is accurate enough for the calculation of $P^{\rm (G)}_{\rm cb,NL}$.  

\section{Beyond EdS kernels\label{app:kernels}}

We consider a universe where the density perturbations $\delta$ and velocity divergence $\theta$ of a self-gravitating fluid evolve according to the continuity and Euler equations (in Fourier space)
\begin{align}
\dot{\delta}(\bm k,\tau)+\theta(\bm k, \tau)
&= -\int_{\bm q}\alpha(\bm q,\bm k{-}\bm q)\theta(\bm q, \tau)\delta(\bm k{-}\bm q, \tau)~,\label{C}\\
\dot{\theta}(\bm k, \tau)+{\mathcal H}(\tau)\, \theta(\bm k, \tau)
&+\frac{3}{2}{\mathcal H}^2(\tau)\,f^2(\tau) \, \nu(\tau) \, \delta(\bm k, \tau)
= -\int_{\bm q}\beta(\bm q,\bm k{-}\bm q)\theta(\bm q, \tau)\theta(\bm k{-}\bm q, \tau)~, \label{E}
\end{align}
where $f(\tau)$ is the linear growth rate (for the gowing mode), $\int_{\bm q}\equiv\int d^{3}q/(2\pi)^{3}$, and the kernels are given by
\begin{align}
\alpha(\bm k_{1},\bm k_{2})&=\frac{(\bm k_{1}{+}\bm k_{2})\cdot\bm k_{1}}{2k_{1}^{2}}~,
\label{eq:alpha}\\
\beta(\bm k_{1},\bm k_{2})&=\frac{|\bm k_{1}{+}\bm k_{2}|^{2}(\bm k_{1}\cdot\bm k_{2})}{2k_{1}^{2}k_{2}^{2}}~.
\label{eq:beta}
\end{align}
These equations describe the growth of structure in various cosmologies, provided that one uses the NM gauge to absorb scale-dependent coefficients. 
The function $\nu(\tau)$ in eq.~(\ref{E}) depends on the model considered: in $\Lambda$CDM with massless neutrinos, $\nu(\tau)=\Omega_{\rm m}(\tau)/f^2(\tau)$, while in presence of massive neutrinos with density fraction $f_\nu$, one has $\nu(\tau)=(1-f_\nu)\,\Omega_{\rm m}(\tau)/f^2(\tau)$. The function $\nu(\tau)$ can be approximated in various way.

\subsection{Constant $\nu$ \label{app:constant_nu}}

In a $\Lambda$CDM cosmology with/without massive neutrinos, the function $\nu(t)$ is constant during matter domination and equal to 1 (massless neutrinos) or $(1-\frac{3}{5}f_\nu)^{-2}$ (massive neutrinos). 
A common approximation consists in assuming that it remains constant also during $\Lambda$ domination. The case $\nu=1$ gives the so-called EdS kernels, but the case $\nu=(1-\frac{3}{5}f_\nu)^{-2}$ is also interesting in order to model the effect of massive neutrinos in the NM gauge (while still treating the effect of $\Lambda$ domination in an approximate way).

When $\nu$ is considered as constant in time, the general solution is given by the separation ansatz
\begin{align}
 \delta(\bm k, \tau)&= \sum_{n=1}^\infty D^n(\tau) \delta^{(n)}(\bm k)~, \nonumber \\
\theta(\bm k, \tau)&= {\mathcal H}(\tau) f(\tau) \sum_{n=1}^\infty  D^n(\tau) \theta^{(n)}(\bm k)~, \label{eq:separation_Ansatz}
\end{align}
with
\begin{align}
\delta^{(n)}(\bm k)&=\int_{\bm q_{1}\dots q_{n}}F_{n}(\bm q_{1},\dots,\bm q_{n})\delta^{(1)}({\bm q}_1)\dots\delta^{(1)}({\bm q}_{n})\delta_{\mathrm D}(\bm k{-}\Sigma\bm q_{i})~,\\
\theta^{(n)}(\bm k)&=-\int_{\bm q_{1}\dots q_{n}}G_{n}(\bm q_{1},\dots,\bm q_{n})\delta^{(1)}({\bm q}_1)\dots\delta^{(1)}({\bm q}_{n})\delta_{\mathrm D}(\bm k{-}\Sigma\bm q_{i})~. 
\end{align}
The fully symmetric kernels $F_n$ and $G_n$ are defined as
\begin{align}
    F_n(\q_1,...,\q_n) = \frac{1}{n !} \sum_{\pi \in S_n}\Tilde{F}_n(\q_{\pi(1)},...,\q_{\pi(n)})~, \\
    G_n(\q_1,...,\q_n) = \frac{1}{n!} \sum_{\pi \in S_n}\Tilde{G}_n(\q_{\pi(1)},...,\q_{\pi(n)})~,
\end{align}
where $\pi \in S_n$ denotes a permutation of the indices $\{1,...,n\}$. The recursion formula for the kernels $\Tilde{F}_n$, $\Tilde{G}_n$ is given by
\begin{align}
\Tilde{F}_n(\q_1, \ldots ,\q_n) &= \sum_{m=1}^{n-1} { \Tilde{G}_m(\q_1, \ldots ,\q_m)
 \over{(2n+3\nu)(n-1)}} \Bigl[(2n+3\nu -2) \alpha(\k_1,\k_2) \Tilde{F}_{n-m}(\q_{m+1},
 \ldots ,\q_n) \nonumber \\ & +2 \beta(\k_1, \k_2)
 \Tilde{G}_{n-m}(\q_{m+1}, \ldots ,\q_n) \Bigr]~, \label{eq:rec_F}\\
\Tilde{G}_n(\q_1, \ldots ,\q_n) &= \sum_{m=1}^{n-1} { \Tilde{G}_m(\q_1, \ldots ,\q_m)
\over{(2n+3\nu)(n-1)}} \Bigl[3\nu \alpha(\k_1,\k_2) \Tilde{F}_{n-m}(\q_{m+1}, \ldots
,\q_n) \nonumber \\ &  +2n \beta(\k_1, \k_2) \Tilde{G}_{n-m}(\q_{m+1},
\ldots ,\q_n) \Bigr]~, \label{eq:rec_G}
\end{align} 
with $\k_1=\q_1+ \ldots +\q_m$ and $\k_2=\q_{m+1}+ \ldots
+\q_n$.
Starting from $F_1=G_1=1$, we find that the symmetrized second-order kernels are given by 
\begin{equation}
   F_2(\q,\k) = \frac{3\nu +2}{3\nu+4} + \frac{1}{2}\frac{\k\cdot \q}{|\k|\,|\q|}\biggl(\frac{|\k|}{|\q|}+\frac{|\q|}{|\k|}\biggr) + \frac{2}{3\nu+4}  \frac{(\k\cdot \q)^2}{|\k|^2|\q|^2}~,
   \label{eq:F2_nu}
\end{equation}
\begin{equation}
   G_2(\q,\k) = \frac{3\nu}{3\nu+4} + \frac{1}{2}\frac{\k\cdot \q}{|\k|\,|\q|}\biggl(\frac{|\k|}{|\q|}+\frac{|\q|}{|\k|}\biggr) + \frac{4}{3\nu+4}  \frac{(\k\cdot \q)^2}{\k^2\q^2}~.
   \label{eq:G2_nu}
\end{equation}
We then derive the symmetrized second-order kernels $F_3(\k_1,\k_2,\k_3)$, $G_3(\k_1,\k_2,\k_3)$. The calculation of the one-loop power spectrum only requires $G_3$ and $F_3$ with the following arguments:
\begin{align}
F_3(\q,-\q, \k) &= \frac{1}{|\k-\q|^2 |\k+\q|^2} \left[ \frac{(2\nu+3\nu^2)\k^4}{(6+3\nu)(4+3\nu)} + \frac{(2\nu+3\nu^2) \k^2\q^2}{(6+3\nu)(4+3\nu)} \right. \nonumber \\  
&\left. - \frac{\k^4 (\k\cdot \q)^2}{6\q^4} + \frac{2(8+8\nu+3\nu^2)(\k\cdot \q)^4}{(6+3\nu)(4+3\nu)\q^4} -\frac{2(4+4\nu+3\nu^2)\k^2(\k\cdot \q)^2}{(6+3\nu)(4+3\nu)\q^2} \right. \nonumber \\
&\left.  - \frac{(8+30\nu+21\nu^2)(\k\cdot \q)^2}{(12+6\nu)(4+3\nu)}+ \frac{2\nu(\k\cdot \q)^4}{(6+3\nu)\k^2\q^2}\right]~, \label{eq:F3_nu}
\end{align}
\begin{align}
G_3(\q,-\q, \k) &= \frac{1}{|\k-\q|^2 |\k+\q|^2} \left[ \frac{(-6\nu+3\nu^2)\k^4}{(6+3\nu)(4+3\nu)} + \frac{(-2\nu+\nu^2) \k^2\q^2}{(2+\nu)(4+3\nu)} \right. \nonumber \\  
&\left. - \frac{\k^4 (\k\cdot \q)^2}{6\q^4} + \frac{2(8+4\nu+3\nu^2)(\k\cdot \q)^4}{(6+3\nu)(4+3\nu)\q^4} +\frac{2(-4+4\nu-3\nu^2)\k^2(\k\cdot \q)^2}{(6+3\nu)(4+3\nu)\q^2} \right. \nonumber \\
&\left.  + \frac{(-8+2\nu-21\nu^2)(\k\cdot \q)^2}{(12+6\nu)(4+3\nu)}+ \frac{2\nu^2(\k\cdot \q)^4}{(2+\nu)(4+3\nu)\k^2\q^2}\right]~. \label{eq:G3_nu}
\end{align}
An independent derivation of the kernels for a time-independent factor $\nu$ has been performed in reference~\cite{Hartmeier:2023brx}. The authors start from the same assumptions as ours, but go through different calculation steps, which explains why their final results seem different from ours at first sight -- see their equations (3.25 - 3.28) with a constant parameter $x_0=\frac{3}{2} \nu$. However, we checked explicitly that, despite a different apparent form, their kernels are exactly equal to ours. Reference \cite{Joyce:2022oal} also derived a set of non-EdS kernel, starting however from slightly different assumptions regarding the background evolution.

\subsection{Time-varying $\nu(\tau)$\label{app:time_nu}}

During $\Lambda$ domination (or more generally dark energy domination), the function $\nu(\tau)$ experiences a small growth as a function of time. As long as this growth leads to small corrections for the evolution of the density and velocity fields, an approximation scheme can be found in which the kernels keep the same structure as in Appendix~\ref{app:constant_nu}, but now with time-dependent $\nu$ coefficients. This leads to a more complex separability ansatz than  
in eq.~(\ref{eq:separation_Ansatz}), with an additional time dependence in $\delta^{(n)}(\k, \tau)$ and $\theta^{(n)}(\k, \tau)$. 

In practice, we substitute the factor $\nu$ with a function $\nu_{F2}(t)$ in eq. (\ref{eq:F2_nu}) and $\nu_{G2}(t)$ in eq. (\ref{eq:G2_nu}):
\begin{align}
   F_2(\q,\k,\tau) &= \frac{3\nu_{F2}(\tau) +2}{3\nu_{F2}(\tau)+4} + \frac{1}{2}\frac{\k\cdot \q}{|\k|\,|\q|}\biggl(\frac{|\k|}{|\q|}+\frac{|\q|}{|\k|}\biggr) + \frac{2}{3\nu_{F2}(\tau)+4}  \frac{(\k\cdot \q)^2}{|\k|^2|\q|^2}~, \label{eq:F2_time}\\
   G_2(\q,\k,\tau) &= \frac{3\nu_{G2}(\tau)}{3\nu_{G2}(\tau)+4} + \frac{1}{2}\frac{\k\cdot \q}{|\k|\,|\q|}\biggl(\frac{|\k|}{|\q|}+\frac{|\q|}{|\k|}\biggr) + \frac{4}{3\nu_{G2}(\tau)+4}  \frac{(\k\cdot \q)^2}{\k^2\q^2}~.
\end{align}
Next, we substitute $\nu$ with $\nu_{F3}(t)$ in the recursion relation providing $F_3$ as a function of ($F_2$, $G_2$), that is, eq. (\ref{eq:rec_F}) with $n=3$. We do the same with a function $\nu_{G3}(t)$ in eq. (\ref{eq:rec_G}). Then, the full expression for ($F_3$, $G_3$) becomes:
\begin{align}
&F_3(\q,-\q, \k, \tau) = \frac{1}{|\k-\q|^2 |\k+\q|^2} \left[ \frac{\nu_{G2}(2+3\nu_{F3})\k^4}{(6+3\nu_{F3})(4+3\nu_{G2})} + \frac{\nu_{G2}(2+3\nu_{F3}) \k^2\q^2}{(6+3\nu_{F3})(4+3\nu_{G2})} \right. \nonumber \\  
&\left. - \frac{\k^4 (\k\cdot \q)^2}{6\q^4} + \frac{2(8+4\nu_{F3}+4\nu_{G2}+3\nu_{F3}\nu_{G2})(\k\cdot \q)^4}{(6+3\nu_{F3})(4+3\nu_{G2})\q^4} -\frac{2(4+2\nu_{F3}+2\nu_{G2}+3\nu_{F3}\nu_{G2})\k^2(\k\cdot \q)^2}{(6+3\nu_{F3})(4+3\nu_{G2})\q^2} \right. \nonumber \\
&\left.  - \frac{(8+26\nu_{G2}+4\nu_{F3}+21\nu_{F3}\nu_{G2})(\k\cdot \q)^2}{(12+6\nu_{F3})(4+3\nu_{G2})}+ \frac{2\nu_{G2}(4+3\nu_{F3})(\k\cdot \q)^4}{(6+3\nu_{F3})(4+3\nu_{G2})\k^2\q^2}\right]~,
\\
&G_3(\q,-\q, \k,\tau) = \frac{1}{|\k-\q|^2 |\k+\q|^2} \left[ \frac{\nu_{G2}(-6+3\nu_{G3})\k^4}{(6+3\nu_{G3})(4+3\nu_{G2})} + \frac{\nu_{G2}(-2+\nu_{G3}) \k^2\q^2}{(2+\nu_{G3})(4+3\nu_{G2})} \right. \nonumber \\  
&\left. - \frac{\k^4 (\k\cdot \q)^2}{6\q^4} + \frac{2(8+4\nu_{G3}+3\nu_{G2}\nu_{G3})(\k\cdot \q)^4}{(6+3\nu_{G3})(4+3\nu_{G2})\q^4} +\frac{2(-4+6\nu_{G2}-2\nu_{G3}-3\nu_{G2}\nu_{G3})\k^2(\k\cdot \q)^2}{(6+3\nu_{G3})(4+3\nu_{G2})\q^2} \right. \nonumber \\
&\left.  + \frac{(-8-4\nu_{G3}+6\nu_{G2}-21\nu_{G2}\nu_{G3})(\k\cdot \q)^2}{(12+6\nu_{G3})(4+3\nu_{G2})}+ \frac{2\nu_{G3}\nu_{G2}(\k\cdot \q)^4}{(2+\nu_{G3})(4+3\nu_{G2})\k^2\q^2}\right]~,
\label{eq:G3_time}
\end{align}
where we omitted the $\tau$ arguments on the right-hand side.
Similarly to reference \cite{Bernardeau:1993qu}, we can define some moments ($l_n$, $m_n$) by averaging the kernels ($F_n$, $G_n$) over all solid angles $\Omega_n={\bf k}_n/|{\bf k}_n|$,
\begin{align}
        l_{n}(\tau) \equiv n! \int \frac{d\Omega_{1}}{4\pi}
        \ldots \frac{d\Omega_{n}}{4\pi}
        F_{n}(\k_{1},\ldots,\k_{n},\tau)~, \label{eq:averageFn}
\\
        m_{n}(\tau) \equiv n! \int \frac{d\Omega_{1}}{4\pi}
        \ldots \frac{d\Omega_{n}}{4\pi}
        G_{n}(\k_{1},\ldots,\k_{n},\tau)~. \label{eq:averageGn}
\end{align}
The first moments read
\begin{align}
        l_{2}(\tau) &= 2 \frac{8/3 + 3\nu_{F2}(\tau)}{4+3\nu_{F2}(\tau)}~, \qquad
        m_{2}(\tau) = 2 \frac{4/3 + 3\nu_{G2}(\tau)}{4+3\nu_{G2}(\tau)}~, \label{eq:l2}\\
        l_{3}(\tau) &= \frac{3}{6+3\nu_{F3}(\tau)} \Biggl( \Bigl(4+3\nu_{F3}(\tau)\Bigr) \frac{l_2(\tau)}{2} + \Bigl(\frac{16}{3} + 3\nu_{F3}(\tau)\Bigr) \frac{m_2(\tau)}{2}  \Biggr)~, \\
        m_{3}(\tau) &= \frac{3}{6+3\nu_{G3}(\tau)} \Biggl( 3\nu_{G3}(\tau) \frac{l_2(\tau)}{2} + \Bigl(4 + 3\nu_{G3}(\tau)\Bigr) \frac{m_2(\tau)}{2}  \Biggr)~. \label{eq:m3}
\end{align}
The angular average of the equations of motion (\ref{C}, \ref{E}) provides well-known differential equations for the moments ($l_n$, $m_n$) \cite{Bernardeau:1993qu,Bernardeau:2001qr}, which read in our case
\begin{align}
\frac{1}{\mathcal{H}f} \frac{d l_n}{d \eta}+ n l_{n} -m_{n} &=
\sum_{k=1}^{n-1} {n \choose k} l_{n-k} m_{k}~,
\label{eq:angle_average_EOM1}
\\
\frac{1}{\mathcal{H}f} \frac{d m_n}{d \eta}+ n m_{n} + 
\left(\frac{3}{2} \nu(\tau) -1\right)m_{n} -\frac{3}{2} \nu(\tau) 
l_{n}&=\frac{1}{3} \sum_{k=1}^{n-1} {n \choose k} m_{n-k} m_{k}~.
\label{eq:angle_average_EOM2}
\end{align}
Inserting out ansatz (\ref{eq:l2} - \ref{eq:m3}) for the moments inside (\ref{eq:angle_average_EOM1}, \ref{eq:angle_average_EOM2}), we obtain differential equations directly for the functions ($\nu_{Fn}$, $\nu_{Gn}$),
\begin{align}
\frac{1}{\mathcal{H}f} \frac{d \nu_{F2}}{d \eta} &=
2\left( \frac{4 + 3\nu_{F2}}{4 + 3\nu_{G2}}\right)(\nu_{G2} - \nu_{F2})~, \label{eq:diff2}\\
\frac{1}{\mathcal{H}f} \frac{d \nu_{G2}}{d \eta} &=
- \left[ \frac{\nu_{G2}(4 + 3\nu_{G2})}{4} + \frac{3}{4} \left( \frac{4 + 3\nu_{G2}}{4 + 3\nu_{F2}} \right) \nu(\tau) \left(-2 \nu_{F2} + \nu_{G2} - \frac{4}{3} \right) \right]~,\\
\frac{1}{\mathcal{H}f} \frac{d \nu_{F3}}{d \eta} &= \frac{\left(6+3\nu_{F3}\right)^2}{9(l_2+\frac{1}{3}m_2)} \left[ m_3 + 3 \left(m_2 + l_2 - l_3 \right) \right. \nonumber\\
&~~\left.
- \frac{3}{2(6+3\nu_{F3})} \left( \left(4 + 3\nu_{F3} \right) \frac{1}{\mathcal{H}f} \frac{d l_2}{d \eta} + \left(\frac{16}{3} + 3\nu_{F3} \right) \frac{1}{\mathcal{H}f} \frac{d m_2}{d \eta}  \right) \right]~,\\
\frac{1}{\mathcal{H}f} \frac{d \nu_{G3}}{d \eta}
&= \frac{\left(6+3\nu_{G3}\right)^2}{9(l_2+2 m_2)} \left[ -\left( \frac{3}{2}\nu(\tau)+2 \right) m_3 + 2 m_2 + \frac{3}{2} \nu(\tau) l_3 \right. \nonumber\\
&~~\left. - \frac{3}{2(6+3\nu_{G3})} \left( 3\nu_{G3}  \frac{1}{\mathcal{H}f} \frac{d l_2}{d \eta} + \left(4 + 3\nu_{G3} \right) \frac{1}{\mathcal{H}f} \frac{d m_2}{d \eta}  \right) \right]~,
\label{eq:diff3}
\end{align}
from which it is easy to check that in the case $\nu(\tau) = 1$ and with the initial conditions $\nu_{Fn} = \nu_{Gn} = 1$ for $n=2,3$, we recover constant solutions that correspond to the EdS kernels. In a $\Lambda$CDM universe with massive/massless neutrinos and in the NM gauge, the initial conditions (during matter domination) are given for $n=2,3$ by
\begin{equation}
\nu_{Fn} = \nu_{Gn} = \frac{\Omega_{\rm m}}{f^2}(1-f_\nu) = \frac{1-f_\nu}{(1-\frac{3}{5} f_\nu)^2}~.
\end{equation}
One can then solve the differential equations numerically to compute $\nu_{Fn}(\tau)$ and $\nu_{Gn}(\tau)$ at any time $\tau$, and use these values in the calculation of the one-loop power spectrum.

We note that this approach is very easy to implement in a code like \classoneloop, with no extra computing cost compared to the standard approach with EdS kernels. Eight differential equations are solved for a given cosmology within the \class{} background module, together with the equation giving $D(\tau)$ and $f(\tau)$: these are (\ref{eq:angle_average_EOM1}, \ref{eq:angle_average_EOM2}) with $n=2,3$, and (\ref{eq:diff2} -- \ref{eq:diff3}). The only other modification consists in substituting the EdS kernels with those in equations (\ref{eq:F2_time}, \ref{eq:G3_time}).

Even if the equations of motion (\ref{eq:angle_average_EOM1}, \ref{eq:angle_average_EOM2}) describe the exact evolution of the angular-averaged kernels, this method is approximate, since the full equations of motion (\ref{C}, \ref{E}) are not solved exactly. However, by comparing with an exact solution, we found that our approach is extremely accurate, as explained in Appendix \ref{app:Bernardeau}.

\subsection{Exact solution\label{app:Bernardeau}}

An exact solution solution up to order $n=3$ was derived by Bernardeau \cite{Bernardeau:1993qu} for any cosmology with a time-dependent $\nu(t)$ in eq. (\ref{E}). This solution is also summarized in sections 2.4.3 - 2.4.5 of \cite{Bernardeau:2001qr} and Appendix A of \cite{Garny:2020ilv}. At order $n=2$, the exact time-dependent kernels read \cite{Garny:2020ilv} 
\begin{align}
   F_2(\q,\k,\tau) &= \Bigl(-\frac{1}{2} + \frac{3}{4} l_2(\tau)\Bigl) \alpha^s(\q, \k) + \Bigl(\frac{3}{2} - \frac{3}{4}l_2(\tau)\Bigl) \beta(\q, \k)~, \label{eq:F2_Garny}\\
   G_2(\q,\k,\tau) &= \Bigl(-\frac{1}{2} + \frac{3}{4} m_2(\tau)\Bigl) \alpha^s(\q, \k) + \Bigl(\frac{3}{2} - \frac{3}{4}m_2(\tau)\Bigl) \beta(\q, \k)~,
\end{align}
where $m_2(\tau)$ and $l_2(\tau)$ are still given by  the solution of the differential system (\ref{eq:angle_average_EOM1}, \ref{eq:angle_average_EOM2}), while $\alpha^s(\q, \k)$ is the symmetrized version of the kernel defined in (\ref{eq:alpha}).
At order $n=3$, the exact kernels take a much more complicated form. They depend on time through $l_3(\tau)$ and another function $\lambda_3(\tau)$ that solves a second-order differential equation. The kernel $F_3$ has a structure of the form
\begin{equation}
   F_3(\q_1,\q_2, \q_3, \tau) = \mathcal{R}_1 + l_2(\tau) \,  \mathcal{R}_2 + l_3(\tau) \,  \mathcal{R}_3 + \lambda_3(\tau) \, \mathcal{R}_4~, \label{eq:F3_Garny}
\end{equation}
with lengthy expressions for the four time-independent terms ${\cal R}_i(\q_1,\q_2, \q_3)$. The complete expression of $\mathcal{R}_i$ and $G_3(\q_1,\q_2, \q_3, \tau)$ can be found in \cite{Bernardeau:1993qu} (p.14-15).

Compared to the approach described in \ref{app:time_nu}, this method has the merit of being exact, but at the expense of much more complicated kernels at order $n=3$. Due to the increasing complexity of the equations with growing $n$, exact solutions have never been derived beyond order 3, while our approximate approach would in principle be easy to generalize to any order. Here, we use the exact kernels of Bernardeau only to validate the method that we presented in \ref{app:time_nu}. Note that reference \cite{Garny:2020ilv} also used the Bernardeau kernels to validate their calculation of the one-loop power spectrum based on a full numerical integration of each kernels over time for every argument $({\bf q}_1, ..., {\bf q}_n)$. The very good agreement between these two exact methods is presented in Appendix A of \cite{Garny:2020ilv}.

Assuming a $\Lambda$CDM cosmology with massless neutrinos, we computed the one-loop power spectrum in redshift space, first using the approximation of \ref{app:time_nu}, and then, replacing our kernels ($F_2$, $G_2$) with the exact ones. We find that both methods recover the correct $\Lambda$-induced correction to the EdS result, which is of the order of 0.8\%, with a relative difference of at most $10^{-10}$ for $k\leq 1\,h{\rm Mpc}^{-1}$. Therefore, our approximate scheme is extremely accurate and can be effectively considered as exact. 

\section{Additional tests\label{app:tests}}

\subsection{Backward versus forward method in the EdS limit\label{app:nu0}}

\begin{figure}
    \centering
    \begin{minipage}{\textwidth}
        \includegraphics[width=\linewidth]{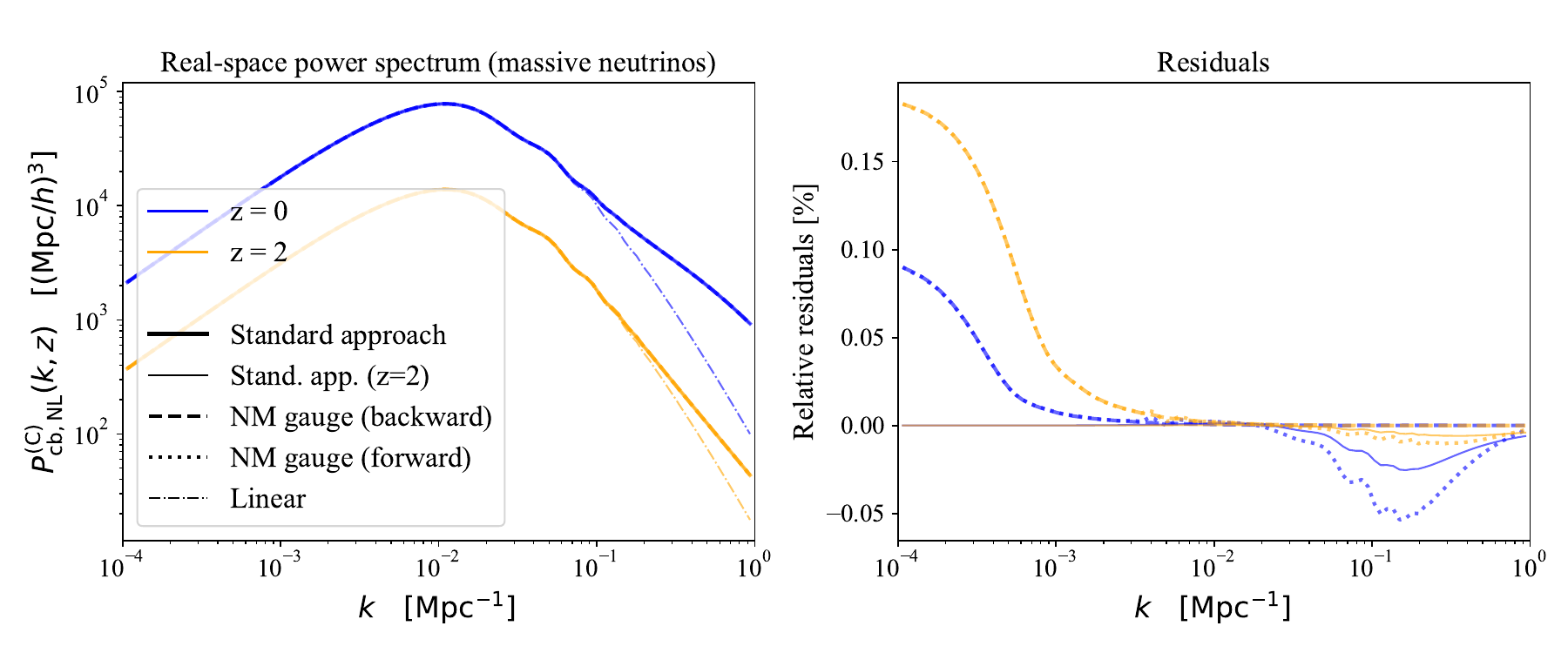}
    \end{minipage}
    \caption{Same as figure~\ref{fig:massive} but with an incomplete version of the NM gauge approach in which the kernels are still the EdS ones.
    \label{fig:massive_oldkernels}}
\end{figure}

In figure~\ref{fig:massive_oldkernels}, we compare an incomplete NM gauge approach in which, instead of $\nu$-dependent kernels, we stick to the same EdS kernels as in the standard approach.

In this case, the standard and (incomplete) NM gauge approaches should coincide, because they rely on exactly the same approximation on small scales: they assume that prior to $z=0$, perturbations grew at the same rate as in a massless neutrino universe. The two methods may differ due to the gauge transformation from the NM to comoving gauge, but this should be negligible on small scales (since we remain in the weak-field regime in which gauge transformations stay deep in the linear regime and have an impact only on large (linear) scales).

We indeed observe that the results of the backward method (dashed lines) and standard approach coincide on small scales, up to negligible ($\sim 0.001\%$) corrections. This confirms that the backward approach is accurate and compatible with the weak-field assumption. Instead, the forward method is off by $\sim 0.05\%$ (dotted lines). This is consistent with the fact that $|H_{\rm T}^{\rm (NM)}|$ becomes large in the forward method (about one order of magnitude larger than in the backward method). Thus, the weak-field approximation is not accurate in this case, and the gauge transformation cannot be described accurately at linear order.

\subsection{Effects of redshift rescaling\label{app:rescaling}}

\begin{figure}
    \centering
    \begin{minipage}{\textwidth}
        \includegraphics[width=\linewidth]{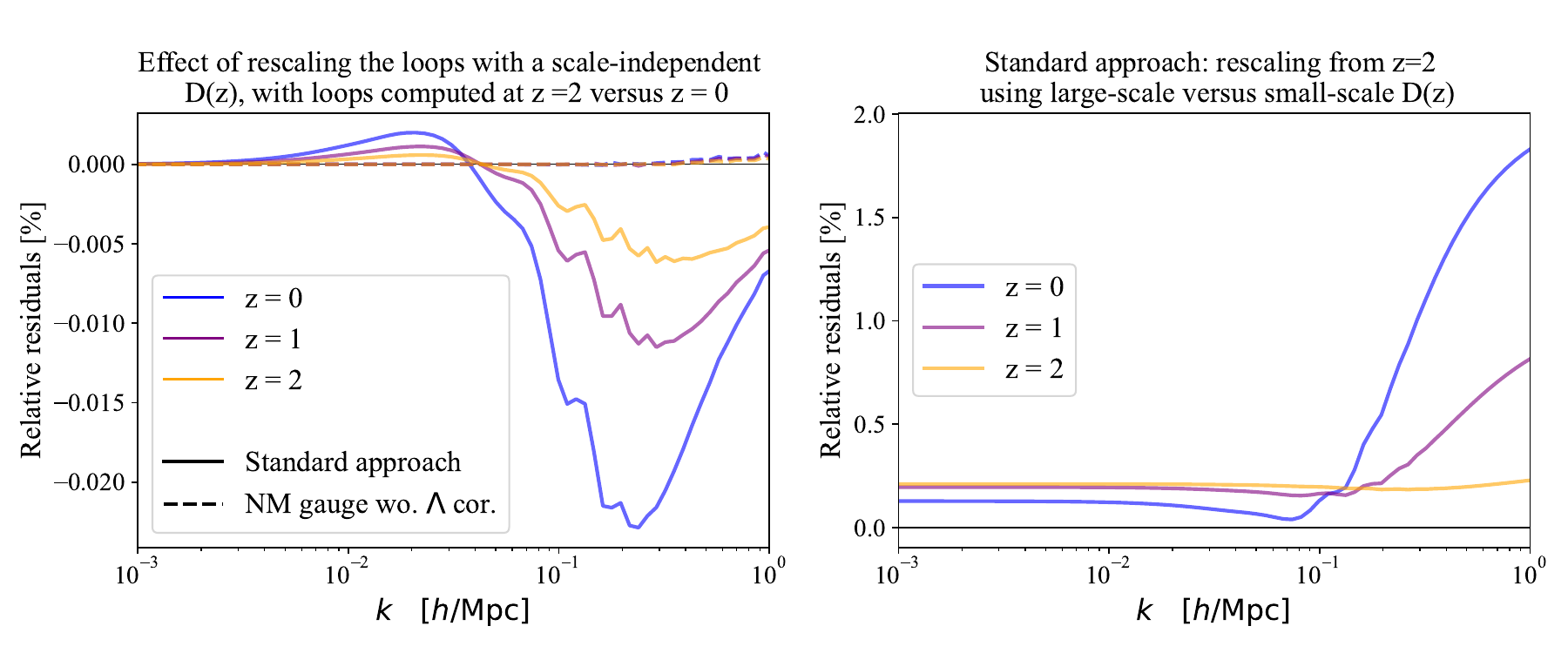}
    \end{minipage}
    \caption{(Left) In a $\Lambda$CDM cosmology with $\Sigma m_\nu=0.12\,{\rm eV}$, ratio of real-space power spectrum $P_{\rm cb,NL}^{\rm (C)}(k,z)$ with loops computed at $z=2$ and rescaled to one of the redshifts $z=0,1,2$ over the same quantity with loops computed at $z=0$ and rescaled to the same redshift, using either the standard (solid lines) or the NM gauge without $\Lambda$ correction (dashed lines) approach. (Right) For the standard approach, ratio of real-space power spectrum $P_{\rm cb,NL}^{\rm (C)}(k,z)$ with loops computed at $z=2$ and rescaled to one of the redshifts $z=0,1,2$ using the large-scale growth factor $D(z)$ (wrong choice) over the same quantity rescaled using the small-scale growth factor (better choice). In these plots, the summed neutrino mass is fixed to $\Sigma m_\nu=0.12\,{\rm eV}$.
    \label{fig:rescaling}}
\end{figure}

The purpose of the left plot in figure~\ref{fig:rescaling} is to show that the standard approach introduces a small error when the one-loop corrections to the comoving-gauge real-space power spectrum are computed at one redshift and rescaled to another redshift using a scale-independent growth factor $D(z)$. This error is however very small (in this example with $\Sigma m_\nu=0.12\,{\rm eV}$, at most 0.02\%). The same plot confirms that rescaling is performed in a fully consistent way in the `NM gauge without $\Lambda$ correction' approach, in which one first rescales the loops in the NM gauge (in which the linear growth is truly scale-independent) and then transforms to the comoving gauge. Note that in the `NM gauge with $\Lambda$ correction' approach, we never rescale the loops from one redshift to another, since the kernels $F_2$, $G_2$, $F_3$, $G_3$ are modified with a factor $\nu$ that applies to a single redshift.

In the standard approach, we always rescale the loops using the scale-independent growth factor $D(z)$ that corresponds to the small-scale limit of the scale-dependent factor $D(k,z)$. Indeed, on the scales that contribute the most to loop corrections, $k\geq 0.05 \, h\,{\rm Mpc}^{-1}$, $D(k,z)$ remains very close to this asymptotic value. This is one of the reasons for which the standard approach is such a good approximation to the more accurate NM gauge result. 

The purpose of the right plot in figure~\ref{fig:rescaling} is to stress the importance of this choice.
There, we always compute the loop corrections at $z=2$ using the standard approach, but then we rescale to $z=0,1$ using either the large-scale or small-scale $D(z)$, and we plot the ratio. The wrong rescaling from $z=0$ to $z=2$ introduces a 1.5\% error at $k= 0.3 \, h\,{\rm Mpc}^{-1}$ in this example with $\Sigma m_\nu=0.12\,{\rm eV}$.

\subsection{Prescription for the total matter power spectrum \label{app:total_matter}}

\begin{figure}
    \centering
    \begin{minipage}{\textwidth}
        \includegraphics[width=\linewidth]{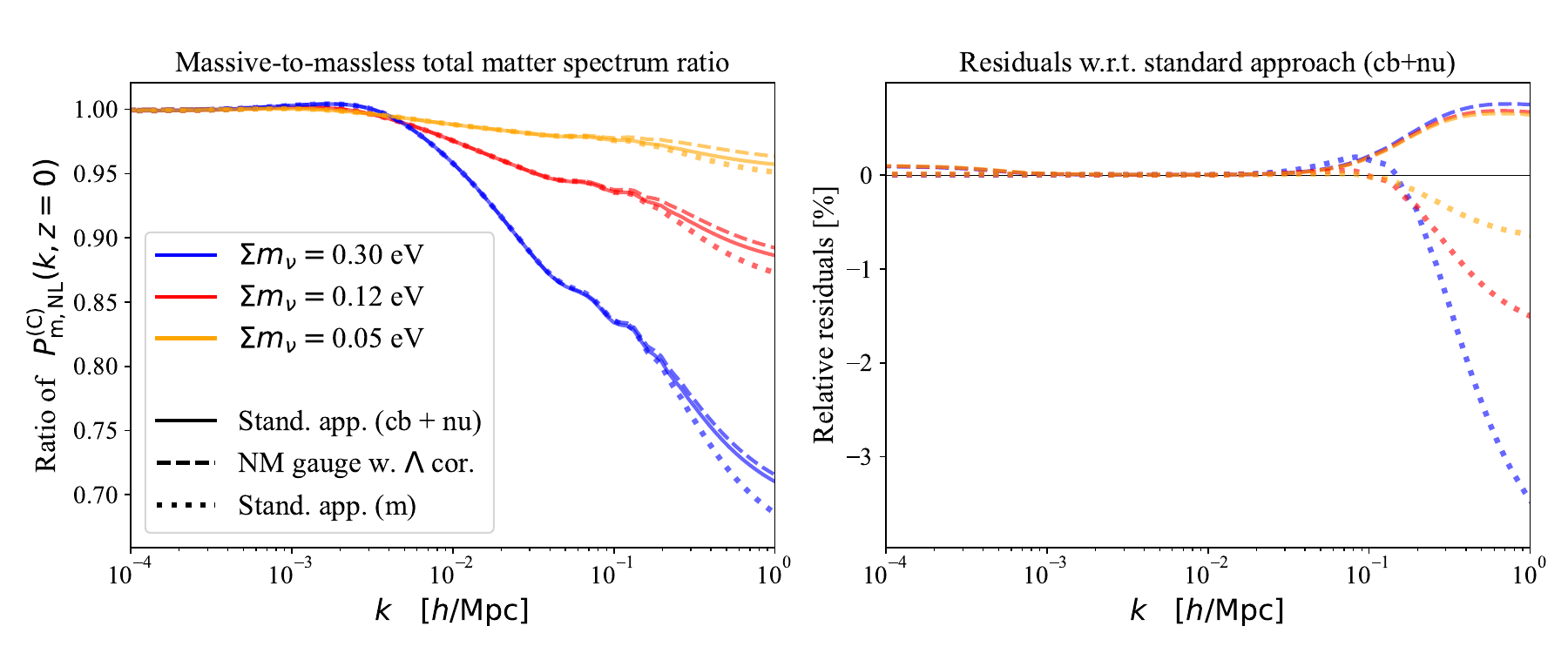}
    \end{minipage}
    \caption{(Left) Ratio of massive-to-massless matter power spectra of the comoving gauge density at $z=0$, $P^{\rm (C)}_{\rm m,NL}(k,z=0)$, in a $\Lambda$CDM cosmology with three degenerate massive neutrinos and $\Sigma m_\nu=0.05, \,0.12, \,0.3\,$eV, computed with different approaches: EFTofLSS applied to the linear comoving gauge power spectrum $P^{\rm (C),L}_{\rm cb}$, with neutrino perturbations added linearly at the end (standard approach, solid); EFTofLSS applied to the NM gauge power spectrum $P^{\rm [Nl]}_{\rm cb,L}$ (with $\Lambda$ correction) and transformed back to the comoving gauge, with neutrino perturbations added linearly at the end (our approach, dashed); and EFTofLSS applied incorrectly to the linear comoving gauge total matter power spectrum, $P^{\rm (C)}_{\rm m,L}$. (Right) Percentage difference between the massive-to-massless ratio of the NM gauge approach or standard approach applied incorrectly to total matter and that in the standard approach with correct neutrino treatment.
    \label{fig:total-matter-wrong}}
\end{figure}

To illustrate that computing the loop contributions from $P_{\rm cb,L}^{\rm (C)}$ rather than $P_{\rm m,L}^{\rm (C)}$ in the standard approach is the right choice, we show in figure \ref{fig:total-matter-wrong} the result obtained when using the latter. We compare this results to the predictions of the standard and NM gauge approaches applied to the `cb' component, with neutrino perturbations added linearly according to eq.~(\ref{eq:P_tot}). All loop calculations are performed at $z=0$, such that rescaling is unnecessary, and no growth factor is involved in these calculations. 
While all approaches show agreement on large scales, dominated by the linear power spectrum, significant discrepancies arise on small scales. For $k>0.1\,h$Mpc$^{-1}$, the power spectrum obtained using the incorrect method is too low (by about 1.2\% at  $k=0.3\,h$Mpc$^{-1}$ for $\Sigma m_\nu=0.30\,{\rm eV}$). The reason is that, on small scales, $\delta_{\rm m}^{\rm (C)}$ does not evolve like a self-gravitating Newtonian fluid, which implies that the use of EdS kernels for this particular field is a bad approximation.

\end{document}